\documentclass[twocolumn]{aastex701}

\usepackage{booktabs}
\usepackage{graphicx}
\usepackage{caption}
\usepackage{rotating}
\usepackage{placeins}
\usepackage{amsmath}

\newcommand{\Gaia}{\textit{Gaia}}

\newcommand{\feh}{\ensuremath{[\rm{Fe/H}]}}

\newcommand{\mh}{{\rm [M/H]}}
\newcommand{\sint}{\sigma_{\rm int}}

\newcommand{\dd}{\,{\rm d}}
\newcommand{\vt}{\boldsymbol{\theta}}

\newcommand{\vb}{\boldsymbol{b}}
\newcommand{\vx}{\boldsymbol{x}}
\newcommand{\vsigma}{\boldsymbol{\Sigma}}
\newcommand{\vd}{\boldsymbol{d}}

\defcitealias{Torres10}{T10}

\newcommand{\ndebcat}{387}
\newcommand{\ndebcatcomponents}{774}

\newcommand{\neb}{321}
\newcommand{\nebcomponents}{605}
\newcommand{\tefferrfill}{136}
\newcommand{\tefferrfloor}{50}
\newcommand{\ntefferrfloor}{40}

\newcommand{\nebmet}{330}
\newcommand{\nebmetsystems}{181}

\newcommand{\nsystemsnomet}{140}

\newcommand{\priorrobustmean}{-0.05}

\newcommand{\deltabic}{20.0}

\newcommand{\afe}{\ensuremath{[\alpha/\mathrm{Fe}]}}

\usepackage{lipsum}

\begin{document}

\title{Empirical Mass and Radius Relations for Main-sequence Stars from Detached Eclipsing Binaries}

\author[0000-0003-2431-981X]{D. M. Rowan}\thanks{NHFP Hubble Fellow}
\email{dmrowan@berkeley.edu}
\affiliation{Department of Astronomy, University of California, Berkeley, CA 94720, USA}

\author[0000-0001-6017-2961]{C. S. Kochanek}
\email{kochanek.1@osu.edu}
\affiliation{Department of Astronomy, The Ohio State University, 140 West 18th Avenue, Columbus, OH 43210, USA}
\affiliation{Center for Cosmology and Astroparticle Physics, The Ohio State University, 191 W. Woodruff Avenue, Columbus, OH, 43210, USA}

\correspondingauthor{D. M. Rowan (dmrowan@berkeley.edu)}

%% Use the \collaboration command to identify collaborations. This command
%% takes an optional argument that is either a number or the word "all"
%% which tells the compiler how many of the authors above the command to
%% show. For example "\collaboration[all]{(DELVE Collaboration)}" wil include
%% all the authors above this command.
%%
%% Mark off the abstract in the ``abstract'' environment. 
\begin{abstract}
Detached eclipsing binaries have long served as fundamental calibrators for stellar models because they provide precise measurements of stellar masses and radii. Given a sample of detached eclipsing binaries, we can derive empirical relations between spectroscopic parameters (effective temperatures, surface gravities and abundances) and stellar masses and radii. We use a sample of \nebcomponents{} stars (\nebmet{} with metallicities) in detached eclipsing binaries to update and refine these empirical models for main-sequence stars. An eight-term polynomial model reproduces the radii and masses of the eclipsing binary calibrators with an RMS scatter of $\sigma_{\log R} = 0.0161$~dex (3.8\%) and $\sigma_{\log M} = 0.0322$~dex (7.7\%), respectively. Most of this scatter is explained by the parameter uncertainties --- the estimated intrinsic scatters in the radius and mass are only $\sigma_{\rm int}=0.0085$ and $0.017$~dex (2.0\% and 4.0\%). The predicted mass and radius of the Sun are $(1.028\pm0.004)\ M_\odot$ and $(1.014\pm0.002)\ R_\odot$. A 14-term model, while formally preferred, explains little more of the intrinsic scatter. We additionally provide the equivalent relations in terms of the mean stellar density instead of the surface gravity for use with exoplanet host stars. If we compare the 8-term model to MIST and PARSEC model stars with closely matched spectroscopic parameters, we find systematic differences.  For example, the radii of hot (cool) stars are systematically over- (under-)estimated in both models by several percent. 
\end{abstract}

%% Keywords should appear after the \end{abstract} command. 
%% PASP uses Unified Astronomy Thesaurus (UAT) concepts:
%% https://astrothesaurus.org
%% You will be asked to selected these concepts during the submission process
%% but this old "keyword" functionality is maintained in case authors want
%% to include these concepts in their preprints.
%%
%% You can use the \uat command to link your UAT concepts back its source.
%\keywords{\uat{Galaxies}{573} --- \uat{Cosmology}{343} --- \uat{High Energy astrophysics}{739} --- \uat{Interstellar medium}{847} --- \uat{Stellar astronomy}{1583} --- \uat{Solar physics}{1476}}


%% From the front matter, we move on to the body of the paper.
%% Sections are demarcated by \section and \subsection, respectively.
%% Observe the use of the LaTeX \label
%% command after the \subsection to give a symbolic KEY to the
%% subsection for cross-referencing in a \ref command.
%% You can use LaTeX's \ref and \label commands to keep track of
%% cross-references to sections, equations, tables, and figures.
%% That way, if you change the order of any elements, LaTeX will
%% automatically renumber them.

\section{Introduction} \label{sec:intro}

\begin{figure*}
    \centering
    \includegraphics[width=\linewidth]{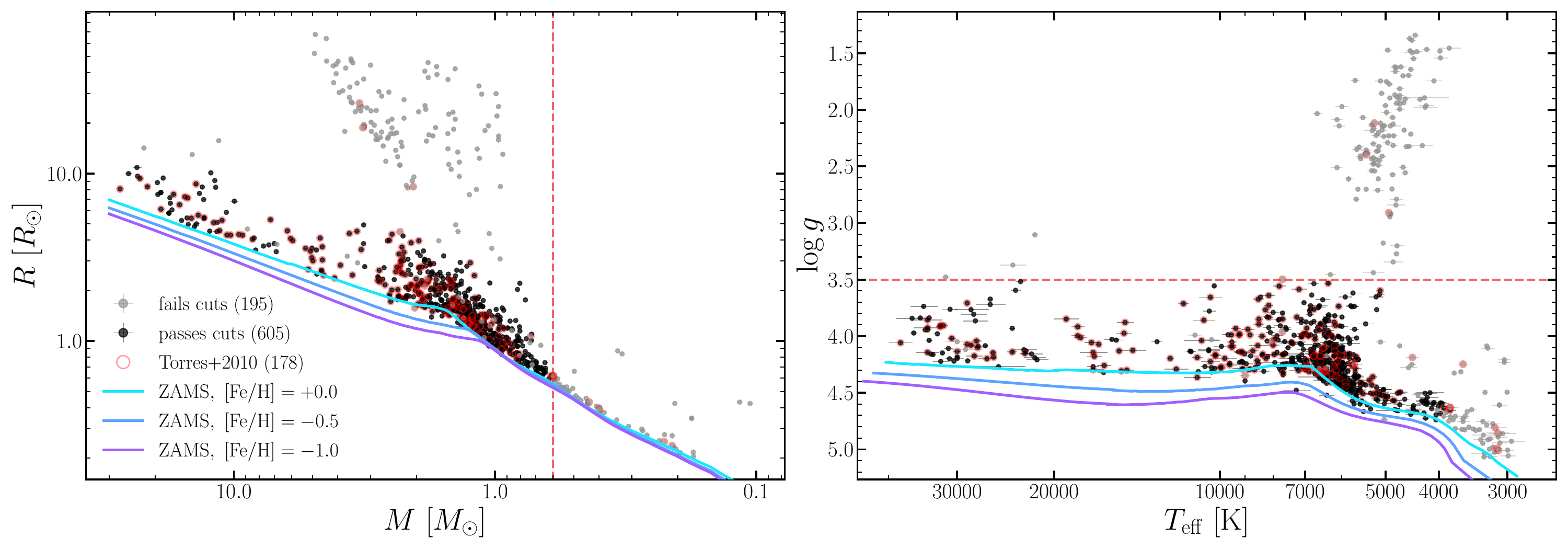}
    \caption{Masses and radii (left) and effective temperatures and surface gravities (right) of the initial sample of 800 stars from 400 detached EBs (gray/black symbols) and the \nebcomponents{} stars (black symbols) in the final EB sample (\S\ref{sec:ebs}). We indicate stars that were included in \citetalias{Torres10} with a red outline. The red lines show the $M < 0.6\ M_\odot$ cut used to remove low-mass stars (left panel) and the $\log g < 3.5$ cut used to remove evolved stars (right panel). In both panels, the colored lines show the MIST zero-age main sequences for metallicities of $\feh{}=0.0$ (top, cyan), $-0.5$ (blue) and $-1.0$ (bottom, purple). }
    \label{fig:mass_radius_kiel}
\end{figure*}

Detached eclipsing binaries (EBs) offer the most effective means of measuring accurate and precise model-independent masses and radii of stars. The light curve constrains the orbital period, eccentricity, inclination and the fractional radii of the binary components. The radial velocity model then determines the masses and the semimajor axis. By combining these constraints, we can measure the physical masses and radii of both stars with fractional uncertainties of $\sim 1$--3\% \citep[e.g.,][hereafter \citetalias{Torres10}]{Andersen91, Torres10}.

Since the stars in detached EBs have evolved effectively in isolation, the mass and radius measurements from EBs are important benchmarks for testing predictions from stellar theory \citep[e.g.,][]{Paxton13,Swayne22, Woody26}. They anchor other methods of mass estimation, such as asteroseismology \citep{Hekker10, Gaulme16}, and calibrate empirical relations between stellar parameters such as mass and luminosity and more easily measured quantities like temperature, surface gravity, and abundances \citep[e.g.,][]{Fernandes21}. These relations can then be used to estimate stellar properties where dynamical masses and radii are not available \citep[e.g.,][]{Serenelli21}, including exoplanet host stars \citep[e.g.,][]{Kuhn16}. 

\citetalias{Torres10} presented fundamental empirical relations for the masses $M$ and radii $R$ of main-sequence stars as a function of effective temperature $\log T_{\rm{eff}}$, surface gravity $\log g$ and metallicity \mh{} using a sample of 95 detached eclipsing binaries with fractional mass and radius uncertainties of $< 3\%$. The uncertainties in the temperatures, gravities and abundances were not included in deriving the relations using weighted least squares fits to $\log M$ and $\log R$, and systems without measured metallicities were assumed to have $\mh{}=0$. Nonetheless, the empirical relations could predict the masses and radii with typical residuals of 6\% and 3\%, respectively, for main-sequence stars more massive than $0.6\ M_\odot$.

Since then, the sample of detached EBs with precise mass and radius measurements has more than tripled. The detached EB catalog DEBCat \citep{Southworth15} includes \ndebcat{} binary systems as of November 2026, all of which have masses and radii with fractional measurement uncertainties of $< 2\%$. The sample of benchmark EBs has also grown considerably more diverse. Whereas there were only three evolved stars with $\log g < 3.0$ in the \citetalias{Torres10} catalog, DEBCat now includes more than 100 evolved stars identified through targeted studies of the Magellanic Clouds \citep{Graczyk18, Graczyk20} and giants in the Galactic field \citep{Rowan25}. The sample of high-mass stars has more than doubled \citep[e.g.,][]{Taormina25, Richardson21} since \citetalias{Torres10}, as has the number of binaries at low metallicity \citep[e.g.,][]{Kaluzny13, Rozyczka22, Rowan26}. There have also been targeted surveys to measure masses for the lowest-mass stars \citep[e.g.,][]{Gill19}, which have provided crucial tests of radius inflation \citep{Swayne22, Davis26}. 

The increased number of benchmark EBs with dynamical mass and radius measurements and their expanded coverage of the Hertzsprung--Russell diagram motivate revisiting the \citetalias{Torres10} empirical relations for the masses and radii of main-sequence stars. In \S\ref{sec:ebs} we define the sample of stars in EBs that we will use for the fits. In \S\ref{sec:fits} we fit the EBs with a Bayesian model that accounts for the uncertainties in the temperatures, gravities and abundances as well as those in the radii and masses. We also apply the Bayesian approach to selecting the preferred model (\S\ref{sec:adopted_models}). In \S\ref{sec:posteriors}, we compare the posteriors on the spectroscopic parameters from the Bayesian model to the measured values. In \S\ref{sec:mist}, we use synthetic populations of stars generated from the MIST \citep{Dotter16, Choi16} and PARSEC \citep{Bressan12, Nguyen22} models to compare the empirical relations to predictions from theory. We summarize the utility and limitations of empirical relations for stellar parameters in \S\ref{sec:conclusions}.  Appendix~\ref{appendix:cov} gives the results for an alternate covariance matrix between the radius and surface gravity, and Appendix~\ref{appendix:rho} provides the equivalent models for the radius in terms of the mean stellar density instead of the surface gravity.

\section{The Eclipsing Binary Sample} \label{sec:ebs}

The objective is to fit empirical models for the stellar radii $R$ and masses $M$ in terms of the effective temperature $\log T_{\rm eff}$, surface gravity $\log g$ and metallicity $\mh{}$. We used the DEBCat \citep{Southworth15} catalog of EBs, which on September 10th, 2026 included \ndebcat{} systems (\ndebcatcomponents{} stars)\footnote{Gaia DR3 4658237043035232256 has two entries in DEBCat and we use the more recent parameters from \citet{Graczyk18}.}. We added the 13 non-giant, metal-poor EBs characterized in \citet{Rowan26}, so we have a total of 400 EBs (800 stars) in the initial sample. Figure \ref{fig:mass_radius_kiel} shows the masses, radii, effective temperatures, and surface gravities of this sample.  Next we removed components with $\log g < 3.5$ or $M < 0.6\ M_\odot$. The surface gravity limit removes evolved stars, which do not follow well-defined $\log R$ and $\log M$ relations, while the mass limit removes low-mass stars where radii are inflated in short-period, rapidly rotating, magnetically active systems \citep{Kraus11}. We also remove nine pre-main-sequence binaries that would have otherwise passed the mass and surface gravity cuts.\footnote{EK Cep \citep{Popper87}; RS Cha \citep{Clausen80}; V1174 Ori \citep{Stassun04}; NP Per \citep{Lacy16}; V1200 Cen \citep{Marcadon20}; 2MASS J12220147$-$5737565 \citep{Stassun22}; ASAS J052821$+$0338.5 \citep{Stempels08}; HD~60637 \citep{Helminiak19_X}; V413~Ser \citep{Cakirli08}.} 

While most binaries in DEBCat have $T_{\rm{eff}}$ reported for both components, two systems, HD 135671 and TYC 459-771-1 \citep{Helminiak24}, do not have temperatures for the secondary, so we do not include the secondaries in our models. Four systems do not have errors on the effective temperatures, two of which are giants and were removed by our $\log g$ cut. For the other two, IR~Cas \citep{Kamenec26} and TIC 71877648 \citep{Rowan25}, we assume an effective temperature uncertainty $\sigma_{T_{\rm{eff}}} = \tefferrfill{}$~K, which is the median $\sigma_{T_{\rm{eff}}}$ of the other binaries in the sample. Unlike stellar masses and radii, which are measured independently of stellar models, effective temperature measurements require some underlying model atmosphere. They are typically measured either spectroscopically, from multi-band light curve fits, or using broadband photometry, and these methods can produce different effective temperature estimates. For example, LL~Aqr is a 20.2~day binary with an F9V and a G3V star. \citet{Graczyk16} measured the primary $T_{\rm{eff}}$ with photometric colors, line depth ratios and atmospheric analyses and found differences of $\sim 100$~K between methods. \citet{Miller26} later report effective temperatures $\sim 100$~K hotter using broadband photometry, and suggest that the difference is due to treatment of interstellar extinction. To account for potentially underestimated uncertainties and systematics between adopted methods, we set an uncertainty floor on $T_{\rm{eff}}$ of $\sigma_{T_{\rm{eff}}} = \tefferrfloor{}$~K. This only impacts \ntefferrfloor{} stars, which have reported uncertainties as low as 24~K \citep[e.g., V785 Cep,][]{Yakut25}. 

While some EBs have measured spectroscopic surface gravities, the dynamical mass and radius measurements allow us to compute $\log g$ directly. In most cases, the $\log g$ tabulated in DEBCat is already the dynamical $\log g$, but not always. For example, for EPIC 211391083 \citep{Chen26}, DEBCat reports a spectroscopic $\log g = 4.06$ for the photometric primary, but the dynamical surface gravity is $\log g = 4.2$. For all stars in DEBCat, we use the dynamical $\log g$ computed from the mass and radius when fitting the empirical relations below. 

In addition to effective temperatures and surface gravities, DEBCat also reports metallicities, when available. In some cases, independent metallicities have been measured for both components. For example, \citet{Torres12} report $\feh{} = 0.49 \pm 0.15$ and $0.61\pm 0.15$ for the two components of SW~CMa, while DEBCat reports the average of the two component metallicities, so that is the value we adopt here as well. DEBCat also reports all metallicities as \mh{}, but for some binaries \feh{} was actually measured. Again following DEBCat, we assume that all metallicities are measured as \mh{}. After making the cuts on $\log g$, low-mass stars, and pre-main-sequence stars, only \nebmetsystems{} systems have metallicity measurements in DEBCat. As discussed in \S\ref{sec:fits}, we use a Bayesian approach with a metallicity prior based on the metallicities of the \nebmetsystems{} systems with measured metallicities for those without. 

Nineteen of the DEBCat binaries have reported metallicities but are missing measured metallicity errors. Of these, fourteen are hot, massive stars in the Milky Way and have $\mh{}=0.0$. Four other systems have $\mh{}=-0.30$, three of which are in the Large Magellanic Cloud. The final one is TIC~71877648 \citep{Rowan25}, which is a low-mass star in the Milky Way with $\mh{}=0.1$. For these nineteen systems where DEBCat reports a metallicity but not a metallicity error, we adopt the median metallicity error of the sample, $\sigma_{\mh{}} = 0.09$. 

Finally, we find that the radius of GK~Dra~A is incorrectly tabulated in DEBCat. We update the radius of GK~Dra~A to the value reported in \citet{Southworth24}. We also removed BD$-$13~4929 \citep{Holgado25} because it is an extreme outlier in all of our fits as well as when using the \citetalias{Torres10} relation, where its $\log R$ residuals are $-5\sigma$ and $-7\sigma$ for the two components. 

\begin{figure}
    \centering
    \includegraphics[width=\linewidth]{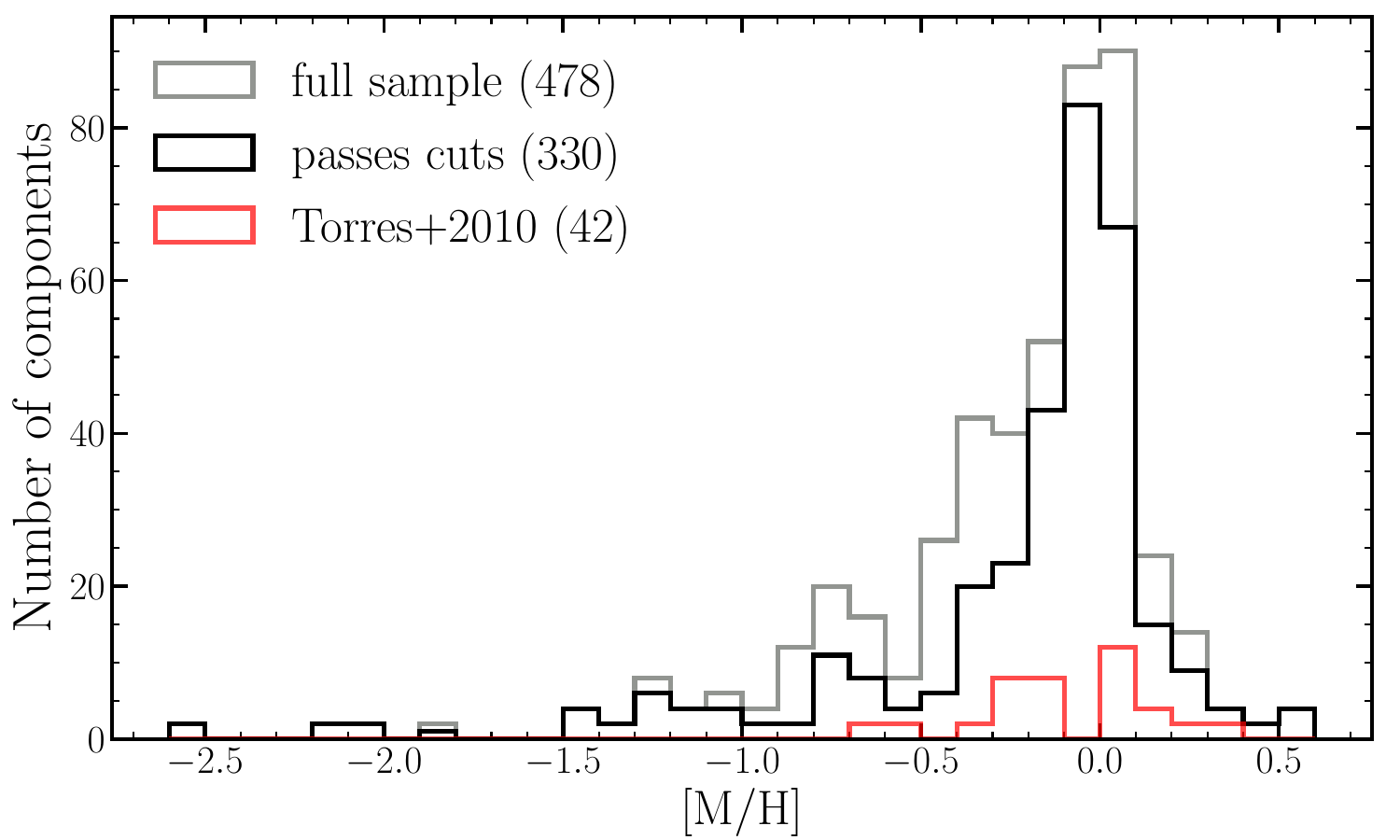}
    \caption{Distribution of metallicities for the initial detached EB sample and those that pass the $\log g$, $M$, and evolutionary state cuts. Stars without metallicity measurements are not included in the histogram. For comparison, we also show the metallicity distribution of binaries included in the \citetalias{Torres10} sample.}
    \label{fig:met_hist}
\end{figure}

After making these cuts, we are left with \nebcomponents{} stars in \neb{} EBs. Figure \ref{fig:mass_radius_kiel} shows the masses, radii, effective temperatures, and surface gravities of the stars that pass these cuts in black. Out of these \nebcomponents{} stars, \nebmet{} are in systems with reported metallicities. Figure \ref{fig:met_hist} shows the distribution of metallicities. For the remainder of this work, we call this sample of \nebcomponents{} stars the EB sample. 

\section{Modeling the Empirical Relation} \label{sec:fits}

\citetalias{Torres10} fit a sample of 95 EBs with empirical relations of the form
\begin{equation} \label{eqn:empirical_torres}
    \begin{split}
        \log M &= a_1 + a_2 X + a_3 X^2 + a_4 X^3 \\
               &\quad + a_5 (\log g)^2 + a_6 (\log g)^3 + a_7 \mh{}, \\
        \log R &= b_1 + b_2 X + b_3 X^2 + b_4 X^3 \\
               &\quad + b_5 (\log g)^2 + b_6 (\log g)^3 + b_7 \mh{},
    \end{split}
\end{equation}
for the mass and radius where $X = \log(T_{\rm{eff}}) - 4.1$. Here the mass and radius are in Solar units. They report scatters in these calibrations of $\sigma_{\log M} = 0.027$ (6.4\%) and $\sigma_{\log R} = 0.014$ (3.3\%). The sample excluded stars with $M < 0.6\ M_\odot$ and pre-main-sequence stars, but did include three evolved stars with $\log g < 3.0$. Stars without metallicities were assigned $\mh{}=0$.  In \citetalias{Torres10}, the $\log M$ and $\log R$ relations are fit independently with weighted least squares, taking into account the uncertainties on $\log M$ and $\log R$ from the eclipsing binary solutions. In \citet{Rowan26}, we updated the fit coefficients using the DEBCat binaries with the same form of Equation \ref{eqn:empirical_torres}. We also included an intrinsic scatter $\sigma_{\rm{int}}$ that is added to the measurement uncertainties in quadrature and determined by maximizing the Gaussian likelihood. These fits had residual root-mean-square scatters of $\sigma_{\log M} = 0.0357$ (8.6\%) and $\sigma_{\log R} = 0.0177$ (4.2\%), respectively.  

There are several issues to address in improving these models. First, the mass, radius and surface gravity are not independent variables because $\log M = 2\log R + \log g - \log g_\odot$ where $\log g_\odot = 4.438$ is the logarithmic surface gravity of the Sun. This is exact for the dynamical surface gravity determined from the EB model masses and radii but must also be true of spectroscopic surface gravities modulo noise and measurement biases. The scatter on the mass and radius relations should then be $\sigma_{\log M} \approx 2\times\sigma_{\log R}$, which is what \citetalias{Torres10} found. Because Equation~\ref{eqn:empirical_torres} lacks a term linear in $\log g$, this relationship between mass, radius and gravity has to be approximated by the other terms. With a linear $\log g$ term added, the terms are trivially related, with $a_i = 2 \times b_i$ except for the term linear in the surface gravity where $a_i = 2\times b_i + 1$ and the constant term where $a_i = 2 b_i - \log g_\odot$. This also means that there is a covariance between the surface gravity, mass and radius. %found, although they suggest that the higher scatter for the $\log M$ relation comes from a more complex dependence of the mass on the input parameters

\newcommand{\tablecaptiontext}{Comparison of fit coefficients for an empirical relation using Equation \ref{eqn:model_torres_plus_linearY} for three choices of priors for stars missing metallicity measurements. The $\Delta\rm{BIC}$ row gives the difference in the Bayesian information criterion between the ``Sample'' prior and the other two priors. \label{tab:prior_comp}}
\begin{deluxetable*}{lr@{}lr@{}lr@{}l}[t]
\tablecaption{\tablecaptiontext}
\tablehead{
\colhead{Parameter} & \multicolumn{2}{c}{Sample} & \multicolumn{2}{c}{Robust sample} & \multicolumn{2}{c}{Solar} \\
\colhead{} & \multicolumn{2}{c}{$\mathcal{N}(-0.21, 0.46)$} & \multicolumn{2}{c}{$\mathcal{N}(-0.05, 0.24)$} & \multicolumn{2}{c}{$\mathcal{N}(+0.00, 0.10)$}
}
\startdata
$b_{1}\;({\rm const})$ & $0.167$ & ${}\pm0.001$ & $0.1681$ & ${}\pm0.0009$ & $0.1686$ & ${}\pm0.0009$ \\
$b_{2}\;(X)$ & $0.68$ & ${}\pm0.02$ & $0.68$ & ${}\pm0.01$ & $0.68$ & ${}\pm0.01$ \\
$b_{3}\;(X^2)$ & $0.00$ & ${}\pm0.06$ & $-0.04$ & ${}\pm0.06$ & $-0.05$ & $^{+0.06}_{-0.05}$ \\
$b_{4}\;(X^3)$ & $0.19$ & ${}\pm0.07$ & $0.24$ & ${}\pm0.07$ & $0.25$ & ${}\pm0.06$ \\
$b_{5}\;(Y)$ & $-0.589$ & ${}\pm0.005$ & $-0.588$ & ${}\pm0.005$ & $-0.587$ & ${}\pm0.005$ \\
$b_{6}\;(Y^2)$ & $0.08$ & ${}\pm0.01$ & $0.09$ & ${}\pm0.01$ & $0.09$ & ${}\pm0.01$ \\
$b_{7}\;(Y^3)$ & $0.06$ & ${}\pm0.03$ & $0.06$ & ${}\pm0.02$ & $0.06$ & $^{+0.02}_{-0.03}$ \\
$b_{8}\;(Z)$ & $0.052$ & $^{+0.002}_{-0.001}$ & $0.056$ & ${}\pm0.002$ & $0.058$ & ${}\pm0.002$ \\
$\sigma_{\rm int}$ [dex] & $0.0100$ & ${}\pm0.0005$ & $0.0091$ & $^{+0.0006}_{-0.0005}$ & $0.0101$ & ${}\pm0.0006$ \\
\hline
RMS $\log R$ [dex] & \multicolumn{2}{c}{$0.0161$} & \multicolumn{2}{c}{$0.0161$} & \multicolumn{2}{c}{$0.0161$} \\
$R_\odot$ predicted [$R_\odot$] & \multicolumn{2}{c}{$1.016\pm0.003$} & \multicolumn{2}{c}{$1.018\pm0.002$} & \multicolumn{2}{c}{$1.019\pm0.002$} \\
$M_\odot$ predicted [$M_\odot$] & \multicolumn{2}{c}{$1.031\pm0.006$} & \multicolumn{2}{c}{$1.036\pm0.005$} & \multicolumn{2}{c}{$1.039\pm0.005$} \\
$\Delta$BIC & \multicolumn{2}{c}{$0.0$} & \multicolumn{2}{c}{$-114.1$} & \multicolumn{2}{c}{$-122.4$} \\
\enddata
\end{deluxetable*}

Second, when we revised the \citetalias{Torres10} fits in \citet{Rowan26}, the fit coefficients showed large changes. This is in large part driven by not centering the $\log g$ terms in the parameter range of the data (i.e., centered at $\log g = 0$ in Equation~\ref{eqn:empirical_torres} instead of near the median of $\log g \simeq 4.2$, see Figure~\ref{fig:mass_radius_kiel}). This also makes it difficult to compare the empirical model to theoretical models and leads to strong covariances between the coefficients. Third, the temperatures, metallicities and surface gravities also have uncertainties which were not accounted for, and for dynamical surface gravities these are also correlated with the mass and radius. Fourth, there is the problem of optimally choosing the form of the polynomial. A general third-order polynomial in three quantities has 20 possible terms while Equation~\ref{eqn:empirical_torres} uses only 7. Here we use a more quantitative approach to selecting the form of the model.

We address these issues with a Bayesian approach. First, we use a model including a term linear in $\log g$ so that we need only fit a model for $\log R$ and the model for $\log M$ simply follows from the relation between radius, mass and surface gravity.  Second, we use polynomials that are ``centered'' in the data, $X \equiv \log T_{\rm{eff}} - X_0$, $Y \equiv \log g - Y_0$, and $Z \equiv \mh{} - Z_0$ with $X_0 = 3.8$, $Y_0 = 4.2$, and $Z_0 = 0.0$ near the sample medians. This will make the values of the coefficients much more robust to changes in the data being modeled and significantly reduce the covariances between parameter values. Third, we distinguish between the measured, noisy input quantities $\vx_i = (X_i,Y_i,Z_i)$ and their ``true'' values $\vt_i = (X_{T_i}, Y_{T_i}, Z_{T_i})$.  This allows us to incorporate the uncertainties in the temperature, metallicity and gravity into the models. Fourth, we use Bayesian model selection methods to select the optimal polynomial model.  

The model consists of the $\vt_i$ for each star, the polynomial coefficients $\vb$ and the intrinsic scatter in the radius $\sint$ beyond the estimated statistical uncertainties. Here we abbreviate the logarithm of the radius as $R_i \pm \sigma_{R_i}$ and the logarithm of the mass as $M_i \pm \sigma_{M_i}$. If we were using spectroscopic gravities, the log-likelihood $\ln P_i$ for a given star would be
\begin{equation}
    \begin{split}
        -2\ln P_i(R_i, \vx_i \mid \vt_i, \vb, \sint)
        &= \biggl[\frac{\big(R_i - R_M(\vt_i,\vb)\big)^2}{\sigma_{R_i}^2+\sint^2} \\
        &\hspace{-10em} + \ln\left(\sigma_{R_i}^2+\sint^2\right)
        + \frac{(X_i - X_{T_i})^2}{\sigma_{X_i}^2} \\
        &\hspace{-10em} + \frac{(Y_i - Y_{T_i})^2}{\sigma_{Y_i}^2}
        + \frac{(Z_i - Z_{T_i})^2}{\sigma_{Z_i}^2}\biggr],
    \end{split}
    \label{eq:star}
\end{equation}

\noindent where $R_M$ is the predicted logarithmic radius given $\vt_i$ and the fit coefficients, $\vb$, and $\sint$ is the intrinsic scatter in the radius. However, since $\log g$ is dynamical, $R_i$ and $Y_i$ are correlated with
\begin{equation} \label{eq:twodgaussian}
    P_i(R_i, Y_i \mid \vt_i, \vb, \sint) = \frac{1}{2\pi \sqrt{\det{\vsigma_i}}} \exp\left[-\frac{1}{2} \vd_i^{\rm{T}} \vsigma_i^{-1} \vd_i\right],
\end{equation}
\noindent where $\vd_i$ is
\begin{equation}
    \vd_i = 
    \begin{pmatrix}
    R_i - R_M(\vt_i, \vb) \\
    Y_i - Y_{T_i}
    \end{pmatrix},
\end{equation}
and $\vsigma_i$ is
\begin{equation} \label{eqn:cov}
    \vsigma_i =
    \begin{pmatrix}
    \sigma_{R_i}^2 + \sint^2 & -2\sigma_{R_i}^2 \\
    -2\sigma_{R_i}^2 & \sigma_{M_i}^2 + 4\sigma_{R_i}^2 
    \end{pmatrix},
\end{equation}
\noindent where $\sigma_{M_i}$ is the uncertainty on the logarithm of the mass. We have assumed that the masses and radii from the eclipsing binary solution are uncorrelated except through $\log g$. However, we note that this is not strictly true in practice, since both depend on the orbital properties and velocity semi-amplitudes. Appendix \ref{appendix:cov} gives the results for an alternate form of the covariance matrix that includes the correlations between the measurements of $R$ and $M$. The log-probability for a single star is then
\begin{equation} \label{eq:starcorr}
    \begin{split}
        -2\ln P_i(R_i, \vx_i \mid \vt_i, \vb, \sint)
        &= \biggl[\boldsymbol{d}_i^{\rm T}\, \boldsymbol{\Sigma}_i^{-1}\, \boldsymbol{d}_i
        + \ln\det\boldsymbol{\Sigma}_i \\
        &\hspace{-10em} + \frac{(X_i - X_{T_i})^2}{\sigma_{X_i}^2}
        + \frac{(Z_i - Z_{T_i})^2}{\sigma_{Z_i}^2}\biggr].
    \end{split}
\end{equation}
We need to marginalize the likelihood over the true values $\vt_i$ by doing the triple integral over $X_T$, $Y_T$, and $Z_T$ for each star
\begin{equation} \label{eq:triplecorr}
    \begin{split}
        P_i(R_i, \vx_i \mid \vb, \sint)
        &= \iiint P_i(R_i, \vx_i \mid \vt_i, \vb, \sint) \\
        &\hspace{-1em} \times \pi(\vt_i)\; \dd X_{T_i}\, \dd Y_{T_i}\, \dd Z_{T_i},
    \end{split}
\end{equation}
\noindent where $\pi (\vt_i)$ is the prior on the stellar parameters. For $N$ stars, the individual stars are independent, so the likelihood is the product of the per-star integrals, 
\begin{equation}
  P({\rm data} \mid \vb, \sint) = \prod_{i=1}^{N} \int P_i(R_i, \vx_i \mid \vt_i, \vb, \sint)\; \pi(\vt_i) \dd^3\vt_i .
  \label{eq:marginal}
\end{equation}
However, while the effective temperatures and surface gravities are independent for the two stars in the binary, the metallicity is shared, assuming that the two stars are co-natal and the surface abundance reflects the true metal content. We can therefore write the latent variable $Z_T$ as a system-level parameter, $Z_{T_s}$, and redefine the per-star log-probability as 
\begin{equation} \label{eq:starshared}
    \begin{split}
        -2\ln P_i(R_i, X_i, Y_i \mid X_{T_i}, Y_{T_i}, Z_{T_s}, \vb, \sint)
        &= \biggl[\boldsymbol{d}_i^{\rm T}\, \boldsymbol{\Sigma}_i^{-1}\, \boldsymbol{d}_i \\
        &\hspace{-10em} + \ln\det\boldsymbol{\Sigma}_i
        + \frac{(X_i - X_{T_i})^2}{\sigma_{X_i}^2}\biggr],
    \end{split}
\end{equation}
\noindent where $Z_{T_s}$ enters through the model radius $R_M(X_{T_i}, Y_{T_i}, Z_{T_s}, \vb)$ inside $\vd_i$. The per-system likelihood is then 
\begin{equation} \label{eqn:likelihood}
    \begin{split}
        \mathcal{L}_s(\vb, \sint)
        ={}& \int \dd Z_{T_s}\; \pi(Z_{T_s})\, P(Z_s \mid Z_{T_s}) \\
        &\times \prod_{i \in s} \int
        \dd X_{T_i}\, \dd Y_{T_i}\;
        \pi(X_{T_i})\,\pi(Y_{T_i}) \\
        &\times
        P_i(R_i, X_i, Y_i \mid X_{T_i}, Y_{T_i}, Z_{T_s}, \vb, \sint),
    \end{split}
\end{equation}
\noindent where the product is for the stars $i$ that are in the binary system, $s$ ($i \in s$), $Z_s$ is the measured metallicity for the system, and 
\begin{equation} \label{eq:zsystem}
    -2\ln P(Z_s \mid Z_{T_s}) = \frac{(Z_s - Z_{T_s})^2}{\sigma_{Z_s}^2}. 
\end{equation}
Treating the latent $Z_{T_s}$ as a shared system-level parameter significantly reduces the number of latent parameters and prevents the fit from reducing the residuals by moving the two components of a binary to different metallicities. The log-likelihood for the sample is then 
\begin{equation} \label{eqn:loglike}
  \ln \mathcal{L}(\vb, \sint) = \sum_{{\rm systems}\ s} \ln \mathcal{L}_s(\vb, \sint).
\end{equation}
Since we marginalize over the latent variables $X_T$, $Y_T$, $Z_{T_s}$, the final log-likelihood of a model only depends on the fit coefficients, $\vb$, and the intrinsic scatter, $\sint$. 

\begin{figure*}
    \centering
    \includegraphics[width=\linewidth]{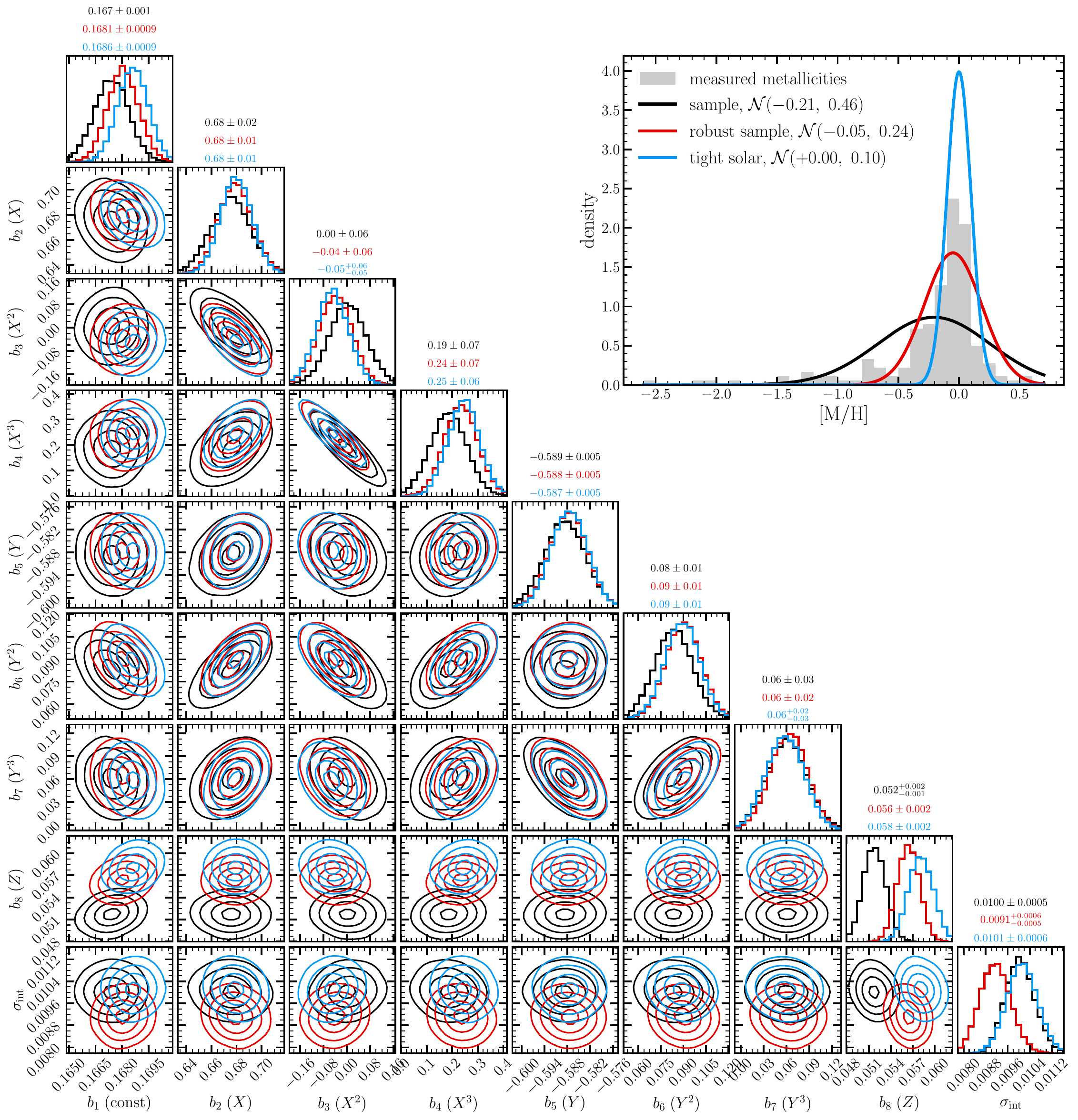}
    \caption{Comparison of the Equation~\ref{eqn:model_torres_plus_linearY} parameter posteriors for the  three choices of priors for stars without metallicity measurements. The top right shows the three priors compared to the sample of EBs with metallicities reported in DEBCat. The choice of metallicity priors most strongly impacts the linear in $Z$ coefficient $b_8$.}
    \label{fig:torres_form_prior_compare}
\end{figure*}

\renewcommand{\tablecaptiontext}{Fit coefficients for the two selected models (Equations \ref{eqn:8term} and \ref{eqn:14term}). We fit the $\log R$ relations and derive the corresponding $\log M$ relations. \label{tab:model_coeffs}}
\newcommand{\figurecaptiontext}{We search for empirical models with up to 17 terms using the marginal likelihood (Equation \ref{eqn:loglike}). Left: we show the difference between the BIC of each model and the BIC of the model with the lowest BIC. Each trial model is shown as a gray point, and the black points connect the model preferred by the data for the different numbers of terms. Right: the $\Delta\rm{BIC}$ cost of adding or removing terms from the 14-term model. $X^2$ and $YZ$ are best supported. The other higher-order terms offer only modest improvements. \label{fig:bic_search}}
\newcommand{\tablefigurepath}{ANC/04_search_summary.pdf}
\begin{deluxetable*}{lr@{}lr@{}lr@{}lr@{}l}[t]
\tablecaption{\tablecaptiontext}
\tablehead{
\colhead{} & \multicolumn{4}{c}{8-term} & \multicolumn{4}{c}{14-term} \\
\cline{2-5} \cline{6-9}
\colhead{Term} & \multicolumn{2}{c}{$\log R$} & \multicolumn{2}{c}{$\log M$} & \multicolumn{2}{c}{$\log R$} & \multicolumn{2}{c}{$\log M$}
}
\startdata
${\rm const}$ & $0.1688$ & ${}\pm0.0009$ & $0.100$ & ${}\pm0.002$ & $0.1685$ & ${}\pm0.0009$ & $0.099$ & ${}\pm0.002$ \\
$X$ & $0.63$ & ${}\pm0.01$ & $1.26$ & ${}\pm0.02$ & $0.69$ & ${}\pm0.02$ & $1.37$ & ${}\pm0.03$ \\
$Y$ & $-0.593$ & ${}\pm0.004$ & $-0.186$ & $^{+0.007}_{-0.008}$ & $-0.595$ & ${}\pm0.004$ & $-0.190$ & ${}\pm0.008$ \\
$Z$ & $0.066$ & ${}\pm0.003$ & $0.133$ & ${}\pm0.007$ & $0.064$ & ${}\pm0.003$ & $0.128$ & ${}\pm0.006$ \\
$X^2$ & $0.19$ & ${}\pm0.02$ & $0.38$ & ${}\pm0.04$ & $-0.13$ & ${}\pm0.07$ & $-0.3$ & ${}\pm0.1$ \\
$Y^2$ & $0.047$ & ${}\pm0.009$ & $0.09$ & ${}\pm0.02$ & $0.07$ & ${}\pm0.01$ & $0.13$ & ${}\pm0.02$ \\
$Z^2$ & $0.007$ & ${}\pm0.002$ & $0.013$ & ${}\pm0.004$ & $0.007$ & ${}\pm0.002$ & $0.014$ & ${}\pm0.004$ \\
$Y Z$ & $-0.028$ & ${}\pm0.005$ & $-0.055$ & $^{+0.009}_{-0.010}$ & $-0.008$ & ${}\pm0.007$ & $-0.02$ & ${}\pm0.01$ \\
$X^3$ & \multicolumn{2}{c}{\nodata} & \multicolumn{2}{c}{\nodata} & $0.35$ & $^{+0.07}_{-0.08}$ & $0.7$ & $^{+0.1}_{-0.2}$ \\
$X Y$ & \multicolumn{2}{c}{\nodata} & \multicolumn{2}{c}{\nodata} & $-0.09$ & $^{+0.04}_{-0.03}$ & $-0.17$ & $^{+0.07}_{-0.06}$ \\
$X Y^2$ & \multicolumn{2}{c}{\nodata} & \multicolumn{2}{c}{\nodata} & $-0.35$ & ${}\pm0.06$ & $-0.7$ & ${}\pm0.1$ \\
$X Z$ & \multicolumn{2}{c}{\nodata} & \multicolumn{2}{c}{\nodata} & $-0.20$ & ${}\pm0.04$ & $-0.40$ & $^{+0.08}_{-0.09}$ \\
$X^2 Z$ & \multicolumn{2}{c}{\nodata} & \multicolumn{2}{c}{\nodata} & $0.6$ & ${}\pm0.1$ & $1.3$ & ${}\pm0.3$ \\
$X Y Z$ & \multicolumn{2}{c}{\nodata} & \multicolumn{2}{c}{\nodata} & $0.8$ & ${}\pm0.1$ & $1.5$ & ${}\pm0.2$ \\
\hline
$\sigma_{\rm int}$ [dex] & \multicolumn{2}{c}{$0.0085\pm0.0005$} & \multicolumn{2}{c}{$0.017\pm0.001$} & \multicolumn{2}{c}{$0.0077\pm0.0005$} & \multicolumn{2}{c}{$0.015\pm0.001$} \\
RMS [dex] & \multicolumn{2}{c}{$0.0161$} & \multicolumn{2}{c}{$0.0322$} & \multicolumn{2}{c}{$0.0155$} & \multicolumn{2}{c}{$0.0310$} \\
Sun & \multicolumn{2}{c}{$1.014\pm0.002$ $R_\odot$} & \multicolumn{2}{c}{$1.028\pm0.004$ $M_\odot$} & \multicolumn{2}{c}{$1.013\pm0.002$ $R_\odot$} & \multicolumn{2}{c}{$1.025\pm0.005$ $M_\odot$} \\
\enddata
\vspace{2em}
\centering
\includegraphics[width=\linewidth]{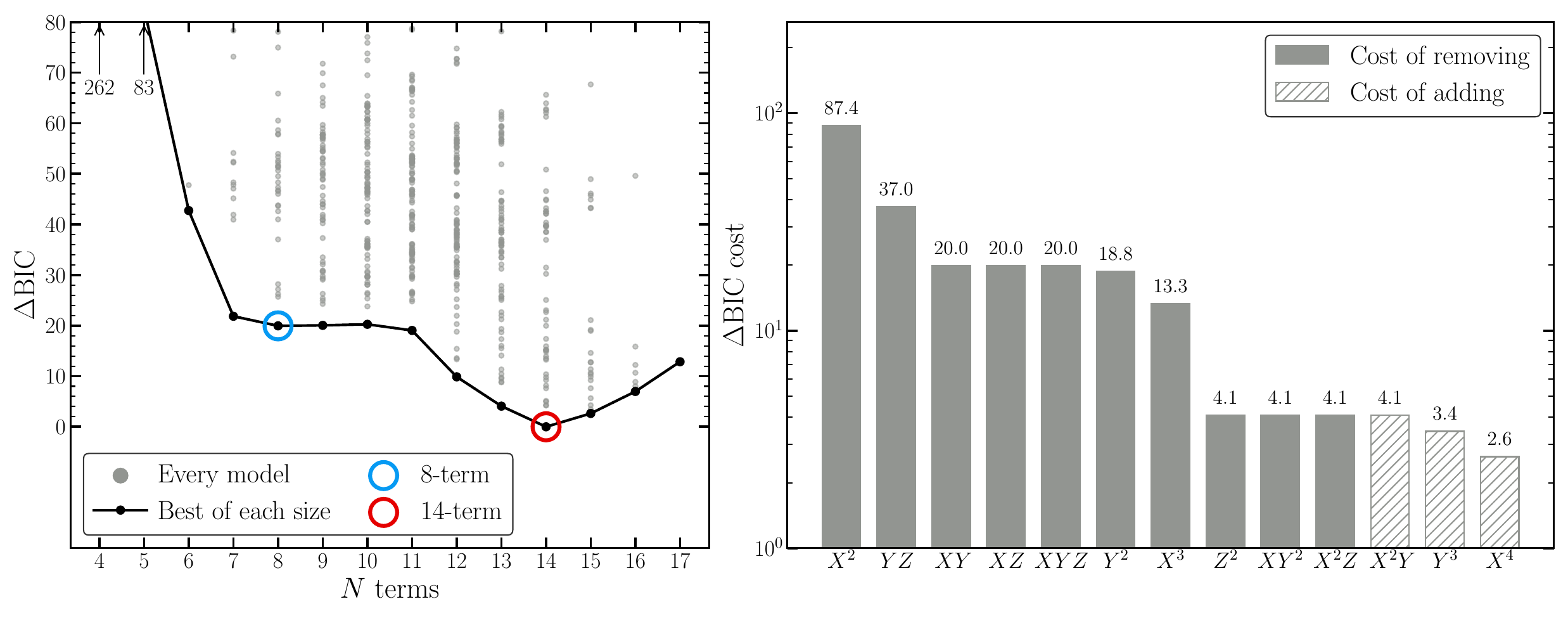}
\captionof{figure}{\figurecaptiontext}
\end{deluxetable*}

We evaluate the integrals for each system numerically using Gauss--Hermite quadratures with nine nodes for each axis. 

For $X_T$ and $Y_T$, we use broad uniform priors, and the Gaussian likelihood factors are much narrower than the prior. We also use a broad flat prior for $Z_{T_s}$ for the \nebmetsystems{} systems with a measured metallicity. For the systems without measured metallicities, we assume that $Z_s$ is unknown ($P(Z_s \mid Z_{T_s}) = 1$) but $Z_T$ is constrained by a Gaussian prior with a mean and standard deviation taken from the population. We tested three Gaussian priors $\mathcal{N}(\mu,\sigma)$ with mean $\mu$ and dispersion $\sigma$ for the metallicities. The ``Sample'' prior uses the measured mean  $\mu = -0.21$ and dispersion $\sigma = 0.46$ of the metallicities. The ``Robust'' prior downweights the tails of the metallicity distribution by using the median metallicity $\mu=\priorrobustmean{}$ and $\sigma = \rm{IQR}/1.349=0.24$, where IQR is the interquartile range (the difference between the 75th and 25th percentiles of the metallicities). In particular, this prior says that it is unlikely that any stars with unmeasured metallicities come from the low-metallicity tail. Finally, the ``Solar'' prior has $\mu = 0$ and $\sigma=0.1$. The majority of stars without metallicity measurements are hot ($T_{\rm{eff}} > 6500$~K) and so are on average likely younger and closer to Solar metallicity. Figure~\ref{fig:torres_form_prior_compare} compares these three priors to the distribution of $\mh{}$ from binaries that have measured metallicities.  

\begin{figure*}
    \centering
    \includegraphics[width=\linewidth]{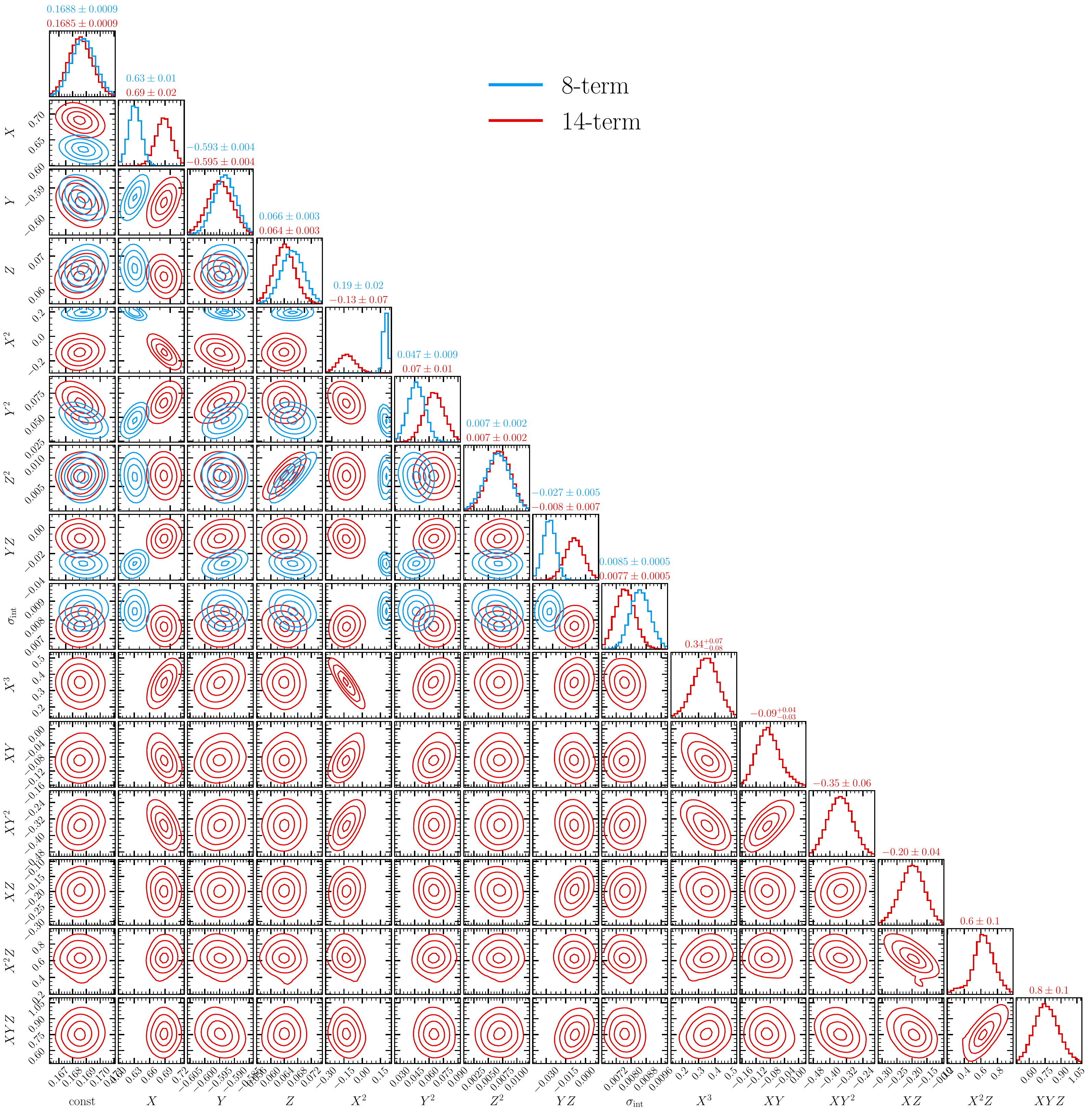}
    \caption{Posteriors for coefficients of the 8- and 14-term models. While the coefficients of some terms like $Z^2$ are consistent between the two models, others, like $X^2$, differ at the few-$\sigma$ level because the less complex model has to model the mean curvature over the full parameter space while the more complex model allows the curvature to vary. }
    \label{fig:corner_overlay}
\end{figure*}

For each model, we first solve for the coefficients $\vb{}$ with Nelder--Mead optimization to find the maximum a posteriori solution using Equation \ref{eqn:loglike}. Then we use the {\tt emcee} \citep{ForemanMackey13} Markov chain Monte Carlo (MCMC) package to sample the likelihood marginalized over the stellar properties for the coefficients $\vb$ and the intrinsic scatter $\sigma_{\rm{int}}$. We use 64 walkers and run the chains for 20{,}000 iterations, discarding the first 1{,}000 as burn-in. We use a prior on the intrinsic scatter
\begin{equation}
    p(\sint) \propto \frac{1}{\sqrt{\bar{\sigma}_R^2 + \sint^2}},
\end{equation}
\noindent where $\bar{\sigma}_R$ is the median of the $\sigma_{R_i}$ for the sample. This naturally shifts from a uniform prior for $\sint$ when it is smaller than $\bar{\sigma}_R$ to a logarithmic prior when it is larger. The resulting chains have integrated autocorrelation times, $\tau$, such that $\tau < N/50 $, where $N$ is the number of post-burn-in iterations.

\section{Adopted Models} \label{sec:adopted_models}

We first fit a model of the form 
\begin{equation} \label{eqn:model_torres_plus_linearY}
    \begin{split}
        \log R &= b_1 + b_2 X + b_3 X^2 + b_4 X^3 \\
               &\quad + b_5 Y + b_6 Y^2 + b_7 Y^3 + b_8 Z,
    \end{split}
\end{equation}
which is the same as the form in \citetalias{Torres10} (Equation \ref{eqn:empirical_torres}) except for the inclusion of the linear $\log g$ term $b_5 Y$. The model for $\log M$ simply requires the coefficient conversions discussed above ($a_i = 2 b_i$ except $a_1 = 2 b_1 +Y_0 - \log g_\odot$ and $a_5 = 2 b_5 + 1$) and will have twice the log residuals. Figure \ref{fig:torres_form_prior_compare} shows the posteriors for this model with the three choices for the metallicity prior $\pi(Z_T)$ and Table \ref{tab:prior_comp} reports the fit coefficients. The choice of the metallicity prior mostly affects the coefficient linear in $Z$, $b_8$, but does not affect the overall RMS scatter in the relation or the estimate of the intrinsic scatter. All three models match the properties of the Sun to $\sim 3\%$ in radius and $\sim 4\%$ in mass. The ``Sample'' prior, which uses the measured mean and dispersion from the full sample of measured metallicities, has a significantly worse Bayesian information criterion (BIC) value (see Table~\ref{tab:prior_comp}),
\begin{equation}
      {\rm BIC} = k \ln n - 2 \ln \mathcal{L},
\end{equation} 
where $k$ is the number of free parameters and $n$ is the number of stars. The other two priors give comparable BICs, with the ``Solar'' prior preferred over the ``Robust'' prior by $\Delta\rm{BIC} \approx 8$. We adopt the slightly broader ``Robust'' prior, which downweights the tails of the metallicity distribution, as the default prior for the rest of the analysis. We motivate this choice further in \S\ref{sec:posteriors}.

Rather than assume the form of the \citetalias{Torres10} relation, we search for the polynomial that is best supported by the data. We start with a simple model $\log R = b_1 + b_2 X + b_3 Y + b_4 Z$ and add terms up to quartic in $X$, cubic in $Y$, and quadratic in $Z$, including cross-terms (e.g., $XY$). For this search, we use Nelder--Mead to maximize the likelihood marginalized over the stellar parameters (Equation \ref{eqn:loglike}). We do not do the full MCMC sampling here because of the  computational cost of integrating over the latent variables at each step. We only consider models that include the relevant parent terms. For example, we only test models with an $X^3$ term if an $X^2$ term is also included. 

Figure \ref{fig:bic_search} shows the BIC for each model. We tested including terms up to the order limits given above. We find the lowest BIC for a 14-term model, but find only small differences $\Delta\rm{BIC} \lesssim 12$ between the 14-, 13- and 12-term models. The models with 7--11 terms also have similar $\rm{BIC}$, and are only $\Delta\rm{BIC}\approx\deltabic{}$ higher than the 14-term model. We therefore also adopt a simpler 8-term model and do the full MCMC sampling for these two models. The 8-term model is
\begin{equation} \label{eqn:8term}
    \begin{split}
        \log R &= b_1 + b_2 X + b_3 Y + b_4 Z \\
               &\quad + b_5 X^2 + b_6 Y^2 + b_7 Z^2 + b_8 YZ,
    \end{split}
\end{equation}
\noindent and the 14-term model is
\begin{equation} \label{eqn:14term}
    \begin{split}
        \log R ={}& c_1 + c_2 X + c_3 Y + c_4 Z + c_5 X^2 \\
                  & + c_6 Y^2 + c_7 Z^2 + c_8 YZ + c_9 X^3 + c_{10} XY \\
                  & + c_{11} XY^2 + c_{12} XZ + c_{13} X^2 Z + c_{14} XYZ.
    \end{split}
\end{equation}
\noindent The 14-term model includes an additional cubic term in $X$, as well as additional cross-terms between $X$, $Y$, and $Z$. Figure \ref{fig:corner_overlay} compares the posteriors between the two models and Table \ref{tab:model_coeffs} reports their coefficients. We find that the coefficients of the shared constant, $Y$, $Z$, $Y^2$, and $Z^2$ terms agree at the $\lesssim 2\sigma$ level. The coefficients of the $X$ and $X^2$ terms do change between the models, but this can be attributed to the cubic term included in the expanded model, which introduces degeneracies between the coefficients of the different powers of $X$. The coefficient of the $YZ$ term also changes between the two models, and this can be explained by the addition of the $XYZ$ term in the 14-term model, which allows for the metallicity dependence to also be temperature-dependent. 

\begin{figure*}
    \centering
    \includegraphics[width=\linewidth]{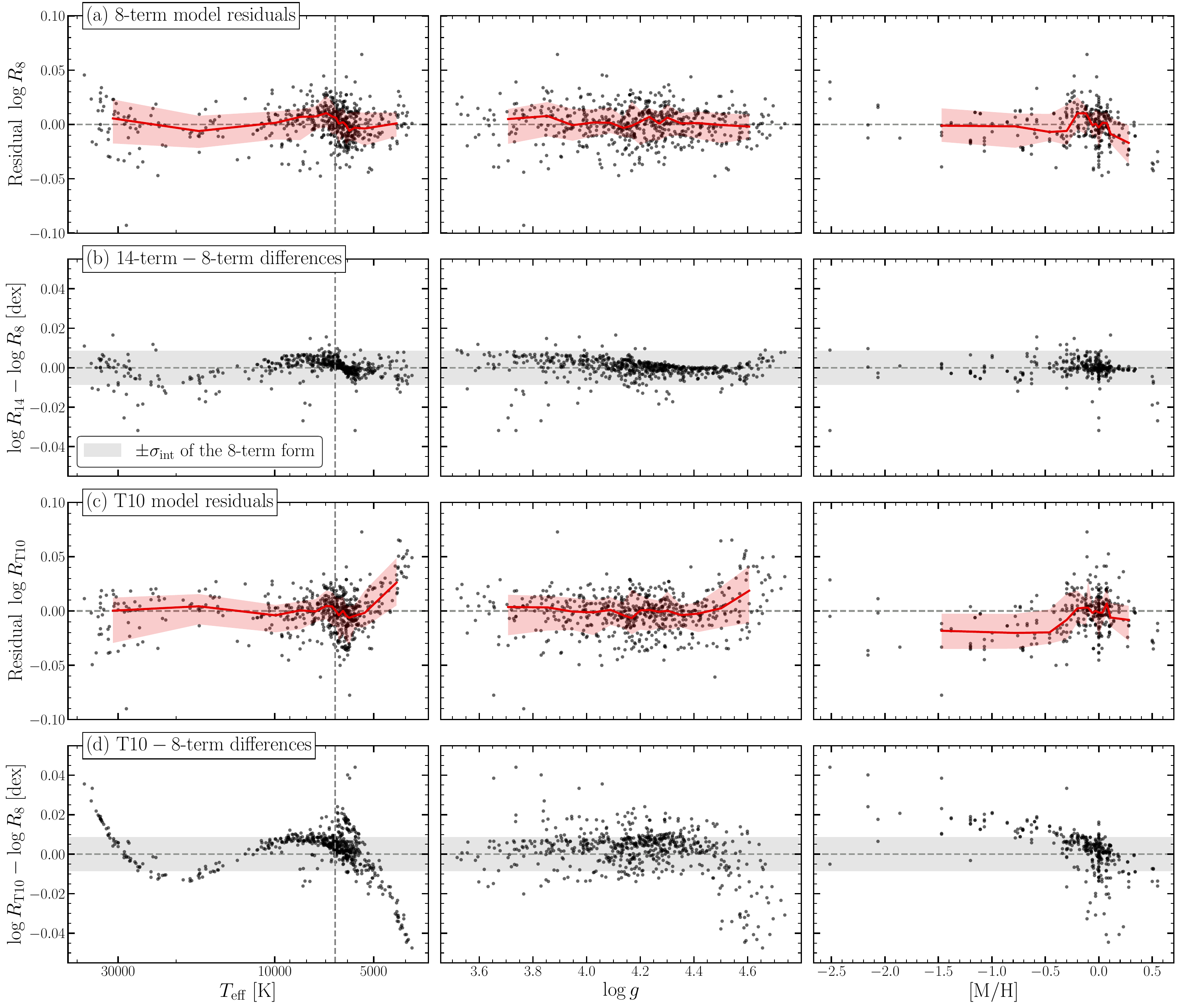}
    \caption{The top row (a) shows the star-by-star residuals for the 8-term model as a function of effective temperature (left), surface gravity (middle) and metallicity (right). The vertical gray line at $T_{\rm{eff}} = 6550$~K corresponds to the Kraft break \citep{Beyer24}, and the red line and shaded band show the binned median and 16th--84th percentile range. The points along the vertical line at $\mh{}=0$ are the stars without metallicity measurements. Row (b) shows the difference between the $\log R$ predicted by the 14-term and 8-term relations for each star. The shaded gray band shows the $\pm\sigma_{\rm int}$ of the 8-term relation. The two relations agree to within $\sigma_{\rm int}$ for most stars and differ most for hot, evolved stars. The third row (c) shows the residuals using the \citetalias{Torres10} model, and row (d) gives the difference between the 8-term $\log R$ relation and the \citetalias{Torres10} relation.}
    \label{fig:residual_figure}
\end{figure*}

Figure \ref{fig:residual_figure} shows the residuals of the 8-term model. We find that they are flat over much of the effective temperature range, but have a ``kink'' near the Kraft break between stars with radiative and convective envelopes \citep{Kraft67,Beyer24}. Both models struggle to model the metallicity dependence, with larger residuals in the highest and lowest metallicity bins. Figure \ref{fig:residual_figure} also shows the difference in the $\log R$ estimates between the simpler 8-term and the more complex 14-term model. The median fractional difference in the radius residuals between the two models is only $0.5\%$, suggesting that there is little gain from using the more complex model given the typical uncertainties in the spectroscopic parameters.

We also compare the 8-term model to the 7- and 9-term models, which have comparable $\rm{BIC}$ (Figure \ref{fig:bic_search}). The 7-term model does not include the $Z^2$ term, and has larger residuals for stars with the highest and lowest metallicities. The best 9-term model replaces the $Z^2$ term with additional $X^3$ and $X^4$ terms, which trades larger residuals for metal-poor and metal-rich stars for lower residuals for a small number of hot stars. 

Finally, we compare the radii predicted by our 8-term model to the radii predicted by the \citetalias{Torres10} empirical relation in the bottom rows of Figure~\ref{fig:residual_figure}. \citetalias{Torres10} used a much smaller sample of $95$ EBs (Figure~\ref{fig:mass_radius_kiel}), most of which were near Solar metallicity (Figure~\ref{fig:met_hist}). We find that the \citetalias{Torres10} relation produces larger residuals than the 8-term model at low metallicity, which is not surprising. Below $\sim 4500$~K (or equivalently at high $\log g$), the \citetalias{Torres10} relation also predicts radii $\sim 9\%$ smaller than the 8-term model.

\section{Posteriors on Stellar Parameters} \label{sec:posteriors}

\begin{figure*}
    \centering
    \includegraphics[width=\linewidth]{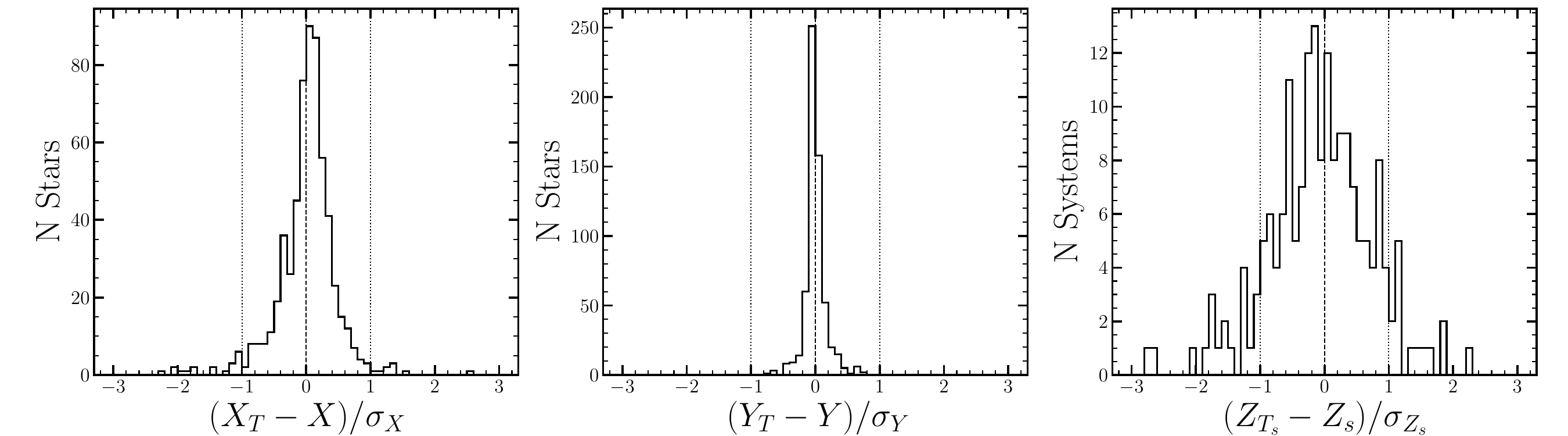}
    \caption{Left: distribution of $X_T - X$ normalized by the measured $\sigma_X$ for each star. We take $X_T$ as the mean of the posterior calculated using the 8-term coefficients and $\sint$. The middle and right panels show the same for $Y$ and $Z_s$.}
    \label{fig:param_posteriors}
\end{figure*}

For model selection and parameter estimation in \S\ref{sec:adopted_models}, we simply marginalized over the ``true'' stellar parameters $X_T$, $Y_T$ and $Z_{T_s}$. However, we can determine the posteriors on these parameters given the coefficients of the 8-term model. If the posteriors on the latent variables are significantly different than the measured values for a given system, this could suggest that its uncertainties are underestimated, or that the system falls far from the expected empirical relation. The joint posterior on the latent variables is the integrand of Equation \ref{eqn:likelihood} divided by $\mathcal{L}_s$,

\begin{equation} \label{eq:joint}
    \begin{split}
        &p\big(\{X_{T_i}, Y_{T_i}\}_{i \in s},\, Z_{T_s} \mid \mathcal{D}_s\big) = \frac{1}{\mathcal{L}_s}\; W_s(Z_{T_s}) \\
        &\quad \times \prod_{i \in s} P_i(R_i, X_i, Y_i \mid X_{T_i}, Y_{T_i}, Z_{T_s}),
    \end{split}
\end{equation}
\noindent where the priors on $X_T$ and $Y_T$ are flat. As in \S\ref{sec:fits}, the logarithm of the radius is abbreviated as $R_i$. $W_s$ is
\begin{equation} \label{eq:W}
    \begin{split}
        W_s(Z_{T_s}) &\equiv \pi(Z_{T_s})\, P(Z_s \mid Z_{T_s}) \\
        &=
        \begin{cases}
            \mathcal{N}(Z_{T_s} \mid Z_s,\, \sigma_{Z_s}) & \text{measured} \\
            \mathcal{N}(Z_{T_s} \mid \mu,\, \sigma) & \text{unmeasured},
        \end{cases}
    \end{split}
\end{equation}
\noindent and $\mathcal{D}_s$ is all of the measurements for that system. We can then determine the posteriors on $X_T$, $Y_T$, or $Z_{T_s}$ by integrating over the other latent variables in the system. For example, the posterior on $Z_{T_s}$ is
\begin{equation} \label{eq:Zpost}
    p(Z_{T_s} \mid \mathcal{D}_s) = \frac{1}{\mathcal{L}_s}\; W_s(Z_{T_s})\; I_i(Z_{T_s})\; I_j(Z_{T_s}),
\end{equation}
where $I_i$ is the per-star integral over $X_{T_i}$ and $Y_{T_i}$,
\begin{equation} \label{eq:I}
    \begin{split}
        I_i(Z_{T_s}) \equiv \iint & P_i(R_i, X_i, Y_i \mid X_{T_i}, Y_{T_i}, Z_{T_s}) \\
        & \times \dd X_{T_i} \dd Y_{T_i},
    \end{split}
\end{equation}
\noindent and $I_j$ is the equivalent for its binary companion. If only one star from a binary is included in the sample (for example, if the secondary has $M_2 < 0.6\ M_\odot$), then $I_j = 1$. We can do the same for $X_{T_i}$ and $Y_{T_i}$, where we still have to include the corresponding integral for the companion, $I_j$, because of the shared metallicity $Z_{T_s}$ between the two stars:
\begin{equation} \label{eq:Xpost}
    \begin{split}
        p(X_{T_i} \mid \mathcal{D}_s) ={}& \frac{1}{\mathcal{L}_s} \iint W_s(Z_{T_s})\; I_j(Z_{T_s}) \\
        &\times P_i(R_i, X_i, Y_i \mid X_{T_i}, Y_{T_i}, Z_{T_s}) \\
        &\times \dd Y_{T_i} \dd Z_{T_s}.
    \end{split}
\end{equation}
\noindent A similar equation can be written for $Y_{T_i}$, integrated over $\dd X_{T_i}$ and $\dd Z_{T_s}$. We compute posteriors on latent variables for all stars by averaging 100 random samples from the $\vb$ and $\sint$ posteriors. As in \S\ref{sec:fits}, we use Gauss--Hermite quadrature for numerical integration. 

Figure \ref{fig:param_posteriors} shows the distributions of the normalized difference between the mean $X_T$, $Y_T$, $Z_{T_s}$ posterior and measured $X$, $Y$, $Z_s$ for each star/system. The metallicity distribution is only for the systems with metallicity measurements. The vast majority of stars have posterior parameter estimates consistent with the measured values, and the residual distributions are all centered at zero. We examine the systems that were significant outliers below. In some cases there appear to be legitimate reasons to drop the systems from the models, but there were a small enough number that we chose not to do so. None of the outliers were systems where we assigned missing uncertainties to be the sample median uncertainties.

There are four stars where the temperature residuals $|(X_T - X)/\sigma_X| > 2$. The first, V~Cir \citep{Pawar24}, has a spectroscopic $T_{\rm{eff}}$ for the secondary that is hotter than the primary, which is in conflict with the surface-brightness ratio from the light curve solution. The second is CC~Cas, a massive O8.5$+$B0.5 binary. The secondary temperature comes from an older light curve solution \citep{Hill94}, which disagrees with the more recent fits \citep{Gorda13, Southworth22} that give the precise masses and radii. The third is NGC~3201~V142, which may be explained by a poor reddening correction \citep{Rozyczka22}. The fourth is KIC~6778289 \citep{ClarkCunningham19}, which has a reported $T_{\rm{eff,2}} > T_{\rm{eff,1}}$ from its APOGEE spectra, in contradiction with the light curve eclipse depths.

%CC~Cas is a massive O8.5$+$B0.5 binary where the secondary temperature was determined by fits to the light curve \citep{Hill94} using a light ratio fixed at the spectroscopic value. \citet{Southworth22} revisited this system using \textit{TESS} light curves to refine mass and radius measurements and fixed $T_{\rm{eff},2}$ at the \citet{Hill94} result. However, they find a different light ratio than the spectroscopic light ratio from \citet{Hill94}. 

The distribution of $(X_T - X)/\sigma_X$ is wider than the $(Y_T - Y)/\sigma_Y$ distribution. Since the surface gravities are dynamical, a shift in $Y_T$ that moves $R_M$ towards the $R_i$ is offset by the covariance term in the likelihood. As a result, the posteriors on $Y_T$ are close to the measured $Y_i$.

%This can be explained by the assumption of independent mass and radius uncertainties from the EB solution. Using the form of the covariance matrix from Equation \ref{eqn:cov}, a shift in $Y_T$ that moves $R_M$ towards the $R_i$ is offset by the covariance term in the likelihood. As a result, the posteriors on $Y_T$ are close to the measured $Y_i$. However, mass and radius measurements are both derived from the joint light curve and radial velocity orbit solution. Appendix \ref{appendix:cov} shows how a more accurate covariance matrix could be used if the uncertainties on the velocity semi-amplitudes and fractional radii were reported from the light curve solutions.

%If the measured radius $R_i$ is different from the model prediction $R_M$, then $X_T$ can be shifted away from $X_i$ to bring the two into agreement. However, this is not true for $Y_T$ since the measured $Y_i$ is dynamical and its error is anticorrelated with the error on $R_i$ (Equation \ref{eqn:cov}). A shift in $Y_T$ that moves $R_M$ towards the $R_i$ is offset by the covariance term in the likelihood. As a result, the posteriors on $Y_T$ are close to the measured $Y_i$. 

There are five systems with large composition residuals, $|(Z_{T_s} - Z_s)/\sigma_{Z_s}| > 2$. The first is TIC~17303250 \citep{Pan2025}, which has a catalog metallicity $\mh{} = -0.05$. Both components of this system have effective temperatures $\sim 600$~K hotter than expected given the masses \citep{Pecaut13}, so the $Z_{T_s}$ posterior shifts to be more metal-poor, with a mean value of $\mu_{Z_{T_s}} = -0.46$. The $Z_{T_s}$ posterior shifts away from the measured value rather than the individual temperature $X_T$ posteriors because of the relative size of the measurement errors.

Two others, GW~Eri and V788~Cen \citep{Graczyk22}, are metallic-line (Am) stars where the surface metallicity does not reflect the bulk composition. In both systems, the posterior mean is $Z_{T_s} \approx 0.1$, as compared to the super-Solar surface abundance $\mh{} \approx 0.5$. 
The fourth is V~Cir, which may have $|(Z_{T_s} - Z_s)/\sigma_{Z_s}| > 2$ because of the effective temperature contradictions described above \citep{Pawar24}. Finally, the fifth is Gaia~DR3~2847197557833077504, the most metal-poor binary in \citet{Rowan26}. It may be the case that the $Z_{T_s}$ posterior shifts from the measured $Z_s$ because the model is only quadratic in $Z$, leading to larger residuals at the edges of the parameter space. But it is also possible that the metallicity is underestimated by $\sim0.2$~dex given the small number of \ion{Fe}{2} lines available to break the $T_{\rm{eff}}$--$\mh{}$ degeneracy in the spectral models.

\begin{figure}
    \centering
    \includegraphics[width=\linewidth]{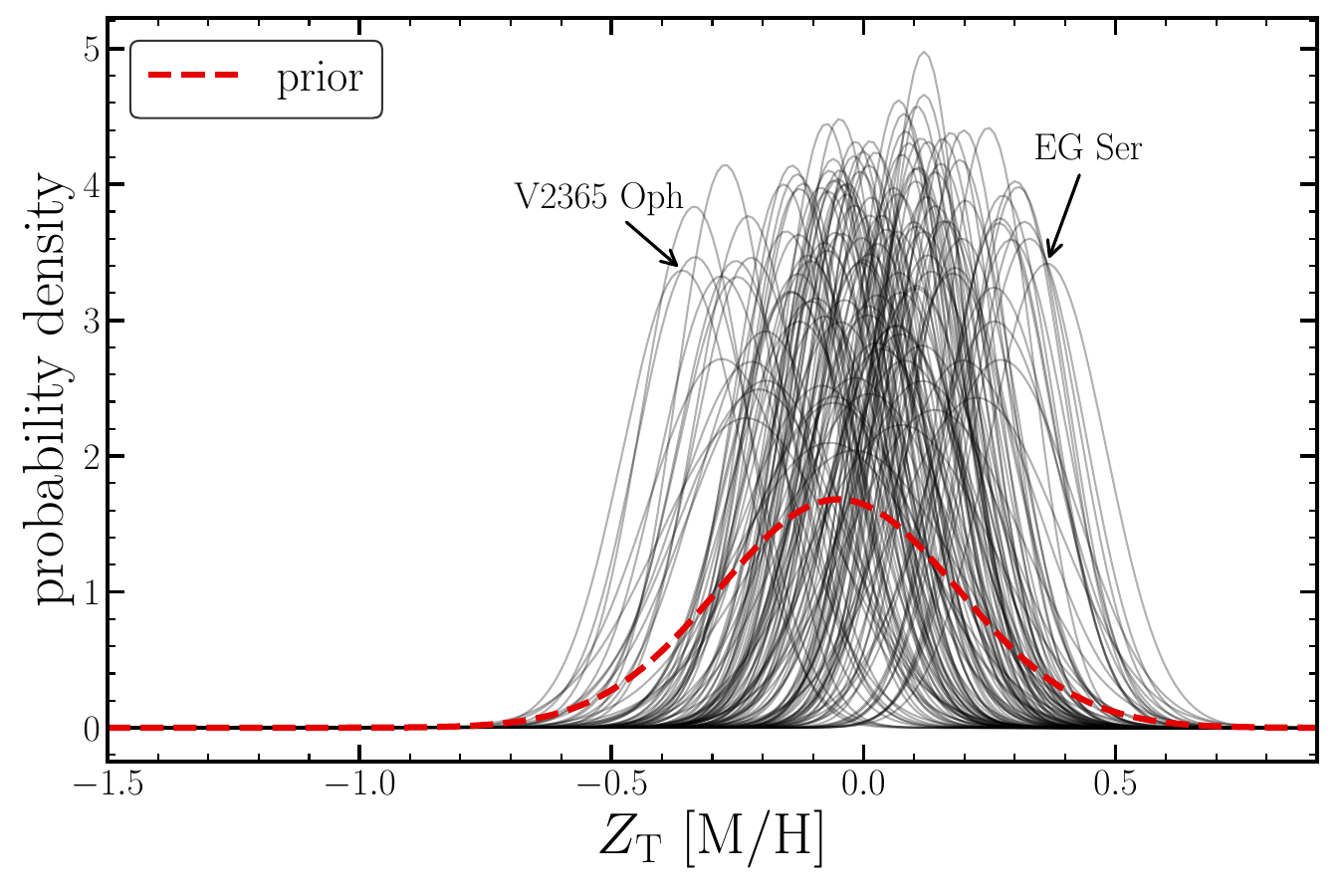}
    \caption{Metallicity posteriors inferred from the 8-term empirical relation for stars without DEBCat metallicities. The red line shows the adopted population prior, and each black line is a single binary system. We highlight the two systems with the highest and lowest posterior means.}
    \label{fig:met_posteriors}
\end{figure}

For systems without measured metallicities, the empirical model gives a posterior on the metallicity $Z_T$ that is a combination of the effects of the metallicity prior, the model and the constraints on the radius, temperature and gravity.  The model constrains the metallicity independent of the prior, so a poor prior should be recognizable by inconsistencies between the metallicity prior and the metallicity posteriors. Figure \ref{fig:met_posteriors} shows the metallicity posteriors for these stars. They all have posterior means in the range $-0.5 < \mh{} < 0.5$~dex. The mean of the posterior means is $\mh{} = +0.015\pm0.02$, which is consistent with the adopted prior and there are no significant outliers. The metallicity posteriors are significantly narrower than the metallicity prior, which means that the radius,  temperature and surface gravity measurements combine through the model to constrain the composition more tightly than the metallicity prior. V2365~Oph has the lowest posterior mean, $\mu_{\mh{}} = -0.37$. While this binary does not have a spectroscopic metallicity, \citet{Ibanoglu08} suggest that it is most likely metal-poor ($\mh{} \sim -0.7$) based on comparisons to evolutionary models. EG~Ser has the highest posterior mean, $\mu_{\mh{}}=0.38$. The masses and radii come from \citet{Thurman23}, which used the radial velocities from the photographic plates of \citet{Popper86}, and more recent spectra are not available. 

We can also make two model-independent checks of these abundances based on their stellar kinematics and abundance estimates from other surveys.
First, we use \Gaia{} DR3 proper motions and parallaxes \citep{GaiaCollaboration2022} and the center-of-mass velocity from the spectroscopic orbits to calculate the velocities, $U$, $V$, $W$, of the binaries in the Galactic coordinate system. We then use the relations from \citet{Ramirez07} to calculate the thin disk, thick disk, and halo membership probabilities. All but one of the \nsystemsnomet{} systems without metallicity measurements used in the models have a thin disk membership probability $p_{\rm{thin}} > 0.9$, consistent with having near-Solar abundances.  The exception, V467~Vel \citep{Michalska13}, is part of the Bochum~7 OB association and therefore also likely to have near-Solar metallicity.  

Next, we cross-matched the stars missing metallicities with large spectroscopic surveys. Even though the abundances measured with these pipelines assume a single-star model, \citet{ElBadry18} estimate that the resulting metallicities for double-lined spectroscopic binaries are biased by $\lesssim 0.2$~dex. We searched for the binaries without metallicities in the Sloan Digital Sky Survey Milky Way Mapper \citep[SDSS-V MWM,][]{Meszaros25} and the Large Sky Area Multi-Object Fiber Spectroscopic Telescope \citep[LAMOST,][]{Cui12} Data Release 8.  We found 33 binaries with spectroscopic parameters measured with the APOGEE Stellar Parameter and Chemical Abundances Pipeline (ASPCAP). However, 15 of them are flagged for having uncertain metallicities due to rapid rotation, and another seven are flagged for having low signal-to-noise ratio spectra or fit parameters falling near the edge of the ASPCAP model grids, leaving only 11 systems. Two of these have unphysically low metallicities. For the first, V414~Pup, the MWM catalog reports $\mh{} = -2.2$, which is not consistent with its spectral classification as a chemically peculiar HgMn star with a metallic-lined secondary \citep{Kochukhov18}. The second, V541~Cyg, is reported as having $\mh{} = -1.9$, well below the expectation from \citet{Torres17} of $-0.5 \lesssim \mh{} \lesssim 0.0$. Excluding these systems, the remaining nine systems have an abundance median and dispersion of  $\mh{} = -0.21 \pm 0.11$.  We matched 23 EBs without metallicities to LAMOST. After removing six systems with effective temperatures outside the LAMOST model grids and stars with close projected companions, we find a median metallicity of $\feh{}=0.0\pm0.19$. Combining the MWM and LAMOST samples, we find a median and dispersion of $\feh{}=-0.07\pm0.19$, which is consistent with our adopted metallicity prior. 

%PV~Cas \citep{Barembaum95} has the highest posterior median, $\mu_{\mh{}}=0.38$. \citet{Pols97} and \citet{Lastennet02} both compare the derived masses and radii and spectroscopic effective temperature to predictions from evolutionary models and find that the primary and secondary, both B9.5~V stars, are $\gtrsim 1000$~K cooler than expected, so they could be metal-rich.

\section{Comparison to MIST and PARSEC Evolutionary Models} \label{sec:mist}

\begin{figure*}
    \centering
    \includegraphics[width=\linewidth]{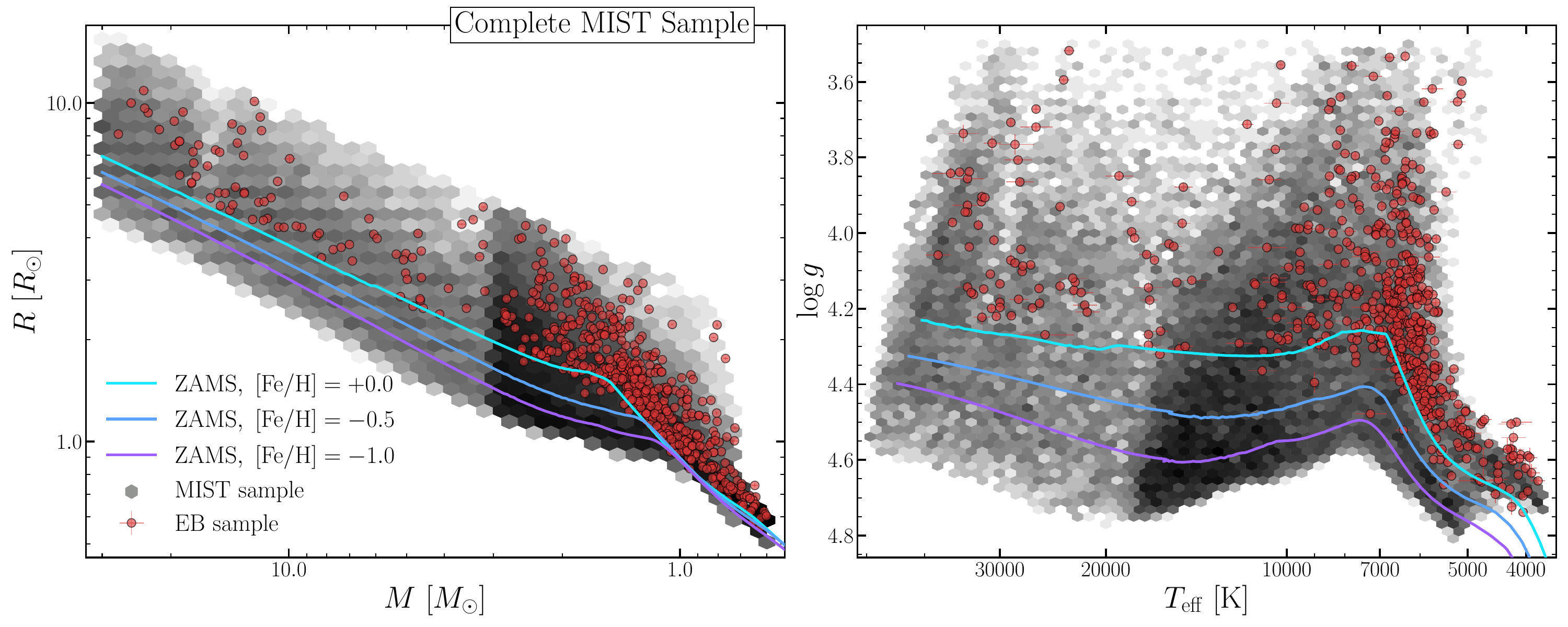}
    \includegraphics[width=\linewidth]{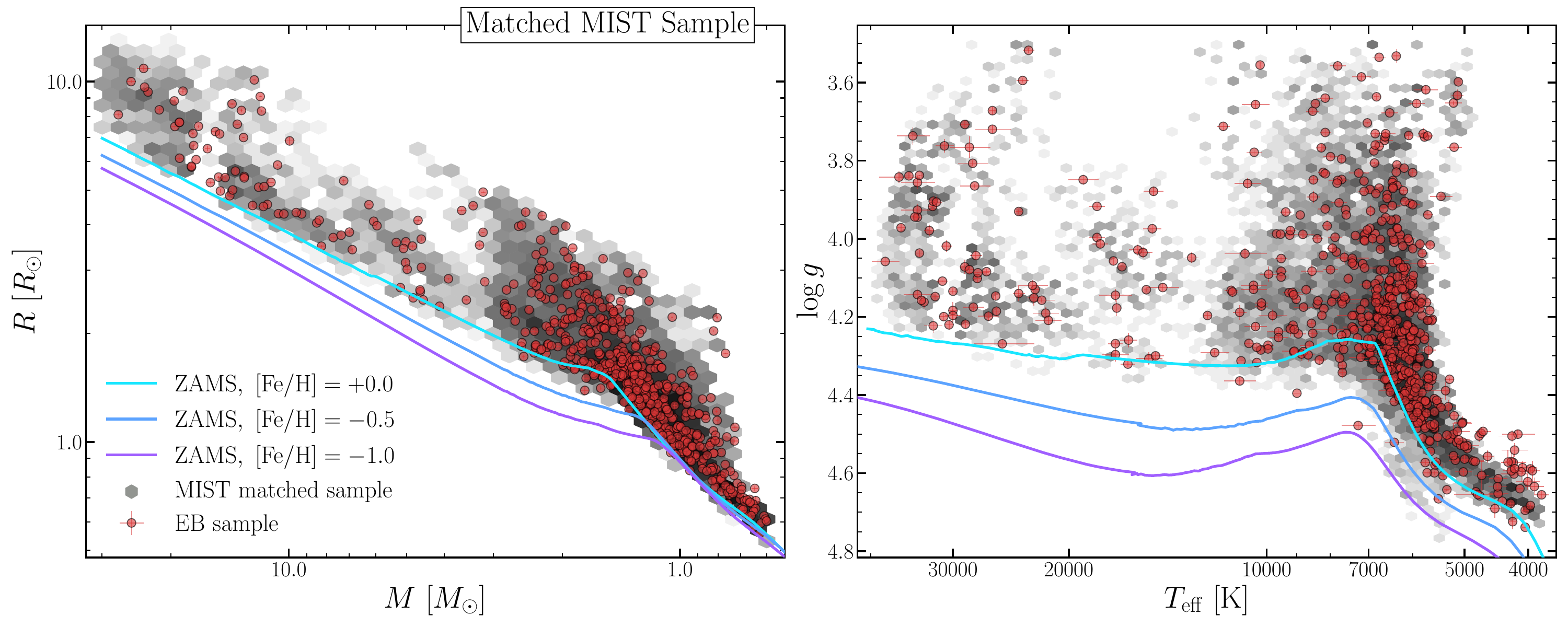}
    \caption{Masses and radii (left) and the Kiel diagram (right) for synthetic stars generated from the MIST evolutionary tracks. The top panels show the ``complete'' MIST sample, which uses a constant star formation history for every mass and metallicity. The bottom panels show the ``matched'' population 
    selected from a KDE model of the distribution of the EB sample in temperature, gravity and metallicity. The EB sample is shown in red for comparison. The colored lines show the MIST zero-age main sequences for metallicities of $\mh{}=0$ (top), $-0.5$ (middle) and $-1$ (bottom).}
    \label{fig:mass_radius_kiel_mist}
\end{figure*}

\begin{figure*}
    \centering
    \includegraphics[width=\linewidth]{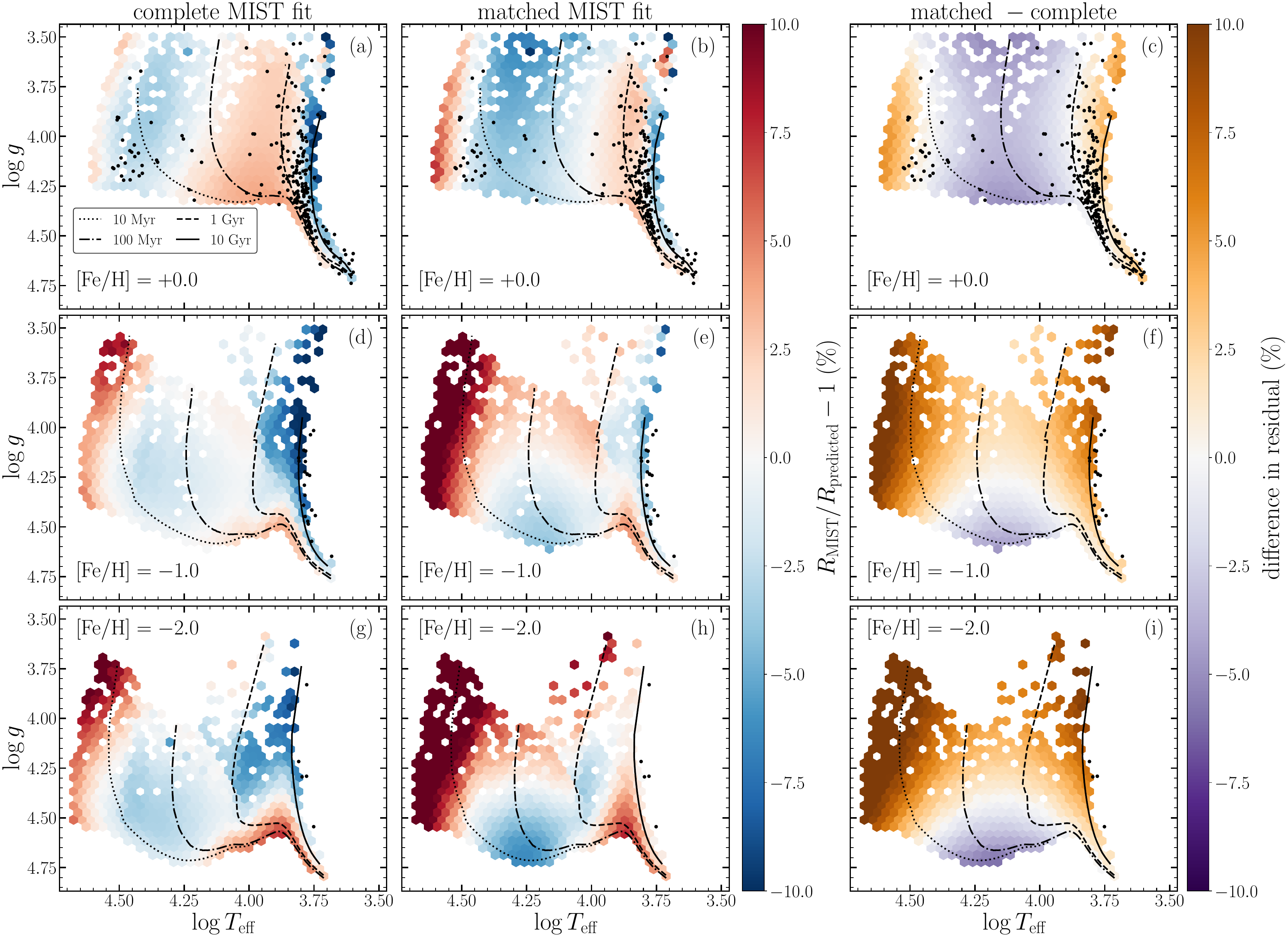}
    \caption{Fractional residuals in the radius predicted by the 8-term relation when it is fit to the full sample of MIST stars (left column) and the KDE-resampled population (middle column). We show the binned mean across the Kiel diagram at Solar metallicity (top), $\feh{}=-1.0$ (middle), and $\feh{} = -2.0$ (bottom). The right column shows the fractional difference between the two models. The black points are the stars in EBs with measured metallicity within $0.25$~dex of that metallicity bin. The black lines show MIST isochrones at four different ages at the metallicity of the corresponding row.}
    \label{fig:mist_map}
\end{figure*}

Empirical stellar relations allow for the direct determination of stellar radii and masses of main-sequence stars given spectroscopic parameters. They can also be used to make direct comparisons to predictions from stellar models. Here we construct samples of synthetic stars to serve as a comparison sample where the mass, radius, $T_{\rm{eff}}$, $\log g$, and $\mh{}$ are exactly consistent with each other. When we fit empirical relations to these synthetic samples, the residuals are due only to the choice of functional form without any contribution from measurement uncertainties. We can then directly compare the fit coefficients from the synthetic sample to those from the EB sample.  

We construct the synthetic star samples using evolutionary tracks from MESA Isochrones \& Stellar Tracks \citep[MIST,][]{Dotter16, Choi16} and PARSEC v2.0 \citep{Bressan12, Nguyen22}. We use {\tt brutus} \citep{Speagle25} to sample the MIST tracks on a grid of 73 masses spanning $0.6$--$30\ M_\odot$. We use steps of $0.05\ M_\odot$ between $0.6$ and $3.0\ M_\odot$, and logarithmic steps from $3.2$ to $30\ M_\odot$. We  use 13 metallicities from $\feh{} = -2.5$ to $+0.5$ in steps of $0.25$~dex. All evolutionary tracks are at $\afe{} = 0.0$. We access the PARSEC models with {\tt ezpadova}\footnote{\url{https://github.com/mfouesneau/ezpadova}}.  Since there is no track interpolator for PARSEC, we build the synthetic sample by downloading isochrones and interpolating over a grid of initial masses.  We use a metallicity
range of $-2.0 < \mh{} < +0.5$.
For both models, we only keep points with $\log g > 3.5$ and ages $\tau < 13.8$~Gyr for each mass and metallicity track.

We constructed two types of synthetic samples. For the first, we select stars over all of the allowed parameter space by drawing 40 random stars from each track at equivalent evolutionary points (EEPs) weighted by the residence time. This is equivalent to assuming a constant star formation rate for every mass and metallicity. Since we use a fixed initial mass grid and not a standard initial mass function, the synthetic samples have a higher fraction of high- and intermediate-mass stars at low metallicity than the real sample. In the old, metal-poor populations of the Milky Way and Magellanic Clouds, such stars would have already evolved off the main sequence. We denote this sample, which includes $\sim 38{,}000$ stars, as the ``complete'' sample. We also construct a second set of ``matched'' samples, where we resample the ``complete'' sample using a kernel density estimate (KDE) in ($\log T_{\rm{eff}}$, $\log g$, $\mh{}$) to statistically match the observed distribution of EB properties. Stars in the observed sample without metallicities were randomly assigned metallicities using the ``Robust'' prior in building the ``matched'' samples. Figure \ref{fig:mass_radius_kiel_mist} compares the masses and radii and Kiel diagrams of the observed EBs to the ``complete'' and  ``matched'' MIST model samples.  

To examine the effects of the observed $T_{\rm{eff}}$, $\log g$, and $\mh{}$ distribution on the fits, we fit both the complete and the matched MIST samples with the 8-term model. This tests both how well the model can fit the synthetic data and how well the fits to the matched samples extrapolate into regions of the parameter space where there are few or no EB benchmarks. When we fit these synthetic samples, we assume that the effective temperatures, surface gravities, and metallicities are known exactly and use a constant nominal $\sigma_{\log R} = 0.001$~dex. 

Figure \ref{fig:mist_map} shows the difference between the two model predictions and the MIST models in bins of $T_{\rm{eff}}$ and $\log g$ at three metallicities. The 8-term model can provide a reasonably good description of the complete sample, although the residuals start to increase rapidly at lower temperatures because of the $M>0.6\ M_\odot$ cutoff on the sample and for hot, low-metallicity stars. Even at the extremes, however, the 8-term model can match the complete MIST sample to better than about 10\% across the full parameter range. Unsurprisingly, the 8-term model of the matched sample fails badly for hot, low-metallicity stars because there are no DEBCat binaries in this part of the parameter space. 

The limited utility of the empirical relations outside the parameter space covered by the EBs should be expected, as it is a general problem of polynomial interpolation functions that they tend to extrapolate badly. The fits to the matched sample do markedly better at the cool edge, which suggests that additional degrees of freedom are required to model the complete sample of MIST stars. The larger fractional residuals for the ``complete'' sample at the cool edge of the parameter space are also likely a consequence of the chosen bounds. If we had instead chosen to generate synthetic stars with masses down to $0.4$ or $0.5\ M_\odot$, we would find smaller fractional residuals at $\log T_{\rm{eff}} = 3.6$ and the larger residual feature at the new $\log T_{\rm{eff}} \approx 3.55$ boundary. 

Table \ref{tab:mist_table} compares the 8-term model fit coefficients between the EB sample and the matched synthetic MIST and PARSEC samples. The largest difference is in the constant term, with a $-9.5\sigma$ ($-12.0\sigma$) shift between the EB model and the matched MIST (PARSEC) model, which says that the observed EBs are $2.2\%$ ($2.8\%$) larger in radius than the models near the centroid of the parameters. The next largest differences are for the $Z$ ($4.0\sigma$) and $X$ ($3.7\sigma$) terms of the MIST models.  Both models have $\sim 2.8\sigma$ differences for the $Y^2$ term compared to the data, followed by a $2.1\sigma$ difference for the $Z^2$ term of the MIST model.

\begin{figure*}
    \centering
    \includegraphics[width=\linewidth]{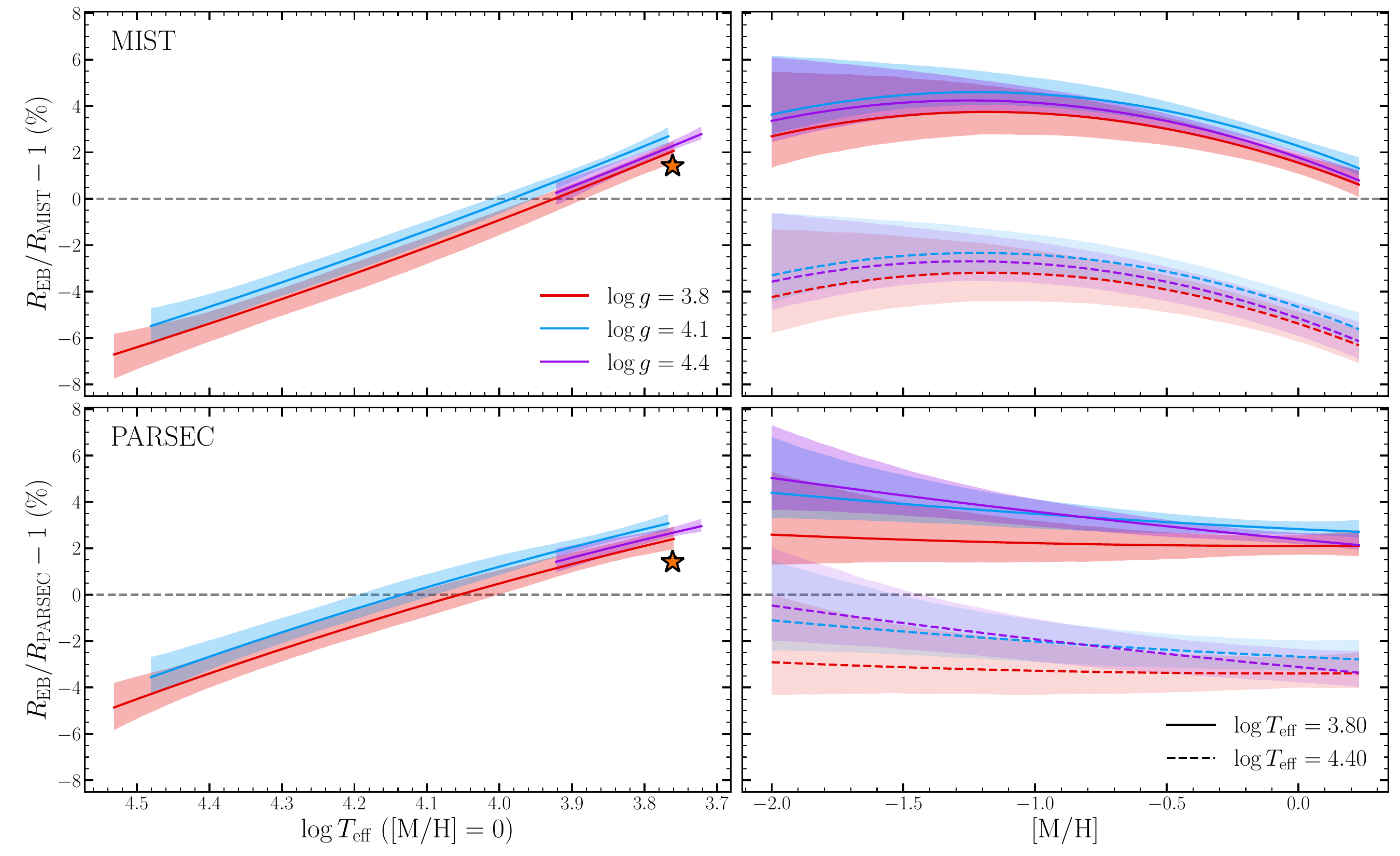}
    \caption{The fractional difference between the radii predicted by the EB relation and by the same model fit to the matched MIST sample (top) and PARSEC sample (bottom). The left panels show  the $\log T_{\rm{eff}}$ dependence for three values of $\log g$ at fixed Solar metallicity. The right panels show the metallicity dependence at fixed $\log T_{\rm{eff}} = 3.8$ and $\log T_{\rm{eff}} = 4.4$. The shaded bands are the 16th to 84th percentile range of the difference. The orange star in the left panels shows $R_{\rm{EB}} / R_\odot - 1$ for the Sun.}
    \label{fig:frac_difference}
\end{figure*}

Figure \ref{fig:frac_difference} compares fractional differences in the radius estimates as a function of $\log T_{\rm eff}$ at fixed metallicity and as a function of metallicity at fixed $\log T_{\rm eff}$ for several values of $\log g$. There are significant but modest systematic trends with both $\log T_{\rm eff}$ and $\mh{}$ while the differences are only weakly dependent on $\log g$. The trend with effective temperature is for high-temperature stars to be $\sim 6\%$ smaller in EBs than predicted by the MIST models, while low-temperature stars are $\sim 2\%$ bigger. We find moderate curvature in the metallicity difference between the EBs and the MIST models, with larger differences at low $\mh{}$. Figure \ref{fig:frac_difference} also shows the radius predicted for the Sun for comparison. The Solar radius is over-predicted by the EB relation by only $\sim 1.4\%$, which is less than $R_{\rm{EB}}/R_{\rm{MIST}} - 1$ and $R_{\rm{EB}}/R_{\rm{PARSEC}} - 1$. This indicates that the difference between the models fit to the EB and synthetic samples cannot be fully attributed to the limitations of a polynomial model.

We find a similar $\log T_{\rm{eff}}$-dependent difference trend in the comparison between the EB model and the PARSEC model. For cool stars, the EB radii are 2--4\% larger than predicted, and $\sim 4\%$ smaller than predicted for hot stars. Whereas $R_{\rm{EB}}/R_{\rm{MIST}} \approx 1$ at $\log T_{\rm{eff}} \approx 3.95$ (at $\mh{} = 0.0$), in the PARSEC models $R_{\rm{EB}}/R_{\rm{PARSEC}} \approx 1$ at $\log T_{\rm{eff}} \approx 4.1$. Unlike for the MIST sample, we do not see any significant metallicity trend in the offset. 

\renewcommand{\tablecaptiontext}{Coefficients of the 8-term $\log R$ relation fit to the EB sample (same as Table \ref{tab:model_coeffs}) and to the MIST and PARSEC stars resampled to match the distribution of EBs across the parameter space using kernel density estimates in ($\log T_{\rm{eff}},\ \log g, \mh{}$). The empirical relation is fit to the synthetic stars with least squares, and the errors on the coefficients are estimated as the standard deviation over 200 bootstrap resamplings. The shift column reports the difference between the MIST/PARSEC coefficients and the EB coefficients divided by their two errors added in quadrature. \label{tab:mist_table}}
\begin{deluxetable*}{lr@{}lr@{}lcr@{}lc}[t]
\tablecaption{\tablecaptiontext}
\tablehead{
\colhead{Term} & \multicolumn{2}{c}{EB} & \multicolumn{2}{c}{MIST} & \colhead{Shift} & \multicolumn{2}{c}{PARSEC} & \colhead{Shift} \\
\colhead{} & \multicolumn{2}{c}{} & \multicolumn{2}{c}{} & \colhead{[$\sigma$]} & \multicolumn{2}{c}{} & \colhead{[$\sigma$]}
}
\startdata
${\rm const}$ & $0.1688$ & ${}\pm0.0009$ & $0.1591$ & ${}\pm0.0005$ & $-9.5$ & $0.1566$ & ${}\pm0.0005$ & $-12.0$ \\
$X$ & $0.63$ & ${}\pm0.01$ & $0.687$ & ${}\pm0.008$ & $+3.7$ & $0.664$ & ${}\pm0.009$ & $+2.1$ \\
$Y$ & $-0.593$ & ${}\pm0.004$ & $-0.589$ & ${}\pm0.002$ & $+1.0$ & $-0.590$ & ${}\pm0.002$ & $+0.8$ \\
$Z$ & $0.066$ & ${}\pm0.003$ & $0.083$ & ${}\pm0.002$ & $+4.0$ & $0.070$ & ${}\pm0.002$ & $+0.8$ \\
$X^2$ & $0.19$ & ${}\pm0.02$ & $0.18$ & ${}\pm0.01$ & $-0.4$ & $0.20$ & ${}\pm0.02$ & $+0.5$ \\
$Y^2$ & $0.047$ & ${}\pm0.009$ & $0.076$ & ${}\pm0.006$ & $+2.8$ & $0.075$ & ${}\pm0.005$ & $+2.7$ \\
$Z^2$ & $0.007$ & ${}\pm0.002$ & $0.013$ & ${}\pm0.002$ & $+2.1$ & $0.006$ & ${}\pm0.002$ & $-0.2$ \\
$Y Z$ & $-0.028$ & ${}\pm0.005$ & $-0.026$ & ${}\pm0.005$ & $+0.2$ & $-0.020$ & ${}\pm0.005$ & $+1.1$ \\
\hline
RMS [dex] & \multicolumn{2}{c}{$0.0161$} & \multicolumn{2}{c}{$0.0104$} & \nodata & \multicolumn{2}{c}{$0.0104$} & \nodata \\
\enddata
\end{deluxetable*}

\section{Discussion and Conclusions} \label{sec:conclusions}

\begin{figure*}
    \centering
    \includegraphics[width=\linewidth]{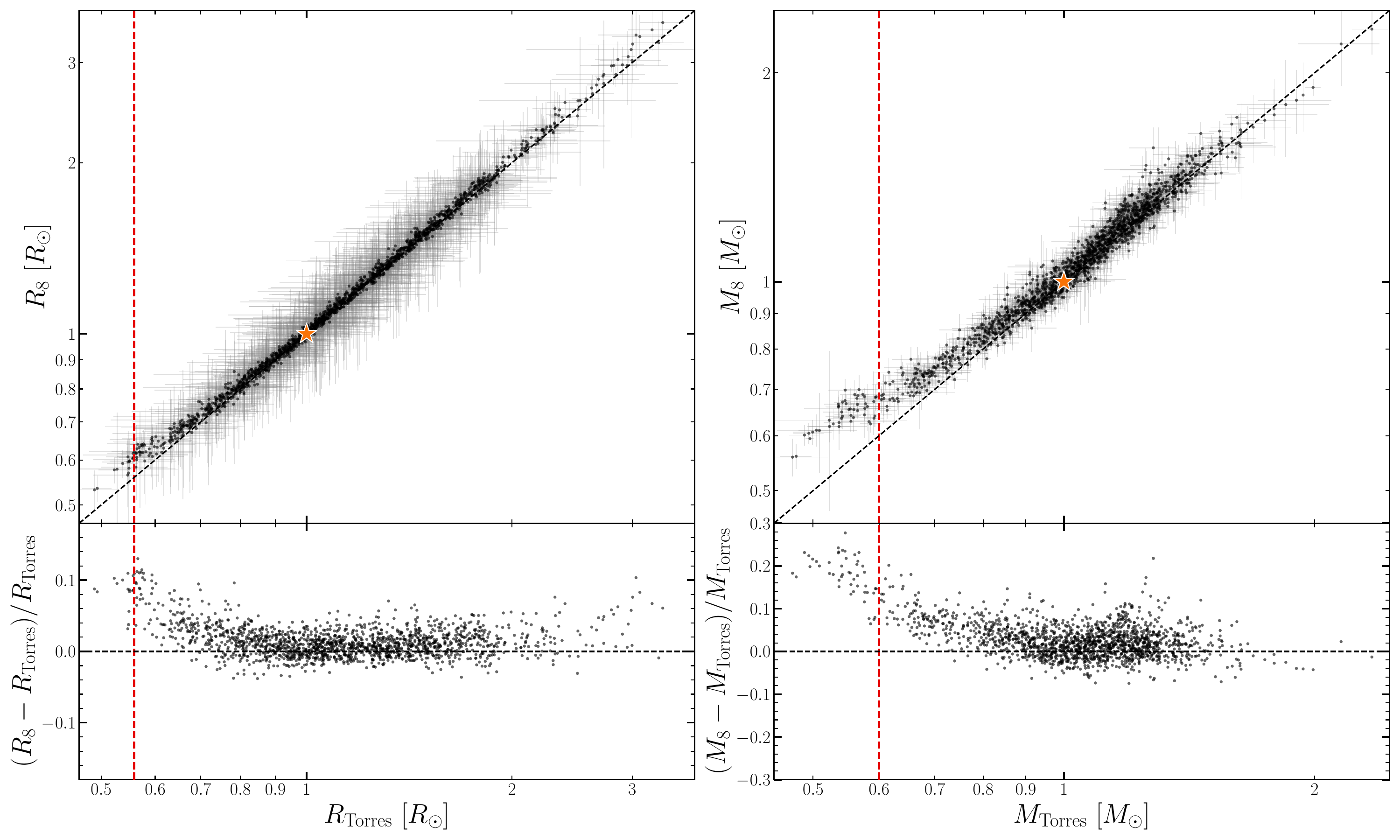}
    \caption{Radii (left) and masses (right) of 1{,}668 exoplanet host stars derived using the 8-term relation compared to those from the \citetalias{Torres10} relation evaluated using the host star's effective temperature, surface gravity, and metallicity. The bottom panels show the fractional differences between the radii and masses from the two relations. The median differences are $+1.2\%$ in radius and $+2.4\%$ in mass. The red line marks the cutoff $M > 0.6\ M_\odot$ used to fit the empirical relations. We use the fit relation to give the corresponding radius cutoff. The orange star marks the position of the Sun.}
    \label{fig:exoplanet_hosts}
\end{figure*}

Eclipsing binaries allow for direct measurement of fundamental stellar parameters. These can be used to construct empirical relations between stellar parameters (\citetalias{Torres10}). We use binaries in DEBCat \citep{Southworth15}, which have masses and radii with fractional uncertainties of $<2\%$, to update the empirical relations presented in \citetalias{Torres10}. Figure \ref{fig:mass_radius_kiel} shows the masses, radii, surface gravities, and effective temperatures of the DEBCat sample and our selection cuts. 

We model the empirical relations within a Bayesian framework, marginalizing over the unknown true spectroscopic parameters, and adopt a prior for the detached EBs without reported metallicities using the properties of the population (Figure \ref{fig:torres_form_prior_compare}). We include the uncertainties in all measured quantities and the correlations between $\log R$, $\log M$ and $\log g$. Instead of assuming a specific polynomial form for the empirical relation, we search for the optimal form (Figure \ref{fig:bic_search}). We report a 14-term model and an 8-term model (Table \ref{tab:model_coeffs}), which have a $\Delta\rm{BIC} = \deltabic{}$ and predicted $\log R$ that agree within the measured intrinsic scatter $\sint$ (Figure \ref{fig:residual_figure}). 

We also experimented with modeling the empirical relations as departures from the theoretical zero-age main sequence (ZAMS). In other words, instead of fitting the relation in terms of $Y = \log g - Y_0$, where $Y_0=4.2$, we tried $Y=\log g - \zeta(X, Z)$, where $\zeta(X, Z)$ is the surface gravity at the ZAMS given $X$ and $Z$. This required a fairly complicated, piecewise model for $\zeta(X,Z)$, which in turn required
a similarly complex model for $\log R$ at $Y \simeq 0$. We concluded it was simpler to use the polynomial approach adopted from \citetalias{Torres10}.  We also experimented with extending the fits to include the full mass range rather than using only the stars with $M > 0.6\ M_\odot$.  These required much higher-order polynomials and had significantly more scatter in the radius residuals.  We suspect that an intermediate approach between exactly referencing $Y$ to the ZAMS $\log g$ and roughly tracking its $\log g$, particularly the change in slope at $T_{\rm eff} \simeq 7000$~K (see Figure~\ref{fig:mass_radius_kiel}), would provide a solution. 

The model can also be used to give posterior estimates of the temperature, gravity and abundances.  In \S\ref{sec:posteriors} we find very few systems with significant disagreements between the measurements and the posteriors.  In several cases, such as the metallic-line Am stars GW~Eri and V788~Cen, the systems should arguably be dropped from the fits. We also examined the posteriors on the abundances of the stars without metallicity measurements, finding that the model combined with the other measured properties constrained the posterior abundances more tightly than the abundance prior and at values consistent with the range of the prior. We also found that the kinematics of these systems and the abundances of those with APOGEE or LAMOST observations were almost all consistent with the range of the ``Robust'' abundance prior.

The coefficients of the empirical model can be compared to predictions from stellar models. We use MIST evolutionary tracks and PARSEC isochrones to construct synthetic populations on a fixed grid of masses and metallicities assuming a constant star formation rate. We resample these population to match the distribution of effective temperatures, surface gravities, and metallicities of the EB sample (Figure \ref{fig:mass_radius_kiel_mist}). We compare the coefficients of the empirical relation derived on this synthetic sample (Table \ref{tab:mist_table}) and find that there is a temperature-dependent difference between the relations fit with the EB sample and the synthetic stellar population (Figure \ref{fig:frac_difference}). We also fit the full sample of synthetic stars to assess the bias imposed by the EB selection function and test how the model performs near and outside the bounds of the EB parameter space (Figure \ref{fig:mist_map}).

The derived empirical relations have RMS scatters of $\sigma_{\log R} = 0.0161$~dex (3.8\%) and $\sigma_{\log M} = 0.0322$~dex (7.7\%), respectively. Most of this, however, is due to measurement uncertainties.  We estimate that the intrinsic scatter in the radius beyond the measurement uncertainties is only $0.0085$~dex (2.0\%).  The fact that the 14-term model reduced this intrinsic scatter little over the 8-term model suggests that modeling the remaining intrinsic scatter requires new physical parameters. Possibilities are rotation velocities (mixing and gravitational settling) or more complex composition models (e.g., \feh{} plus \afe{} affecting opacities).  

These new models have quantitative consequences beyond stellar models.  For example, planetary transits determine the ratio $R_P/R_*$ (e.g., \citealt{Winn2010}) between the radius of the planet ($R_P$) and that of the star ($R_*$).  Figure \ref{fig:exoplanet_hosts} shows the masses and radii derived for $1{,}668$ exoplanet host stars from the NASA Exoplanet Archive \citep{Christiansen25} with the necessary stellar parameters. As compared to the \citetalias{Torres10} relation, the updated empirical relation finds stellar radii that are $1.2\%$ larger on average, but with significantly larger differences for the lowest and highest mass stars. Both relations are extrapolations for $M < 0.6\ M_\odot$, indicating the need for a more global relation. In the case of transiting exoplanets the mean density can be directly measured from the light curve, so we report empirical model coefficients using the mean density in place of surface gravity, following \citet{Enoch10}, in Appendix \ref{appendix:rho}.

Empirical relations for stellar radii and masses serve as an observationally calibrated alternative to stellar models. The masses and radii in detached EBs are directly measured from the combination of the photometry and the spectroscopy, but an underlying assumption in benchmark EB systems is that the stars have evolved effectively independently. While this is generally a safe assumption for long-period binaries with low-mass stars, some DEBCat binaries have stars that are close enough that their stellar surfaces are tidally distorted. For example, IR~Cas \citep{Kamenec26} and AH~Cep \citep{Pavlovski18} both have out-of-eclipse variability from ellipsoidal modulations. Forty-nine of the \nebcomponents{} stars (8.1\%) used in our fits have fractional radii $R/a > 0.25$, which is a typical cutoff above which stars cannot be assumed to be spherical \citep{North04, Helminiak24}. The majority ($\sim 60\%$) of the components here with $R/a > 0.25$ have $M > 2\ M_\odot$. 

The light curve models of these distorted stars typically use a full Roche potential or some approximation of it, so the measured radii themselves are unlikely to be biased. However, tidal distortion lowers the effective surface gravity, so we may expect these stars to be larger than if they were isolated and spherical stars. \citet{Fabry22} estimate that the discrepancy can be as much as 5\% for the most distorted stars. Interestingly, the empirical relations give radii smaller than the stellar models for hot stars (Figure \ref{fig:frac_difference}), so the distortion of massive stars in close EBs is unlikely to be responsible for the temperature-dependent radius differences we observe. 

The limited sample of benchmark EBs with mass and radius measurements and their non-uniform sampling of parameter space remain a significant limitation. Figure \ref{fig:mist_map} shows that our simple 8-term polynomial is effective for stars within the parameter space covered by the detached EBs. However, if we instead had EBs with more equal coverage across the parameter space, similar to the ``complete'' MIST sample shown in Figure \ref{fig:mass_radius_kiel_mist}, an 8-term polynomial would not have enough degrees of freedom. Despite dedicated observational efforts to target sparse regions of the parameter space \citep[e.g.,][]{Rowan26}, the detached EB sample remains dominated by Solar-metallicity main-sequence stars with $0.8 < M/M_\odot < 3$. 

Some of these problems can be addressed with \Gaia{} DR4, which will include multi-epoch medium-resolution spectra for thousands of eclipsing binaries. Joint fits of the \Gaia{} radial velocities with the EB light curves will yield greatly improved solutions because the light curves so tightly constrain the period, eccentricity and orbital phase independent of the velocity measurements. \citet{Rowan23} used a subset of such orbits in \Gaia{} DR3 and light curves from the All-Sky Automated Survey for Supernovae \citep{Shappee14, Kochanek17} to measure masses and radii with fractional uncertainties of 5--10\%. While these have much greater uncertainties than what is achievable through dedicated ground-based spectroscopic follow-up, \Gaia{} DR4 has the potential to increase the sample of benchmark EBs by an order of magnitude when combined with an approach like that used here to fully account for the measurement uncertainties.

Such a larger sample, better covering the parameter space occupied by stars, should make it much more feasible to incorporate additional physical parameters like $\alpha$-abundances and rotational velocities into the empirical relations. With fractional mass and radius uncertainties of $\sim 5$--$10\%$, it would be challenging to directly tune detailed physics like convective overshoot \citep[e.g.,][]{Claret16} and mixing \citep[e.g.,][]{Woody26} by comparing individual observations to theoretical models. The empirical relations essentially provide a means of averaging over stars without requiring them to be truly identical, so it may be more feasible to test models by matching the empirical relations rather than individual stars. Finally, in addition to better probing the physics of stellar evolution, larger samples will enable tests for systematic biases. For example, we could model the dependence of the model on the Roche lobe filling factors, orbital period or estimates of the Rossby number as a proxy for the effects of magnetic fields.

%We found that higher order polynomials in $\log T_e$, $\log g$ and $[M/H]$ were not easily
%able to explain the intrinsic scatter of $0.0086$~dex (2.0\%) we found for $\log R$.  This
%suggests that additional physical variables are required.  ??WHAT MIGHT THEY BE?? 
%$v \sin i$,  $[Fe/H]$+$[\alpha/Fe]$ SPECULATE?

\begin{acknowledgments}

We thank M. Pinsonneault, K. Stanek, and T.~A. Thompson for extensive discussions and G. Torres for clarifying aspects of their models in \citetalias{Torres10}. DMR is supported by NASA Hubble Fellowship grant HST-HF2-51588.001-A awarded by the Space Telescope Science Institute, which is operated by the Association of Universities for Research in Astronomy, Inc., for NASA, under contract NAS5-26555. CSK is partially supported by NSF grants AST-2307385 and AST-2407206. This work has benefited from using Claude Code for coding development. 

\end{acknowledgments}

\software{{\tt ebrelations} (\url{https://github.com/dmrowan/eb-relations}), {\tt emcee} \citep{ForemanMackey13}, {\tt brutus} \citep{Speagle25}, {\tt ezpadova} (\url{https://github.com/mfouesneau/ezpadova})}

%% Appendix material should be preceded with a single \appendix command.
%% There should be a \section command for each appendix. Mark appendix
%% subsections with the same markup you use in the main body of the paper.
%%
%% Each Appendix (indicated with \section) will be lettered A, B, C, etc.
%% The equation counter will reset when it encounters the \appendix
%% command and will number appendix equations (A1), (A2), etc. The
%% Figure and Table counter will not reset.

\appendix

\section{Covariance Matrix} \label{appendix:cov}

In \S\ref{sec:fits}, we use a covariance matrix where the mass and radius uncertainties are assumed to be independent (Equation \ref{eqn:cov}). We can write a more accurate form using the velocity semi-amplitudes, $K_1$ and $K_2$, and their uncertainties, $\sigma_{K_1}$ and $\sigma_{K_2}$, and the fractional radius $\hat{r}_i=R_i/a$, where $a$ is the semimajor axis and $R_i$ is the linear radius (as compared to the abbreviated logarithmic radius in \S\ref{sec:fits}). The radius, mass, and surface gravity for the primary are then

\begin{align}
    \log R_1 &= \log(K_1 + K_2) + \log \hat{r}_1 + \text{const}, \\
    \log M_1 &= \log K_2 + 2\log(K_1 + K_2) + \text{const}, \\
    \log g_1 &= \log K_2 - 2\log \hat{r}_1 + \text{const},
\end{align}
\noindent where we have assumed that the orbital period is measured exactly, $\sin(i)\approx1$, and that $e=0$. The covariance matrix is then
\begin{equation}
    \vsigma_1 =
    \begin{pmatrix}
    \dfrac{\sigma_{K_1}^2 + \sigma_{K_2}^2}{(K_1 + K_2)^2} + \sigma_{\hat{r}_1}^2 + \sint^2 & \dfrac{\sigma_{K_2}^2}{K_2\,(K_1 + K_2)} - 2\sigma_{\hat{r}_1}^2 \\[12pt]
    \dfrac{\sigma_{K_2}^2}{K_2\,(K_1 + K_2)} - 2\sigma_{\hat{r}_1}^2 & \dfrac{\sigma_{K_2}^2}{K_2^2} + 4\sigma_{\hat{r}_1}^2
    \end{pmatrix}.
\end{equation}
The velocity semi-amplitudes and the fractional radii are measured directly from the radial velocities and light curve, but their posteriors and uncertainties are not consistently reported. If we assume that $K_1 = K_2$ and $\sigma_{K_1}=\sigma_{K_2}$, this can be simplified further, 
\begin{equation}
    \vsigma_1 =
    \begin{pmatrix}
    \tfrac{1}{2}f^2 + \sigma_{\hat{r}_1}^2 + \sint^2 & \tfrac{1}{2}f^2 - 2\sigma_{\hat{r}_1}^2 \\[6pt]
    \tfrac{1}{2}f^2 - 2\sigma_{\hat{r}_1}^2 & f^2 + 4\sigma_{\hat{r}_1}^2
    \end{pmatrix},
\end{equation}
where $f \equiv \frac{\sigma_K}{K}$.
The uncertainty on the fractional radius can be written in terms of $\sigma_{\log R_1}$ and $\sigma_{\log M_1}$ as 
\begin{align}
    \sigma_{\log R_1}^2 &= \tfrac{1}{2}f^2 + \sigma_{\hat{r}_1}^2, \qquad \sigma_{\log M_1}^2 = 5f^2, \\
    \sigma_{\hat{r}_1}^2 &= \sigma_{\log R_1}^2 - \tfrac{1}{2}f^2 = \sigma_{\log R_1}^2 - \tfrac{1}{10}\sigma_{\log M_1}^2.
\end{align}
\noindent This assumption of $K_1 = K_2$ is therefore only valid when $\sigma_{\log M_1}^2 \leq 10\,\sigma_{\log R_1}^2$. The covariance matrix for the primary can then be written in terms of its mass and radius uncertainties as
\begin{equation}
    \vsigma_1 =
    \begin{pmatrix}
    \sigma_{\log R_1}^2 + \sint^2 & \tfrac{3}{10}\sigma_{\log M_1}^2 - 2\sigma_{\log R_1}^2 \\[6pt]
    \tfrac{3}{10}\sigma_{\log M_1}^2 - 2\sigma_{\log R_1}^2 & 4\sigma_{\log R_1}^2 - \tfrac{1}{5}\sigma_{\log M_1}^2
    \end{pmatrix}.
\end{equation}
We re-fit the 8-term empirical model in $\log R$ using this form of the covariance matrix. For targets with $\sigma_{\log M_1}^2 > 10\,\sigma_{\log R_1}^2$ ($\sim 10\%$ of the sample) we use the standard Equation \ref{eqn:cov} form of the covariance. The fit coefficients agree with those reported in Table \ref{tab:model_coeffs} to within $\sim 0.1\sigma$. The posteriors on the latent parameter shifts agree with those shown in Figure \ref{fig:param_posteriors}.

\section{Fits to Stellar Density} \label{appendix:rho}

For stars characterized spectroscopically, $T_{\rm{eff}}$, $\log g$, and $\mh{}$ are a natural parameterization since these are measured directly from the spectra. However, for exoplanet host stars, an alternate parameterization is to replace $\log g$ with the stellar density, $\log \rho$, because the stellar density can be measured directly from the transiting exoplanet light curve \citep{Seager03}. Following the empirical relations presented in \citetalias{Torres10} in terms of $\log g$, \citet{Enoch10} reported the equivalent relations in terms of $\log \rho$. 

We repeat our fit to $\log R$ using the 8- and 14-term models (Equations \ref{eqn:8term} and \ref{eqn:14term}), but using $Y=\log \rho$, where 
\begin{equation}
    \log \rho = \log M - 3 \log R.
\end{equation}
\noindent The fit follows directly from \S\ref{sec:fits}, using a pivot $Y_0 = 0$. 
Following Equation~\ref{eqn:cov}, the covariance matrix for the radius and the density is
\begin{equation}
    \vsigma_i =
    \begin{pmatrix}
    \sigma_{\log R_i}^2 + \sigma_{\rm int}^2 & -3\,\sigma_{\log R_i}^2 \\
    -3\,\sigma_{\log R_i}^2 & \sigma_{\log M_i}^2 + 9\,\sigma_{\log R_i}^2
    \end{pmatrix},
\end{equation}
and we use the same priors, quadrature nodes, and sampling as for the $\log g$ models described in \S\ref{sec:fits}. Table \ref{tab:logrho} reports the coefficients of the $\log R$ 8- and 14-term relations fit with $\log \rho$. The $\log M$ coefficients are again derived directly from the $\log R$ relation. 

The relations fit with $\log \rho$ perform similarly to the relations using $\log g$. The overall $\log R$ fit RMS is $\sim 1/3$ smaller using $\log \rho$ instead of $\log g$, but this is just a consequence of the $\rho \propto R^{-3}$ and $g \propto R^{-2}$ dependencies. The RMS in the $\log M$ is comparable, so both parameterizations are equally effective, and the utility depends on which of $\log g$ or $\log \rho$ can be measured more precisely.

\renewcommand{\tablecaptiontext}{Same as Table \ref{tab:model_coeffs}, but for models using $Y = \log \rho$ instead of $Y = \log g - Y_0$. \label{tab:logrho}}
\begin{deluxetable*}{lr@{}lr@{}lr@{}lr@{}l}[t]
\tablecaption{\tablecaptiontext}
\tablehead{
\colhead{} & \multicolumn{4}{c}{8-term} & \multicolumn{4}{c}{14-term} \\
\cline{2-5} \cline{6-9}
\colhead{Term} & \multicolumn{2}{c}{$\log R$} & \multicolumn{2}{c}{$\log M$} & \multicolumn{2}{c}{$\log R$} & \multicolumn{2}{c}{$\log M$}
}
\startdata
${\rm const}$ & $0.0182$ & ${}\pm0.0007$ & $0.054$ & ${}\pm0.002$ & $0.0184$ & ${}\pm0.0009$ & $0.055$ & ${}\pm0.003$ \\
$X$ & $0.386$ & ${}\pm0.008$ & $1.16$ & ${}\pm0.03$ & $0.40$ & ${}\pm0.01$ & $1.19$ & ${}\pm0.03$ \\
$Y$ & $-0.368$ & ${}\pm0.003$ & $-0.105$ & ${}\pm0.008$ & $-0.365$ & ${}\pm0.003$ & $-0.10$ & ${}\pm0.01$ \\
$Z$ & $0.038$ & ${}\pm0.002$ & $0.113$ & ${}\pm0.007$ & $0.041$ & ${}\pm0.002$ & $0.124$ & ${}\pm0.007$ \\
$X^2$ & $0.12$ & ${}\pm0.01$ & $0.37$ & ${}\pm0.04$ & $-0.06$ & ${}\pm0.05$ & $-0.2$ & ${}\pm0.1$ \\
$Y^2$ & $0.008$ & ${}\pm0.002$ & $0.023$ & ${}\pm0.005$ & $0.012$ & ${}\pm0.003$ & $0.037$ & ${}\pm0.008$ \\
$Z^2$ & $0.004$ & ${}\pm0.001$ & $0.012$ & ${}\pm0.004$ & $0.005$ & ${}\pm0.001$ & $0.014$ & ${}\pm0.004$ \\
$Y Z$ & $-0.011$ & ${}\pm0.002$ & $-0.032$ & ${}\pm0.006$ & $-0.000$ & ${}\pm0.003$ & $-0.001$ & ${}\pm0.009$ \\
$X^3$ & \multicolumn{2}{c}{\nodata} & \multicolumn{2}{c}{\nodata} & $0.22$ & ${}\pm0.05$ & $0.7$ & ${}\pm0.1$ \\
$X Y$ & \multicolumn{2}{c}{\nodata} & \multicolumn{2}{c}{\nodata} & $-0.07$ & ${}\pm0.02$ & $-0.22$ & ${}\pm0.07$ \\
$X Y^2$ & \multicolumn{2}{c}{\nodata} & \multicolumn{2}{c}{\nodata} & $-0.06$ & ${}\pm0.01$ & $-0.18$ & ${}\pm0.03$ \\
$X Z$ & \multicolumn{2}{c}{\nodata} & \multicolumn{2}{c}{\nodata} & $0.02$ & $^{+0.02}_{-0.03}$ & $0.06$ & $^{+0.07}_{-0.08}$ \\
$X^2 Z$ & \multicolumn{2}{c}{\nodata} & \multicolumn{2}{c}{\nodata} & $0.41$ & $^{+0.09}_{-0.08}$ & $1.2$ & $^{+0.3}_{-0.2}$ \\
$X Y Z$ & \multicolumn{2}{c}{\nodata} & \multicolumn{2}{c}{\nodata} & $0.30$ & $^{+0.05}_{-0.04}$ & $0.9$ & ${}\pm0.1$ \\
\hline
$\sigma_{\rm int}$ [dex] & \multicolumn{2}{c}{$0.0054\pm0.0003$} & \multicolumn{2}{c}{$0.016\pm0.001$} & \multicolumn{2}{c}{$0.0050\pm0.0003$} & \multicolumn{2}{c}{$0.015\pm0.001$} \\
RMS [dex] & \multicolumn{2}{c}{$0.0101$} & \multicolumn{2}{c}{$0.0302$} & \multicolumn{2}{c}{$0.0100$} & \multicolumn{2}{c}{$0.0301$} \\
Sun & \multicolumn{2}{c}{$1.008\pm0.001$ $R_\odot$} & \multicolumn{2}{c}{$1.024\pm0.004$ $M_\odot$} & \multicolumn{2}{c}{$1.007\pm0.002$ $R_\odot$} & \multicolumn{2}{c}{$1.021\pm0.005$ $M_\odot$} \\
\enddata
\end{deluxetable*}

\bibliography{main}{}

@ARTICLE{Andersen91,
       author = {{Andersen}, J.},
        title = "{Accurate masses and radii of normal stars}",
      journal = {\aapr},
         year = 1991,
        month = jan,
       volume = {3},
       number = {2},
        pages = {91-126},
          doi = {10.1007/BF00873538},
       adsurl = {https://ui.adsabs.harvard.edu/abs/1991A&ARv...3...91A}
}

@ARTICLE{Beyer24,
       author = {{Beyer}, Alexa C. and {White}, Russel J.},
        title = "{The Kraft Break Sharply Divides Low-mass and Intermediate-mass Stars}",
      journal = {\apj},
         year = 2024,
        month = sep,
       volume = {973},
       number = {1},
          eid = {28},
        pages = {28},
          doi = {10.3847/1538-4357/ad6b0d},
archivePrefix = {arXiv},
       eprint = {2408.02638},
 primaryClass = {astro-ph.SR},
       adsurl = {https://ui.adsabs.harvard.edu/abs/2024ApJ...973...28B}
}

@ARTICLE{Bressan12,
       author = {{Bressan}, Alessandro and {Marigo}, Paola and {Girardi}, L{\'e}o. and {Salasnich}, Bernardo and {Dal Cero}, Claudia and {Rubele}, Stefano and {Nanni}, Ambra},
        title = "{PARSEC: stellar tracks and isochrones with the PAdova and TRieste Stellar Evolution Code}",
      journal = {\mnras},
         year = 2012,
        month = nov,
       volume = {427},
       number = {1},
        pages = {127-145},
          doi = {10.1111/j.1365-2966.2012.21948.x},
archivePrefix = {arXiv},
       eprint = {1208.4498},
 primaryClass = {astro-ph.SR},
       adsurl = {https://ui.adsabs.harvard.edu/abs/2012MNRAS.427..127B}
}

@ARTICLE{Cakirli08,
       author = {{{\c{C}}ak{\i}rl{\i}}, {\"O}. and {Ibano{\v{g}}lu}, C. and {Southworth}, J. and {Frasca}, A. and {Hernandez}, J.},
        title = "{Spectroscopic and photometric analysis of NSV24512: an early-type eclipsing binary embedded in a dust cloud}",
      journal = {\mnras},
         year = 2008,
        month = sep,
       volume = {389},
       number = {1},
        pages = {205-212},
          doi = {10.1111/j.1365-2966.2008.13528.x},
       adsurl = {https://ui.adsabs.harvard.edu/abs/2008MNRAS.389..205C}
}

@ARTICLE{Chen26,
       author = {{Chen}, Yue and {Wang}, Yifan and {Fu}, Yanning and {Yuan}, Ye and {Xia}, Fang and {Ye}, Tong},
        title = "{Fundamental parameters of the double-lined spectroscopic and detached eclipsing binary EPIC 211391083}",
      journal = {\mnras},
         year = 2026,
        month = sep,
       volume = {551},
       number = {2},
          eid = {stag1440},
        pages = {stag1440},
          doi = {10.1093/mnras/stag1440},
       adsurl = {https://ui.adsabs.harvard.edu/abs/2026MNRAS.551g1440C}
}

@ARTICLE{Choi16,
       author = {{Choi}, Jieun and {Dotter}, Aaron and {Conroy}, Charlie and {Cantiello}, Matteo and {Paxton}, Bill and {Johnson}, Benjamin D.},
        title = "{Mesa Isochrones and Stellar Tracks (MIST). I. Solar-scaled Models}",
      journal = {\apj},
         year = 2016,
        month = jun,
       volume = {823},
       number = {2},
          eid = {102},
        pages = {102},
          doi = {10.3847/0004-637X/823/2/102},
archivePrefix = {arXiv},
       eprint = {1604.08592},
 primaryClass = {astro-ph.SR},
       adsurl = {https://ui.adsabs.harvard.edu/abs/2016ApJ...823..102C}
}

@ARTICLE{Christiansen25,
       author = {{Christiansen}, Jessie L. and {McElroy}, Douglas L. and {Harbut}, Marcy and {Ciardi}, David R. and {Crane}, Megan and {Good}, John and {Hardegree-Ullman}, Kevin K. and {Kesseli}, Aurora Y. and {Lund}, Michael B. and {Lynn}, Meca and et al.},
        title = "{The NASA Exoplanet Archive and Exoplanet Follow-up Observing Program: Data, Tools, and Usage}",
      journal = {\psj},
         year = 2025,
        month = aug,
       volume = {6},
       number = {8},
          eid = {186},
        pages = {186},
          doi = {10.3847/PSJ/ade3c2},
archivePrefix = {arXiv},
       eprint = {2506.03299},
 primaryClass = {astro-ph.EP},
       adsurl = {https://ui.adsabs.harvard.edu/abs/2025PSJ.....6..186C}
}

@ARTICLE{Claret16,
       author = {{Claret}, A. and {Torres}, G.},
        title = "{The dependence of convective core overshooting on stellar mass}",
      journal = {\aap},
         year = 2016,
        month = jul,
       volume = {592},
          eid = {A15},
        pages = {A15},
          doi = {10.1051/0004-6361/201628779},
       adsurl = {https://ui.adsabs.harvard.edu/abs/2016A&A...592A..15C}
}

@ARTICLE{ClarkCunningham19,
       author = {{Clark Cunningham}, Joni Marie and {Rawls}, Meredith L. and {Windemuth}, Diana and {Ali}, Aleezah and {Jackiewicz}, Jason and {Agol}, Eric and {Stassun}, Keivan G.},
        title = "{APOGEE/Kepler Overlap Yields Orbital Solutions for a Variety of Eclipsing Binaries}",
      journal = {\aj},
         year = 2019,
        month = sep,
       volume = {158},
       number = {3},
          eid = {106},
        pages = {106},
          doi = {10.3847/1538-3881/ab2d2b},
archivePrefix = {arXiv},
       eprint = {1907.00095},
 primaryClass = {astro-ph.SR},
       adsurl = {https://ui.adsabs.harvard.edu/abs/2019AJ....158..106C}
}

@ARTICLE{Clausen80,
       author = {{Clausen}, J.~V. and {Nordstrom}, B.},
        title = "{Four-colour photometry of eclipsing binaries. XIa.}",
      journal = {\aap},
         year = 1980,
        month = mar,
       volume = {83},
        pages = {339-347},
       adsurl = {https://ui.adsabs.harvard.edu/abs/1980A&A....83..339C}
}

@ARTICLE{Cui12,
       author = {{Cui}, Xiang-Qun and {Zhao}, Yong-Heng and {Chu}, Yao-Quan and {Li}, Guo-Ping and {Li}, Qi and {Zhang}, Li-Ping and {Su}, Hong-Jun and {Yao}, Zheng-Qiu and {Wang}, Ya-Nan and {Xing}, Xiao-Zheng and {Li}, Xin-Nan and {Zhu}, Yong-Tian and {Wang}, Gang and {Gu}, Bo-Zhong and {Luo}, A. -Li and {Xu}, Xin-Qi and {Zhang}, Zhen-Chao and {Liu}, Gen-Rong and {Zhang}, Hao-Tong and {Yang}, De-Hua and {Cao}, Shu-Yun and {Chen}, Hai-Yuan and {Chen}, Jian-Jun and {Chen}, Kun-Xin and {Chen}, Ying and {Chu}, Jia-Ru and {Feng}, Lei and {Gong}, Xue-Fei and {Hou}, Yong-Hui and {Hu}, Hong-Zhuan and {Hu}, Ning-Sheng and {Hu}, Zhong-Wen and {Jia}, Lei and {Jiang}, Fang-Hua and {Jiang}, Xiang and {Jiang}, Zi-Bo and {Jin}, Ge and {Li}, Ai-Hua and {Li}, Yan and {Li}, Ye-Ping and {Liu}, Guan-Qun and {Liu}, Zhi-Gang and {Lu}, Wen-Zhi and {Mao}, Yin-Dun and {Men}, Li and {Qi}, Yong-Jun and {Qi}, Zhao-Xiang and {Shi}, Huo-Ming and {Tang}, Zheng-Hong and {Tao}, Qing-Sheng and {Wang}, Da-Qi and {Wang}, Dan and {Wang}, Guo-Min and {Wang}, Hai and {Wang}, Jia-Ning and {Wang}, Jian and {Wang}, Jian-Ling and {Wang}, Jian-Ping and {Wang}, Lei and {Wang}, Shu-Qing and {Wang}, You and {Wang}, Yue-Fei and {Xu}, Ling-Zhe and {Xu}, Yan and {Yang}, Shi-Hai and {Yu}, Yong and {Yuan}, Hui and {Yuan}, Xiang-Yan and {Zhai}, Chao and {Zhang}, Jing and {Zhang}, Yan-Xia and {Zhang}, Yong and {Zhao}, Ming and {Zhou}, Fang and {Zhou}, Guo-Hua and {Zhu}, Jie and {Zou}, Si-Cheng},
        title = "{The Large Sky Area Multi-Object Fiber Spectroscopic Telescope (LAMOST)}",
      journal = {Research in Astronomy and Astrophysics},
         year = 2012,
        month = sep,
       volume = {12},
       number = {9},
        pages = {1197-1242},
          doi = {10.1088/1674-4527/12/9/003},
       adsurl = {https://ui.adsabs.harvard.edu/abs/2012RAA....12.1197C}
}

@ARTICLE{Davis26,
       author = {{Davis}, Yasmin T. and {Triaud}, Amaury H.~M.~J. and {Martin}, David V. and {Freckelton}, Alix V. and {French}, Jenni R. and {Stevenson}, Adam T. and {Mortier}, Annelies and {Collins}, Karen and {Maxted}, Pierre F.~L. and {Sebastian}, Daniel and et al.},
        title = "{The EBLM Project - XIX. The absence of radius inflation in a homogeneous analysis of 66 Northern low-mass eclipsing binaries}",
      journal = {arXiv e-prints},
         year = 2026,
        month = oct,
          eid = {arXiv:2610.18338},
        pages = {arXiv:2610.18338},
          doi = {10.48550/arXiv.2610.18338},
archivePrefix = {arXiv},
       eprint = {2610.18338},
 primaryClass = {astro-ph.SR},
       adsurl = {https://ui.adsabs.harvard.edu/abs/2026arXiv261018338D}
}

@ARTICLE{Dotter16,
       author = {{Dotter}, Aaron},
        title = "{MESA Isochrones and Stellar Tracks (MIST) 0: Methods for the Construction of Stellar Isochrones}",
      journal = {\apjs},
         year = 2016,
        month = jan,
       volume = {222},
       number = {1},
          eid = {8},
        pages = {8},
          doi = {10.3847/0067-0049/222/1/8},
archivePrefix = {arXiv},
       eprint = {1601.05144},
 primaryClass = {astro-ph.SR},
       adsurl = {https://ui.adsabs.harvard.edu/abs/2016ApJS..222....8D}
}

@ARTICLE{Enoch10,
       author = {{Enoch}, B. and {Collier Cameron}, A. and {Parley}, N.~R. and {Hebb}, L.},
        title = "{An improved method for estimating the masses of stars with transiting planets}",
      journal = {\aap},
         year = 2010,
        month = jun,
       volume = {516},
          eid = {A33},
        pages = {A33},
          doi = {10.1051/0004-6361/201014326},
archivePrefix = {arXiv},
       eprint = {1004.1991},
 primaryClass = {astro-ph.EP},
       adsurl = {https://ui.adsabs.harvard.edu/abs/2010A&A...516A..33E}
}

@ARTICLE{ElBadry18,
       author = {{El-Badry}, Kareem and {Rix}, Hans-Walter and {Ting}, Yuan-Sen and {Weisz}, Daniel R. and {Bergemann}, Maria and {Cargile}, Phillip and {Conroy}, Charlie and {Eilers}, Anna-Christina},
        title = "{Signatures of unresolved binaries in stellar spectra: implications for spectral fitting}",
      journal = {\mnras},
         year = 2018,
        month = feb,
       volume = {473},
       number = {4},
        pages = {5043-5049},
          doi = {10.1093/mnras/stx2758},
archivePrefix = {arXiv},
       eprint = {1709.03983},
 primaryClass = {astro-ph.SR},
       adsurl = {https://ui.adsabs.harvard.edu/abs/2018MNRAS.473.5043E}
}

@ARTICLE{Fabry22,
       author = {{Fabry}, M. and {Marchant}, P. and {Sana}, H.},
        title = "{Modeling overcontact binaries. I. The effect of tidal deformation}",
      journal = {\aap},
         year = 2022,
        month = may,
       volume = {661},
          eid = {A123},
        pages = {A123},
          doi = {10.1051/0004-6361/202243094},
archivePrefix = {arXiv},
       eprint = {2202.08852},
 primaryClass = {astro-ph.SR},
       adsurl = {https://ui.adsabs.harvard.edu/abs/2022A&A...661A.123F}
}

@ARTICLE{ForemanMackey13,
       author = {{Foreman-Mackey}, Daniel and {Hogg}, David W. and {Lang}, Dustin and {Goodman}, Jonathan},
        title = "{emcee: The MCMC Hammer}",
      journal = {\pasp},
         year = 2013,
        month = mar,
       volume = {125},
       number = {925},
        pages = {306},
          doi = {10.1086/670067},
archivePrefix = {arXiv},
       eprint = {1202.3665},
 primaryClass = {astro-ph.IM},
       adsurl = {https://ui.adsabs.harvard.edu/abs/2013PASP..125..306F}
}

@ARTICLE{Fernandes21,
       author = {{Fernandes}, Jo{\~a}o and {Gafeira}, Ricardo and {Andersen}, Johannes},
        title = "{Revisiting the mass- and radius-luminosity relations for FGK main-sequence stars}",
      journal = {\aap},
         year = 2021,
        month = mar,
       volume = {647},
          eid = {A90},
        pages = {A90},
          doi = {10.1051/0004-6361/202040035},
archivePrefix = {arXiv},
       eprint = {2103.11044},
 primaryClass = {astro-ph.SR},
       adsurl = {https://ui.adsabs.harvard.edu/abs/2021A&A...647A..90F}
}

@ARTICLE{GaiaCollaboration2022,
       author = {{Gaia Collaboration} and {Vallenari}, A. and {Brown}, A.~G.~A. and {Prusti}, T. and {de Bruijne}, J.~H.~J. and {Arenou}, F. and {Babusiaux}, C. and {Biermann}, M. and {Creevey}, O.~L. and {Ducourant}, C. and {Evans}, D.~W. and {Eyer}, L. and {Guerra}, R. and {Hutton}, A. and {Jordi}, C. and {Klioner}, S.~A. and {Lammers}, U.~L. and {Lindegren}, L. and {Luri}, X. and {Mignard}, F. and {Panem}, C. and {Pourbaix}, D. and {Randich}, S. and {Sartoretti}, P. and {Soubiran}, C. and {Tanga}, P. and {Walton}, N.~A. and {Bailer-Jones}, C.~A.~L. and {Bastian}, U. and {Drimmel}, R. and {Jansen}, F. and {Katz}, D. and {Lattanzi}, M.~G. and {van Leeuwen}, F. and {Bakker}, J. and {Cacciari}, C. and {Casta{\~n}eda}, J. and {De Angeli}, F. and {Fabricius}, C. and {Fouesneau}, M. and {Fr{\'e}mat}, Y. and {Galluccio}, L. and {Guerrier}, A. and {Heiter}, U. and {Masana}, E. and {Messineo}, R. and {Mowlavi}, N. and {Nicolas}, C. and {Nienartowicz}, K. and {Pailler}, F. and {Panuzzo}, P. and {Riclet}, F. and {Roux}, W. and {Seabroke}, G.~M. and {Sordo}, R. and {Th{\'e}venin}, F. and {Gracia-Abril}, G. and {Portell}, J. and {Teyssier}, D. and {Altmann}, M. and {Andrae}, R. and {Audard}, M. and {Bellas-Velidis}, I. and {Benson}, K. and {Berthier}, J. and {Blomme}, R. and {Burgess}, P.~W. and {Busonero}, D. and {Busso}, G. and {C{\'a}novas}, H. and {Carry}, B. and {Cellino}, A. and {Cheek}, N. and {Clementini}, G. and {Damerdji}, Y. and {Davidson}, M. and {de Teodoro}, P. and {Nu{\~n}ez Campos}, M. and {Delchambre}, L. and {Dell'Oro}, A. and {Esquej}, P. and {Fern{\'a}ndez-Hern{\'a}ndez}, J. and {Fraile}, E. and {Garabato}, D. and {Garc{\'\i}a-Lario}, P. and {Gosset}, E. and {Haigron}, R. and {Halbwachs}, J.-L. and {Hambly}, N.~C. and {Harrison}, D.~L. and {Hern{\'a}ndez}, J. and {Hestroffer}, D. and {Hodgkin}, S.~T. and {Holl}, B. and {Jan{\ss}en}, K. and {Jevardat de Fombelle}, G. and {Jordan}, S. and {Krone-Martins}, A. and {Lanzafame}, A.~C. and {L{\"o}ffler}, W. and {Marchal}, O. and {Marrese}, P.~M. and {Moitinho}, A. and {Muinonen}, K. and {Osborne}, P. and {Pancino}, E. and {Pauwels}, T. and {Recio-Blanco}, A. and {Reyl{\'e}}, C. and {Riello}, M. and {Rimoldini}, L. and {Roegiers}, T. and {Rybizki}, J. and {Sarro}, L.~M. and {Siopis}, C. and {Smith}, M. and {Sozzetti}, A. and {Utrilla}, E. and {van Leeuwen}, M. and {Abbas}, U. and {{\'A}brah{\'a}m}, P. and {Abreu Aramburu}, A. and {Aerts}, C. and {Aguado}, J.~J. and {Ajaj}, M. and {Aldea-Montero}, F. and {Altavilla}, G. and {{\'A}lvarez}, M.~A. and {Alves}, J. and {Anders}, F. and {Anderson}, R.~I. and {Anglada Varela}, E. and {Antoja}, T. and {Baines}, D. and {Baker}, S.~G. and {Balaguer-N{\'u}{\~n}ez}, L. and {Balbinot}, E. and {Balog}, Z. and {Barache}, C. and {Barbato}, D. and {Barros}, M. and {Barstow}, M.~A. and {Bartolom{\'e}}, S. and {Bassilana}, J.-L. and {Bauchet}, N. and {Becciani}, U. and {Bellazzini}, M. and {Berihuete}, A. and {Bernet}, M. and {Bertone}, S. and {Bianchi}, L. and {Binnenfeld}, A. and {Blanco-Cuaresma}, S. and {Blazere}, A. and {Boch}, T. and {Bombrun}, A. and {Bossini}, D. and {Bouquillon}, S. and {Bragaglia}, A. and {Bramante}, L. and {Breedt}, E. and {Bressan}, A. and {Brouillet}, N. and {Brugaletta}, E. and {Bucciarelli}, B. and {Burlacu}, A. and {Butkevich}, A.~G. and {Buzzi}, R. and {Caffau}, E. and {Cancelliere}, R. and {Cantat-Gaudin}, T. and {Carballo}, R. and {Carlucci}, T. and {Carnerero}, M.~I. and {Carrasco}, J.~M. and {Casamiquela}, L. and {Castellani}, M. and {Castro-Ginard}, A. and {Chaoul}, L. and {Charlot}, P. and {Chemin}, L. and {Chiaramida}, V. and {Chiavassa}, A. and {Chornay}, N. and {Comoretto}, G. and {Contursi}, G. and {Cooper}, W.~J. and {Cornez}, T. and {Cowell}, S. and {Crifo}, F. and {Cropper}, M. and {Crosta}, M. and {Crowley}, C. and {Dafonte}, C. and {Dapergolas}, A. and {David}, M. and {David}, P. and {de Laverny}, P. and {De Luise}, F. and {De March}, R.},
        title = "{Gaia Data Release 3. Summary of the content and survey properties}",
      journal = {\aap},
         year = 2023,
        month = jun,
       volume = {674},
          eid = {A1},
        pages = {A1},
          doi = {10.1051/0004-6361/202243940},
archivePrefix = {arXiv},
       eprint = {2208.00211},
 primaryClass = {astro-ph.GA},
       adsurl = {https://ui.adsabs.harvard.edu/abs/2023A&A...674A...1G}
}

@ARTICLE{Gaulme16,
       author = {{Gaulme}, P. and {McKeever}, J. and {Jackiewicz}, J. and {Rawls}, M.~L. and {Corsaro}, E. and {Mosser}, B. and {Southworth}, J. and {Mahadevan}, S. and {Bender}, C. and {Deshpande}, R.},
        title = "{Testing the Asteroseismic Scaling Relations for Red Giants with Eclipsing Binaries Observed by Kepler}",
      journal = {\apj},
         year = 2016,
        month = dec,
       volume = {832},
       number = {2},
          eid = {121},
        pages = {121},
          doi = {10.3847/0004-637X/832/2/121},
archivePrefix = {arXiv},
       eprint = {1609.06645},
 primaryClass = {astro-ph.SR},
       adsurl = {https://ui.adsabs.harvard.edu/abs/2016ApJ...832..121G}
}

@ARTICLE{Gill19,
       author = {{Gill}, S. and {Maxted}, P.~F.~L. and {Evans}, J.~A. and {Evans}, D.~F. and {Southworth}, J. and {Smalley}, B. and {Gary}, B.~L. and {Anderson}, D.~R. and {Bouchy}, F. and {Cameron}, A.~C. and et al.},
        title = "{The EBLM project. VI. Mass and radius of five low-mass stars in F+M binaries discovered by the WASP survey}",
      journal = {\aap},
         year = 2019,
        month = jun,
       volume = {626},
          eid = {A119},
        pages = {A119},
          doi = {10.1051/0004-6361/201833054},
archivePrefix = {arXiv},
       eprint = {1904.12695},
 primaryClass = {astro-ph.SR},
       adsurl = {https://ui.adsabs.harvard.edu/abs/2019A&A...626A.119G}
}

@ARTICLE{Gorda13,
       author = {{Gorda}, S. Yu.},
        title = "{CCD spectrophotometry of CC Cas. I. Radial velocity curves}",
      journal = {Astrophysical Bulletin},
         year = 2013,
        month = jan,
       volume = {68},
       number = {1},
        pages = {101-106},
          doi = {10.1134/S1990341313010094},
       adsurl = {https://ui.adsabs.harvard.edu/abs/2013AstBu..68..101G}
}

@ARTICLE{Graczyk16,
       author = {{Graczyk}, D. and {Smolec}, R. and {Pavlovski}, K. and {Southworth}, J. and {Pietrzy{\'n}ski}, G. and {Maxted}, P.~F.~L. and {Konorski}, P. and {Gieren}, W. and {Pilecki}, B. and {Taormina}, M. and {Suchomska}, K. and {Karczmarek}, P. and {G{\'o}rski}, M. and {Wielg{\'o}rski}, P. and {Anderson}, R.~I.},
        title = "{A solar twin in the eclipsing binary LL Aquarii}",
      journal = {\aap},
         year = 2016,
        month = oct,
       volume = {594},
          eid = {A92},
        pages = {A92},
          doi = {10.1051/0004-6361/201628918},
archivePrefix = {arXiv},
       eprint = {1608.01000},
 primaryClass = {astro-ph.SR},
       adsurl = {https://ui.adsabs.harvard.edu/abs/2016A&A...594A..92G}
}

@ARTICLE{Graczyk18,
       author = {{Graczyk}, Dariusz and {Pietrzy{\'n}ski}, Grzegorz and {Thompson}, Ian B. and {Gieren}, Wolfgang and {Pilecki}, Bogumi{\l} and {Konorski}, Piotr and {Villanova}, Sandro and {G{\'o}rski}, Marek and {Suchomska}, Ksenia and {Karczmarek}, Paulina and {Stepie{\'n}}, Kazimierz and {Storm}, Jesper and {Taormina}, M{\'o}nica and {Ko{\l}aczkowski}, Zbigniew and {Wielg{\'o}rski}, Piotr and {Narloch}, Weronika and {Zgirski}, Bart{\l}omiej and {Gallenne}, Alexandre and {Ostrowski}, Jakub and {Smolec}, Rados{\l}aw and {Udalski}, Andrzej and {Soszy{\'n}ski}, Igor and {Kervella}, Pierre and {Nardetto}, Nicolas and {Szyma{\'n}ski}, Micha{\l} K. and {Wyrzykowski}, {\L}ukasz and {Ulaczyk}, Krzysztof and {Poleski}, Rados{\l}aw and {Pietrukowicz}, Pawe{\l} and {Koz{\l}owski}, Szymon and {Skowron}, Jan and {Mr{\'o}z}, Przemys{\l}aw},
        title = "{The Late-type Eclipsing Binaries in the Large Magellanic Cloud: Catalog of Fundamental Physical Parameters}",
      journal = {\apj},
         year = 2018,
        month = jun,
       volume = {860},
       number = {1},
          eid = {1},
        pages = {1},
          doi = {10.3847/1538-4357/aac2bf},
archivePrefix = {arXiv},
       eprint = {1805.04952},
 primaryClass = {astro-ph.SR},
       adsurl = {https://ui.adsabs.harvard.edu/abs/2018ApJ...860....1G}
}

@ARTICLE{Graczyk20,
       author = {{Graczyk}, Dariusz and {Pietrzy{\'n}ski}, Grzegorz and {Thompson}, Ian B. and {Gieren}, Wolfgang and {Zgirski}, Bart{\l}omiej and {Villanova}, Sandro and {G{\'o}rski}, Marek and {Wielg{\'o}rski}, Piotr and {Karczmarek}, Paulina and {Narloch}, Weronika and {Pilecki}, Bogumi{\l} and {Taormina}, Monica and {Smolec}, Rados{\l}aw and {Suchomska}, Ksenia and {Gallenne}, Alexandre and {Nardetto}, Nicolas and {Storm}, Jesper and {Kudritzki}, Rolf-Peter and {Ka{\l}uszy{\'n}ski}, Miko{\l}aj and {Pych}, Wojciech},
        title = "{A Distance Determination to the Small Magellanic Cloud with an Accuracy of Better than Two Percent Based on Late-type Eclipsing Binary Stars}",
      journal = {\apj},
         year = 2020,
        month = nov,
       volume = {904},
       number = {1},
          eid = {13},
        pages = {13},
          doi = {10.3847/1538-4357/abbb2b},
archivePrefix = {arXiv},
       eprint = {2010.08754},
 primaryClass = {astro-ph.GA},
       adsurl = {https://ui.adsabs.harvard.edu/abs/2020ApJ...904...13G}
}

@ARTICLE{Graczyk22,
       author = {{Graczyk}, D. and {Pietrzy{\'n}ski}, G. and {Galan}, C. and {Southworth}, J. and {Gieren}, W. and {Ka{\l}uszy{\'n}ski}, M. and {Zgirski}, B. and {Gallenne}, A. and {G{\'o}rski}, M. and {Hajdu}, G. and et al.},
        title = "{Surface brightness-colour relations of dwarf stars from detached eclipsing binaries. I. Calibrating sample}",
      journal = {\aap},
         year = 2022,
        month = oct,
       volume = {666},
          eid = {A128},
        pages = {A128},
          doi = {10.1051/0004-6361/202244122},
archivePrefix = {arXiv},
       eprint = {2208.07257},
 primaryClass = {astro-ph.SR},
       adsurl = {https://ui.adsabs.harvard.edu/abs/2022A&A...666A.128G}
}

@ARTICLE{Hekker10,
       author = {{Hekker}, S. and {Debosscher}, J. and {Huber}, D. and {Hidas}, M.~G. and {De Ridder}, J. and {Aerts}, C. and {Stello}, D. and {Bedding}, T.~R. and {Gilliland}, R.~L. and {Christensen-Dalsgaard}, J. and {Brown}, T.~M. and {Kjeldsen}, H. and {Borucki}, W.~J. and {Koch}, D. and {Jenkins}, J.~M. and {Van Winckel}, H. and {Beck}, P.~G. and {Blomme}, J. and {Southworth}, J. and {Pigulski}, A. and {Chaplin}, W.~J. and {Elsworth}, Y.~P. and {Stevens}, I.~R. and {Dreizler}, S. and {Kurtz}, D.~W. and {Maceroni}, C. and {Cardini}, D. and {Derekas}, A. and {Suran}, M.~D.},
        title = "{Discovery of a Red Giant with Solar-like Oscillations in an Eclipsing Binary System from Kepler Space-based Photometry}",
      journal = {\apjl},
         year = 2010,
        month = apr,
       volume = {713},
       number = {2},
        pages = {L187-L191},
          doi = {10.1088/2041-8205/713/2/L187},
archivePrefix = {arXiv},
       eprint = {1001.0399},
 primaryClass = {astro-ph.SR},
       adsurl = {https://ui.adsabs.harvard.edu/abs/2010ApJ...713L.187H}
}

@ARTICLE{Helminiak19_X,
       author = {{He{\l}miniak}, K.~G. and {Tokovinin}, A. and {Niemczura}, E. and {Paw{\l}aszek}, R. and {Yanagisawa}, K. and {Brahm}, R. and {Espinoza}, N. and {Ukita}, N. and {Kambe}, E. and {Ratajczak}, M. and {Hempel}, M. and {Jord{\'a}n}, A. and {Konacki}, M. and {Sybilski}, P. and {Koz{\l}owski}, S.~K. and {Litwicki}, M. and {Tamura}, M.},
        title = "{Orbital and physical parameters of eclipsing binaries from the All-Sky Automated Survey catalogue. X. Three high-contrast systems with secondaries detected with IR spectroscopy}",
      journal = {\aap},
         year = 2019,
        month = feb,
       volume = {622},
          eid = {A114},
        pages = {A114},
          doi = {10.1051/0004-6361/201732482},
archivePrefix = {arXiv},
       eprint = {1812.04319},
 primaryClass = {astro-ph.SR},
       adsurl = {https://ui.adsabs.harvard.edu/abs/2019A&A...622A.114H}
}

@ARTICLE{Helminiak24,
       author = {{He{\l}miniak}, K.~G. and {Olszewska}, J.~M. and {Puciata-Mroczynska}, M. and {Pawar}, T.},
        title = "{High-resolution spectroscopy of detached eclipsing binaries during total eclipses}",
      journal = {\aap},
         year = 2024,
        month = nov,
       volume = {691},
          eid = {A170},
        pages = {A170},
          doi = {10.1051/0004-6361/202450607},
archivePrefix = {arXiv},
       eprint = {2409.20144},
 primaryClass = {astro-ph.SR},
       adsurl = {https://ui.adsabs.harvard.edu/abs/2024A&A...691A.170H}
}

@ARTICLE{Hill94,
       author = {{Hill}, G. and {Hilditch}, R.~W. and {Aikman}, G.~C.~L. and {Khalesseh}, B.},
        title = "{Studies of early-type variable stars. VIII. The massive binary system CC Cassiopeiae.}",
      journal = {\aap},
         year = 1994,
        month = feb,
       volume = {282},
        pages = {455-466},
       adsurl = {https://ui.adsabs.harvard.edu/abs/1994A&A...282..455H}
}

@ARTICLE{Holgado25,
       author = {{Holgado}, G. and {Ma{\'\i}z Apell{\'a}niz}, J. and {Gamen}, R.~C.},
        title = "{MONOS: Multiplicity Of Northern O-type Spectroscopic systems: III. New orbits and Gaia─TESS analysis for ten SB2E systems}",
      journal = {\aap},
         year = 2025,
        month = sep,
       volume = {701},
          eid = {A246},
        pages = {A246},
          doi = {10.1051/0004-6361/202556068},
archivePrefix = {arXiv},
       eprint = {2508.04272},
 primaryClass = {astro-ph.SR},
       adsurl = {https://ui.adsabs.harvard.edu/abs/2025A&A...701A.246H}
}

@ARTICLE{Ibanoglu08,
       author = {{Ibano{\v{g}}lu}, C. and {Evren}, S. and {Ta{\textcommabelow s}}, G. and {{\c{C}}ak{\i}rl{\i}}, {\"O}. and {Bozkurt}, Z. and {Af{\textcommabelow s}ar}, M. and {Frasca}, A. and {Sipahi}, E. and {Dal}, H.~A. and {{\"O}zdarcan}, O. and {{\c{C}}amurdan}, D. Zengin and {{\c{C}}amurdan}, M.},
        title = "{Spectroscopic and photometric observations of the selected Algol-type binaries - II. V2080 Cygni and V2365 Ophiuchi}",
      journal = {\mnras},
         year = 2008,
        month = feb,
       volume = {384},
       number = {1},
        pages = {331-342},
          doi = {10.1111/j.1365-2966.2007.12701.x},
       adsurl = {https://ui.adsabs.harvard.edu/abs/2008MNRAS.384..331I}
}

@ARTICLE{Kaluzny13,
       author = {{Kaluzny}, J. and {Thompson}, I.~B. and {Rozyczka}, M. and {Dotter}, A. and {Krzeminski}, W. and {Pych}, W. and {Rucinski}, S.~M. and {Burley}, G.~S. and {Shectman}, S.~A.},
        title = "{The Cluster AgeS Experiment (CASE). V. Analysis of Three Eclipsing Binaries in the Globular Cluster M4}",
      journal = {\aj},
         year = 2013,
        month = feb,
       volume = {145},
       number = {2},
          eid = {43},
        pages = {43},
          doi = {10.1088/0004-6256/145/2/43},
archivePrefix = {arXiv},
       eprint = {1301.2946},
 primaryClass = {astro-ph.SR},
       adsurl = {https://ui.adsabs.harvard.edu/abs/2013AJ....145...43K}
}

@ARTICLE{Kamenec26,
       author = {{Kamenec}, Mat{\'u}{\v{s}} and {Gajdo{\v{s}}}, Pavol and {Va{\v{n}}ko}, Martin and {Munagala}, Reddy Charan Reddy and {Skarka}, Marek and {Parimucha}, {\v{S}}tefan},
        title = "{Revision of the Detached Eclipsing System IR Cas from TESS Observations, Ground-based Photometry, and Spectroscopy}",
      journal = {\aj},
         year = 2026,
        month = aug,
       volume = {172},
       number = {2},
          eid = {126},
        pages = {126},
          doi = {10.3847/1538-3881/ae87a2},
archivePrefix = {arXiv},
       eprint = {2607.02076},
 primaryClass = {astro-ph.SR},
       adsurl = {https://ui.adsabs.harvard.edu/abs/2026AJ....172..126K}
}

@ARTICLE{Kochanek17,
       author = {{Kochanek}, C.~S. and {Shappee}, B.~J. and {Stanek}, K.~Z. and {Holoien}, T.~W. -S. and {Thompson}, Todd A. and {Prieto}, J.~L. and {Dong}, Subo and {Shields}, J.~V. and {Will}, D. and {Britt}, C. and {Perzanowski}, D. and {Pojma{\'n}ski}, G.},
        title = "{The All-Sky Automated Survey for Supernovae (ASAS-SN) Light Curve Server v1.0}",
      journal = {\pasp},
         year = 2017,
        month = oct,
       volume = {129},
       number = {980},
        pages = {104502},
          doi = {10.1088/1538-3873/aa80d9},
archivePrefix = {arXiv},
       eprint = {1706.07060},
 primaryClass = {astro-ph.SR},
       adsurl = {https://ui.adsabs.harvard.edu/abs/2017PASP..129j4502K}
}

@ARTICLE{Kochukhov18,
       author = {{Kochukhov}, O. and {Johnston}, C. and {Alecian}, E. and {Wade}, G.~A. and {BinaMIcS Collaboration}},
        title = "{HD 66051: the first eclipsing binary hosting an early-type magnetic star}",
      journal = {\mnras},
         year = 2018,
        month = aug,
       volume = {478},
       number = {2},
        pages = {1749-1762},
          doi = {10.1093/mnras/sty1118},
archivePrefix = {arXiv},
       eprint = {1804.10537},
 primaryClass = {astro-ph.SR},
       adsurl = {https://ui.adsabs.harvard.edu/abs/2018MNRAS.478.1749K}
}

@ARTICLE{Kraft67,
       author = {{Kraft}, Robert P.},
        title = "{Studies of Stellar Rotation. V. The Dependence of Rotation on Age among Solar-Type Stars}",
      journal = {\apj},
         year = 1967,
        month = nov,
       volume = {150},
        pages = {551},
          doi = {10.1086/149359},
       adsurl = {https://ui.adsabs.harvard.edu/abs/1967ApJ...150..551K}
}

@ARTICLE{Kraus11,
       author = {{Kraus}, Adam L. and {Tucker}, Roy A. and {Thompson}, Michael I. and {Craine}, Eric R. and {Hillenbrand}, Lynne A.},
        title = "{The Mass-Radius(-Rotation?) Relation for Low-mass Stars}",
      journal = {\apj},
         year = 2011,
        month = feb,
       volume = {728},
       number = {1},
          eid = {48},
        pages = {48},
          doi = {10.1088/0004-637X/728/1/48},
archivePrefix = {arXiv},
       eprint = {1011.2757},
 primaryClass = {astro-ph.SR},
       adsurl = {https://ui.adsabs.harvard.edu/abs/2011ApJ...728...48K}
}

@ARTICLE{Kuhn16,
       author = {{Kuhn}, Rudolf B. and {Rodriguez}, Joseph E. and {Collins}, Karen A. and {Lund}, Michael B. and {Siverd}, Robert J. and {Col{\'o}n}, Knicole D. and {Pepper}, Joshua and {Stassun}, Keivan G. and {Cargile}, Phillip A. and {James}, David J. and et al.},
        title = "{KELT-10b: the first transiting exoplanet from the KELT-South survey - a hot sub-Jupiter transiting a V = 10.7 early G-star}",
      journal = {\mnras},
         year = 2016,
        month = jul,
       volume = {459},
       number = {4},
        pages = {4281-4298},
          doi = {10.1093/mnras/stw880},
archivePrefix = {arXiv},
       eprint = {1509.02323},
 primaryClass = {astro-ph.EP},
       adsurl = {https://ui.adsabs.harvard.edu/abs/2016MNRAS.459.4281K}
}

@ARTICLE{Lacy16,
       author = {{Lacy}, Claud H. Sandberg and {Fekel}, Francis C. and {Pavlovski}, Kre{\v{s}}imir and {Torres}, Guillermo and {Muterspaugh}, Matthew W.},
        title = "{Absolute Properties of the Pre-main-sequence Eclipsing Binary Star NP Persei}",
      journal = {\aj},
         year = 2016,
        month = jul,
       volume = {152},
       number = {1},
          eid = {2},
        pages = {2},
          doi = {10.3847/0004-6256/152/1/2},
       adsurl = {https://ui.adsabs.harvard.edu/abs/2016AJ....152....2L}
}

@ARTICLE{Marcadon20,
       author = {{Marcadon}, F. and {He{\l}miniak}, K.~G. and {Marques}, J.~P. and {Paw{\l}aszek}, R. and {Sybilski}, P. and {Koz{\l}owski}, S.~K. and {Ratajczak}, M. and {Konacki}, M.},
        title = "{Analysis of eclipsing binaries in multiple stellar systems: the case of V1200 Centauri}",
      journal = {\mnras},
         year = 2020,
        month = dec,
       volume = {499},
       number = {2},
        pages = {3019-3033},
          doi = {10.1093/mnras/staa3040},
archivePrefix = {arXiv},
       eprint = {2009.13496},
 primaryClass = {astro-ph.SR},
       adsurl = {https://ui.adsabs.harvard.edu/abs/2020MNRAS.499.3019M}
}

@ARTICLE{Michalska13,
       author = {{Michalska}, G. and {Niemczura}, E. and {Pigulski}, A. and {Ste{\'s}licki}, M. and {Williams}, A.},
        title = "{The massive eclipsing system ALS 1135 and variable stars in the field of the distant OB association Bochum 7}",
      journal = {\mnras},
         year = 2013,
        month = feb,
       volume = {429},
       number = {2},
        pages = {1354-1366},
          doi = {10.1093/mnras/sts419},
archivePrefix = {arXiv},
       eprint = {1211.3587},
 primaryClass = {astro-ph.SR},
       adsurl = {https://ui.adsabs.harvard.edu/abs/2013MNRAS.429.1354M}
}

@ARTICLE{Miller26,
       author = {{Miller}, N.~J. and {Maxted}, P.~F.~L. and {Hahlin}, A. and {Graczyk}, D.},
        title = "{Fundamental effective temperature measurements for eclipsing binary stars ─ VII. The solar twin in LL Aquarii}",
      journal = {\mnras},
         year = 2026,
        month = jan,
       volume = {545},
       number = {3},
          eid = {staf2078},
        pages = {staf2078},
          doi = {10.1093/mnras/staf2078},
archivePrefix = {arXiv},
       eprint = {2511.16519},
 primaryClass = {astro-ph.SR},
       adsurl = {https://ui.adsabs.harvard.edu/abs/2026MNRAS.545f2078M}
}

@ARTICLE{Meszaros25,
       author = {{M{\'e}sz{\'a}ros}, Szabolcs and {Jofr{\'e}}, Paula and {Johnson}, Jennifer A. and {Bird}, Jonathan C. and {Bovy}, Jo and {Casey}, Andrew R. and {Chanam{\'e}}, Julio and {Cunha}, Katia and {De Lee}, Nathan and {Frinchaboy}, Peter and {Guiglion}, Guillaume and {Heged{\H{u}}s}, Viola and {Ji}, Alex P. and {Kollmeier}, Juna A. and {Ness}, Melissa K. and {Otto}, Jonah and {Pinsonneault}, Marc H. and {Roman-Lopes}, Alexandre and {Saydjari}, Andrew and {Sinha}, Amaya and {Song}, Ying-Yi and {Stringfellow}, Guy S. and {Stassun}, Keivan G. and {Tayar}, Jamie and {Tkachenko}, Andrew and {Valentini}, Marica and {Way}, Zachary and {Weingrill}, J{\"o}rg},
        title = "{SDSS-V Milky Way Mapper (MWM): ASPCAP Stellar Parameters and Abundances in SDSS-V Data Release 19}",
      journal = {\aj},
         year = 2025,
        month = aug,
       volume = {170},
       number = {2},
          eid = {96},
        pages = {96},
          doi = {10.3847/1538-3881/ade4b9},
archivePrefix = {arXiv},
       eprint = {2506.07845},
 primaryClass = {astro-ph.SR},
       adsurl = {https://ui.adsabs.harvard.edu/abs/2025AJ....170...96M}
}

@ARTICLE{North04,
       author = {{North}, P. and {Zahn}, J.-P.},
        title = "{Circularization in B-type eclipsing binaries in both Magellanic Clouds}",
      journal = {\nar},
         year = 2004,
        month = jul,
       volume = {48},
       number = {9},
        pages = {741-748},
          doi = {10.1016/j.newar.2004.03.009},
       adsurl = {https://ui.adsabs.harvard.edu/abs/2004NewAR..48..741N}
}

@ARTICLE{Nguyen22,
       author = {{Nguyen}, C.~T. and {Costa}, G. and {Girardi}, L. and {Volpato}, G. and {Bressan}, A. and {Chen}, Y. and {Marigo}, P. and {Fu}, X. and {Goudfrooij}, P.},
        title = "{PARSEC V2.0: Stellar tracks and isochrones of low- and intermediate-mass stars with rotation}",
      journal = {\aap},
         year = 2022,
        month = sep,
       volume = {665},
          eid = {A126},
        pages = {A126},
          doi = {10.1051/0004-6361/202244166},
archivePrefix = {arXiv},
       eprint = {2207.08642},
 primaryClass = {astro-ph.SR},
       adsurl = {https://ui.adsabs.harvard.edu/abs/2022A&A...665A.126N}
}

@ARTICLE{Pan2025,
       author = {{Pan}, Yang and {Frasca}, Antonio and {Wang}, Jiaxin and {Fu}, Jianning and {Zhang}, Xiaobin},
        title = "{Analysis of Six Long-period Low-mass Eclipsing Binaries}",
      journal = {\aj},
         year = 2025,
        month = nov,
       volume = {170},
       number = {5},
          eid = {252},
        pages = {252},
          doi = {10.3847/1538-3881/ae04ed},
       adsurl = {https://ui.adsabs.harvard.edu/abs/2025AJ....170..252P}
}

@ARTICLE{Pavlovski18,
       author = {{Pavlovski}, K. and {Southworth}, J. and {Tamajo}, E.},
        title = "{Physical properties and CNO abundances for high-mass stars in four main-sequence detached eclipsing binaries: V478 Cyg, AH Cep, V453 Cyg, and V578 Mon}",
      journal = {\mnras},
         year = 2018,
        month = dec,
       volume = {481},
       number = {3},
        pages = {3129-3147},
          doi = {10.1093/mnras/sty2516},
archivePrefix = {arXiv},
       eprint = {1809.04061},
 primaryClass = {astro-ph.SR},
       adsurl = {https://ui.adsabs.harvard.edu/abs/2018MNRAS.481.3129P}
}

@ARTICLE{Pawar24,
       author = {{Pawar}, T. and {He{\l}miniak}, K.~G. and {Moharana}, A. and {Pawar}, G. and {Pyatnytskyy}, M. and {Lala}, H.~N. and {Konacki}, M.},
        title = "{A comprehensive study of five candidate {\ensuremath{\delta}} Scuti-type pulsators in detached eclipsing binaries}",
      journal = {\aap},
         year = 2024,
        month = nov,
       volume = {691},
          eid = {A101},
        pages = {A101},
          doi = {10.1051/0004-6361/202451126},
archivePrefix = {arXiv},
       eprint = {2410.00590},
 primaryClass = {astro-ph.SR},
       adsurl = {https://ui.adsabs.harvard.edu/abs/2024A&A...691A.101P}
}

@ARTICLE{Paxton13,
       author = {{Paxton}, Bill and {Cantiello}, Matteo and {Arras}, Phil and {Bildsten}, Lars and {Brown}, Edward F. and {Dotter}, Aaron and {Mankovich}, Christopher and {Montgomery}, M.~H. and {Stello}, Dennis and {Timmes}, F.~X. and {Townsend}, Richard},
        title = "{Modules for Experiments in Stellar Astrophysics (MESA): Planets, Oscillations, Rotation, and Massive Stars}",
      journal = {\apjs},
         year = 2013,
        month = sep,
       volume = {208},
       number = {1},
          eid = {4},
        pages = {4},
          doi = {10.1088/0067-0049/208/1/4},
archivePrefix = {arXiv},
       eprint = {1301.0319},
 primaryClass = {astro-ph.SR},
       adsurl = {https://ui.adsabs.harvard.edu/abs/2013ApJS..208....4P}
}

@ARTICLE{Pecaut13,
       author = {{Pecaut}, Mark J. and {Mamajek}, Eric E.},
        title = "{Intrinsic Colors, Temperatures, and Bolometric Corrections of Pre-main-sequence Stars}",
      journal = {\apjs},
         year = 2013,
        month = sep,
       volume = {208},
       number = {1},
          eid = {9},
        pages = {9},
          doi = {10.1088/0067-0049/208/1/9},
archivePrefix = {arXiv},
       eprint = {1307.2657},
 primaryClass = {astro-ph.SR},
       adsurl = {https://ui.adsabs.harvard.edu/abs/2013ApJS..208....9P}
}

@ARTICLE{Popper86,
       author = {{Popper}, Daniel M.},
        title = "{Spectroscopic orbits of two early-type eclipsing binaries : EG Serpentis and DN Ursae Majoris.}",
      journal = {\pasp},
         year = 1986,
        month = dec,
       volume = {98},
        pages = {1312-1316},
          doi = {10.1086/131935},
       adsurl = {https://ui.adsabs.harvard.edu/abs/1986PASP...98.1312P}
}

@ARTICLE{Popper87,
       author = {{Popper}, Daniel M.},
        title = "{A Pre--Main-Sequence Star in the Detached Binary EK Cephei}",
      journal = {\apjl},
         year = 1987,
        month = feb,
       volume = {313},
        pages = {L81},
          doi = {10.1086/184835},
       adsurl = {https://ui.adsabs.harvard.edu/abs/1987ApJ...313L..81P}
}

@ARTICLE{Ramirez07,
       author = {{Ram{\'\i}rez}, I. and {Allende Prieto}, C. and {Lambert}, D.~L.},
        title = "{Oxygen abundances in nearby stars. Clues to the formation and evolution of the Galactic disk}",
      journal = {\aap},
         year = 2007,
        month = apr,
       volume = {465},
       number = {1},
        pages = {271-289},
          doi = {10.1051/0004-6361:20066619},
       adsurl = {https://ui.adsabs.harvard.edu/abs/2007A&A...465..271R}
}

@INPROCEEDINGS{Richardson21,
       author = {{Richardson}, Noel D. and {Pablo}, Herbert and {Barclay}, Krister and {Beltran}, Marina and {Daly}, Andrea and {Elliott}, Ashley and {Lane}, Alexis and {Medina}, Emelito and {Pavao}, Clarissa and {Peatt}, Megan and {Perez}, Lauren and {Pickett}, Connor and {Williams}, Amara and {Rachford}, Brian},
        title = "{Binary Parameters for the Massive Eclipsing Binary CC Cassiopeiae}",
    booktitle = {Posters from the TESS Science Conference II (TSC2)},
         year = 2021,
        month = jul,
          eid = {110},
        pages = {110},
          doi = {10.5281/zenodo.5129293},
       adsurl = {https://ui.adsabs.harvard.edu/abs/2021tsc2.confE.110R}
}

@ARTICLE{Rowan23,
       author = {{Rowan}, D.~M. and {Jayasinghe}, T. and {Stanek}, K.~Z. and {Kochanek}, C.~S. and {Thompson}, Todd A. and {Shappee}, B.~J. and {Giles}, W.},
        title = "{The value-added catalogue of ASAS-SN eclipsing binaries - III. Masses and radii of Gaia spectroscopic binaries}",
      journal = {\mnras},
         year = 2023,
        month = aug,
       volume = {523},
       number = {2},
        pages = {2641-2650},
          doi = {10.1093/mnras/stad1560},
archivePrefix = {arXiv},
       eprint = {2212.02509},
 primaryClass = {astro-ph.SR},
       adsurl = {https://ui.adsabs.harvard.edu/abs/2023MNRAS.523.2641R}
}

@ARTICLE{Rowan25,
       author = {{Rowan}, Dominick M. and {Stanek}, Krzysztof Z. and {Kochanek}, Christopher S. and {Thompson}, Todd A. and {Jayasinghe}, Tharindu and {Blaum}, Jacqueline and {Fulton}, Benjamin J. and {Ilyin}, Ilya and {Isaacson}, Howard and {LeBaron}, Natalie and {Lu}, Jessica R. and {Martin}, David V.},
        title = "{Precise and Accurate Mass and Radius Measurements of Fifteen Galactic Red Giants in Detached Eclipsing Binaries}",
      journal = {The Open Journal of Astrophysics},
         year = 2025,
        month = feb,
       volume = {8},
          eid = {18},
        pages = {18},
          doi = {10.33232/001c.129962},
archivePrefix = {arXiv},
       eprint = {2409.02983},
 primaryClass = {astro-ph.SR},
       adsurl = {https://ui.adsabs.harvard.edu/abs/2025OJAp....8E..18R}
}

@ARTICLE{Rowan26,
       author = {{Rowan}, D.~M. and {Stanek}, K.~Z. and {Kochanek}, C.~S. and {Fulton}, Benjamin J. and {Ilyin}, Ilya and {Isaacson}, Howard and {Martin}, David V.},
        title = "{Precise and Accurate Mass and Radius Measurements of Thirty Galactic Metal-Poor Stars in Detached Eclipsing Binaries}",
      journal = {arXiv e-prints},
         year = 2026,
        month = sep,
          eid = {arXiv:2609.19248},
        pages = {arXiv:2609.19248},
          doi = {10.48550/arXiv.2609.19248},
archivePrefix = {arXiv},
       eprint = {2609.19248},
 primaryClass = {astro-ph.SR},
       adsurl = {https://ui.adsabs.harvard.edu/abs/2026arXiv260919248R}
}

@ARTICLE{Rozyczka22,
       author = {{Rozyczka}, M. and {Thompson}, I.~B. and {Dotter}, A. and {Mazur}, B. and {Narloch}, W. and {Pych}, W. and {He{\l}miniak}, K.~G. and {Moharana}, A. and {Beletsky}, Y. and {Burley}, G.~S. and {Marshall}, J.~L. and {Morrell}, N. and {Osip}, D. and {Shectman}, S.~A. and {Bernstein}, R. and {Pilecki}, B. and {Zgirski}, B.},
        title = "{The Cluster Ages Experiment (CASE) - IX. Analysis of four detached eclipsing binaries in the globular cluster NGC 3201}",
      journal = {\mnras},
         year = 2022,
        month = dec,
       volume = {517},
       number = {2},
        pages = {2485-2501},
          doi = {10.1093/mnras/stac2751},
       adsurl = {https://ui.adsabs.harvard.edu/abs/2022MNRAS.517.2485R}
}

@ARTICLE{Seager03,
       author = {{Seager}, S. and {Mall{\'e}n-Ornelas}, G.},
        title = "{A Unique Solution of Planet and Star Parameters from an Extrasolar Planet Transit Light Curve}",
      journal = {\apj},
         year = 2003,
        month = mar,
       volume = {585},
       number = {2},
        pages = {1038-1055},
          doi = {10.1086/346105},
archivePrefix = {arXiv},
       eprint = {astro-ph/0206228},
 primaryClass = {astro-ph},
       adsurl = {https://ui.adsabs.harvard.edu/abs/2003ApJ...585.1038S}
}

@ARTICLE{Serenelli21,
       author = {{Serenelli}, Aldo and {Weiss}, Achim and {Aerts}, Conny and {Angelou}, George C. and {Baroch}, David and {Bastian}, Nate and {Beck}, Paul G. and {Bergemann}, Maria and {Bestenlehner}, Joachim M. and {Czekala}, Ian and {Elias-Rosa}, Nancy and {Escorza}, Ana and {Van Eylen}, Vincent and {Feuillet}, Diane K. and {Gandolfi}, Davide and {Gieles}, Mark and {Girardi}, L{\'e}o and {Lebreton}, Yveline and {Lodieu}, Nicolas and {Martig}, Marie and {Miller Bertolami}, Marcelo M. and {Mombarg}, Joey S.~G. and {Morales}, Juan Carlos and {Moya}, Andr{\'e}s and {Nsamba}, Benard and {Pavlovski}, Kre{\v{s}}imir and {Pedersen}, May G. and {Ribas}, Ignasi and {Schneider}, Fabian R.~N. and {Silva Aguirre}, Victor and {Stassun}, Keivan G. and {Tolstoy}, Eline and {Tremblay}, Pier-Emmanuel and {Zwintz}, Konstanze},
        title = "{Weighing stars from birth to death: mass determination methods across the HRD}",
      journal = {\aapr},
         year = 2021,
        month = dec,
       volume = {29},
       number = {1},
          eid = {4},
        pages = {4},
          doi = {10.1007/s00159-021-00132-9},
archivePrefix = {arXiv},
       eprint = {2006.10868},
 primaryClass = {astro-ph.SR},
       adsurl = {https://ui.adsabs.harvard.edu/abs/2021A&ARv..29....4S}
}

@ARTICLE{Shappee14,
       author = {{Shappee}, B.~J. and {Prieto}, J.~L. and {Grupe}, D. and {Kochanek}, C.~S. and {Stanek}, K.~Z. and {De Rosa}, G. and {Mathur}, S. and {Zu}, Y. and {Peterson}, B.~M. and {Pogge}, R.~W. and {Komossa}, S. and {Im}, M. and {Jencson}, J. and {Holoien}, T.~W. -S. and {Basu}, U. and {Beacom}, J.~F. and {Szczygie{\l}}, D.~M. and {Brimacombe}, J. and {Adams}, S. and {Campillay}, A. and {Choi}, C. and {Contreras}, C. and {Dietrich}, M. and {Dubberley}, M. and {Elphick}, M. and {Foale}, S. and {Giustini}, M. and {Gonzalez}, C. and {Hawkins}, E. and {Howell}, D.~A. and {Hsiao}, E.~Y. and {Koss}, M. and {Leighly}, K.~M. and {Morrell}, N. and {Mudd}, D. and {Mullins}, D. and {Nugent}, J.~M. and {Parrent}, J. and {Phillips}, M.~M. and {Pojmanski}, G. and {Rosing}, W. and {Ross}, R. and {Sand}, D. and {Terndrup}, D.~M. and {Valenti}, S. and {Walker}, Z. and {Yoon}, Y.},
        title = "{The Man behind the Curtain: X-Rays Drive the UV through NIR Variability in the 2013 Active Galactic Nucleus Outburst in NGC 2617}",
      journal = {\apj},
         year = 2014,
        month = jun,
       volume = {788},
       number = {1},
          eid = {48},
        pages = {48},
          doi = {10.1088/0004-637X/788/1/48},
archivePrefix = {arXiv},
       eprint = {1310.2241},
 primaryClass = {astro-ph.HE},
       adsurl = {https://ui.adsabs.harvard.edu/abs/2014ApJ...788...48S}
}

@INPROCEEDINGS{Southworth15,
       author = {{Southworth}, J.},
        title = "{DEBCat: A Catalog of Detached Eclipsing Binary Stars}",
    booktitle = {Living Together: Planets, Host Stars and Binaries},
         year = 2015,
       editor = {{Rucinski}, S.~M. and {Torres}, G. and {Zejda}, M.},
       series = {Astronomical Society of the Pacific Conference Series},
       volume = {496},
        month = jul,
        pages = {164},
          doi = {10.48550/arXiv.1411.1219},
archivePrefix = {arXiv},
       eprint = {1411.1219},
 primaryClass = {astro-ph.SR},
       adsurl = {https://ui.adsabs.harvard.edu/abs/2015ASPC..496..164S}
}

@ARTICLE{Southworth22,
       author = {{Southworth}, John and {Bowman}, Dominic M.},
        title = "{High-mass pulsators in eclipsing binaries observed using TESS}",
      journal = {\mnras},
         year = 2022,
        month = jul,
       volume = {513},
       number = {3},
        pages = {3191-3209},
          doi = {10.1093/mnras/stac875},
archivePrefix = {arXiv},
       eprint = {2203.15365},
 primaryClass = {astro-ph.SR},
       adsurl = {https://ui.adsabs.harvard.edu/abs/2022MNRAS.513.3191S}
}

@ARTICLE{Southworth24,
       author = {{Southworth}, John},
        title = "{Rediscussion of Eclipsing Binaries. Paper 16: The {\ensuremath{\delta}}Scuti/{\ensuremath{\gamma}}Doradus Hybrid Pulsator GK Draconis}",
      journal = {The Observatory},
         year = 2024,
        month = feb,
       volume = {144},
        pages = {24-36},
          doi = {10.48550/arXiv.2401.13396},
archivePrefix = {arXiv},
       eprint = {2401.13396},
 primaryClass = {astro-ph.SR},
       adsurl = {https://ui.adsabs.harvard.edu/abs/2024Obs...144...24S}
}

@ARTICLE{Speagle25,
       author = {{Speagle}, Joshua S. and {Zucker}, Catherine and {Beane}, Angus and {Cargile}, Phillip A. and {Dotter}, Aaron and {Finkbeiner}, Douglas P. and {Green}, Gregory M. and {Johnson}, Benjamin D. and {Schlafly}, Edward F. and {Bonaca}, Ana and {Conroy}, Charlie and {Eadie}, Gwendolyn and {Eisenstein}, Daniel J. and {Goodman}, Alyssa A. and {Han}, Jiwon Jesse and {Kamdar}, Harshil M. and {Naidu}, Rohan and {Rix}, Hans-Walter and {Saydjari}, Andrew K. and {Ting}, Yuan-Sen and {Zelko}, Ioana A.},
        title = "{Deriving Stellar Properties, Distances, and Reddenings using Photometry and Astrometry with BRUTUS}",
      journal = {arXiv e-prints},
         year = 2025,
        month = mar,
          eid = {arXiv:2503.02227},
        pages = {arXiv:2503.02227},
          doi = {10.48550/arXiv.2503.02227},
archivePrefix = {arXiv},
       eprint = {2503.02227},
 primaryClass = {astro-ph.SR},
       adsurl = {https://ui.adsabs.harvard.edu/abs/2025arXiv250302227S}
}

@ARTICLE{Stassun04,
       author = {{Stassun}, Keivan G. and {Mathieu}, Robert D. and {Vaz}, Luiz Paulo R. and {Stroud}, Nicholas and {Vrba}, Frederick J.},
        title = "{Dynamical Mass Constraints on Low-Mass Pre-Main-Sequence Stellar Evolutionary Tracks: An Eclipsing Binary in Orion with a 1.0 M$_{solar}$ Primary and a 0.7 M$_{solar}$ Secondary}",
      journal = {\apjs},
         year = 2004,
        month = apr,
       volume = {151},
       number = {2},
        pages = {357-385},
          doi = {10.1086/382353},
archivePrefix = {arXiv},
       eprint = {astro-ph/0312575},
 primaryClass = {astro-ph},
       adsurl = {https://ui.adsabs.harvard.edu/abs/2004ApJS..151..357S}
}

@ARTICLE{Stempels08,
       author = {{Stempels}, H.~C. and {Hebb}, L. and {Stassun}, K.~G. and {Holtzman}, J. and {Dunstone}, N. and {Glowienka}, L. and {Frandsen}, S.},
        title = "{The pre-main-sequence eclipsing binary ASAS J052821+0338.5}",
      journal = {\aap},
         year = 2008,
        month = apr,
       volume = {481},
       number = {3},
        pages = {747-755},
          doi = {10.1051/0004-6361:20078585},
archivePrefix = {arXiv},
       eprint = {0801.3927},
 primaryClass = {astro-ph},
       adsurl = {https://ui.adsabs.harvard.edu/abs/2008A&A...481..747S}
}

@ARTICLE{Stassun22,
       author = {{Stassun}, Keivan G. and {Torres}, Guillermo and {Kounkel}, Marina and {Feliz}, Dax L. and {Bouma}, Luke G. and {Howell}, Steve B. and {Gnilka}, Crystal L. and {Furlan}, E.},
        title = "{A Low-mass Pre-main-sequence Eclipsing Binary in Lower Centaurus Crux Discovered with TESS}",
      journal = {\apj},
         year = 2022,
        month = dec,
       volume = {941},
       number = {2},
          eid = {125},
        pages = {125},
          doi = {10.3847/1538-4357/aca32e},
archivePrefix = {arXiv},
       eprint = {2211.07899},
 primaryClass = {astro-ph.SR},
       adsurl = {https://ui.adsabs.harvard.edu/abs/2022ApJ...941..125S}
}

@ARTICLE{Swayne22,
       author = {{Swayne}, M.~I. and {Maxted}, P.~F.~L. and {Triaud}, A.~H.~M.~J. and {Sousa}, S.~G. and {Deline}, A. and {Ehrenreich}, D. and {Hoyer}, S. and {Olofsson}, G. and {Boisse}, I. and {Duck}, A. and {Gill}, S. and {Martin}, D. and {McCormac}, J. and {Persson}, C.~M. and {Santerne}, A. and {Sebastian}, D. and {Standing}, M.~R. and {Acu{\~n}a}, L. and {Alibert}, Y. and {Alonso}, R. and {Anglada}, G. and {B{\'a}rczy}, T. and {Barrado Navascues}, D. and {Barros}, S.~C.~C. and {Baumjohann}, W. and {Baycroft}, T.~A. and {Beck}, M. and {Beck}, T. and {Benz}, W. and {Billot}, N. and {Bonfils}, X. and {Borsato}, L. and {Bourrier}, V. and {Brandeker}, A. and {Broeg}, C. and {Carmona}, A. and {Charnoz}, S. and {Collier Cameron}, A. and {Cort{\'e}s-Zuleta}, P. and {Csizmadia}, Sz and {Cubillos}, P.~E. and {Davies}, M.~B. and {Deleuil}, M. and {Delfosse}, X. and {Delrez}, L. and {Demangeon}, O.~D.~S. and {Demory}, B.-O. and {Dransfield}, G. and {Erikson}, A. and {Fortier}, A. and {Forveille}, T. and {Fossati}, L. and {Fridlund}, M. and {Gandolfi}, D. and {Gillon}, M. and {G{\"u}del}, M. and {G{\"u}nther}, M.~N. and {Hara}, N. and {H{\'e}brard}, G. and {Heidari}, N. and {Hellier}, C. and {Helling}, Ch and {Isaak}, K.~G. and {Kerschbaum}, F. and {Kiefer}, F. and {Kiss}, L.~L. and {Kunovac}, V. and {Lalitha}, S. and {Lam}, K.~W.~F. and {Laskar}, J. and {Lecavelier des Etangs}, A. and {Lendl}, M. and {Magrin}, D. and {Marafatto}, L. and {Martioli}, E. and {Miller}, N.~J. and {Mordasini}, C. and {Moutou}, C. and {Nascimbeni}, V. and {Ottensamer}, R. and {Pagano}, I. and {Pall{\'e}}, E. and {Peter}, G. and {Piazza}, D. and {Piotto}, G. and {Pollacco}, D. and {Queloz}, D. and {Ragazzoni}, R. and {Rando}, N. and {Rauer}, H. and {Ribas}, I. and {Santos}, N.~C. and {Scandariato}, G. and {S{\'e}gransan}, D. and {Simon}, A.~E. and {Smith}, A.~M.~S. and {Southworth}, R. and {Stalport}, M. and {Szab{\'o}}, Gy M. and {Thomas}, N. and {Udry}, S. and {Ulmer}, B. and {Van Grootel}, V. and {Venturini}, J. and {Walton}, N.~A. and {Willett}, E. and {Wilson}, T.~G.},
        title = "{The EBLM Project- XI. Mass, radius, and effective temperature measurements for 23 M-dwarf companions to solar-type stars observed with CHEOPS}",
      journal = {\mnras},
         year = 2024,
        month = mar,
       volume = {528},
       number = {4},
        pages = {5703-5722},
          doi = {10.1093/mnras/stad3866},
archivePrefix = {arXiv},
       eprint = {2312.11339},
 primaryClass = {astro-ph.SR},
       adsurl = {https://ui.adsabs.harvard.edu/abs/2024MNRAS.528.5703S}
}

@ARTICLE{Taormina25,
       author = {{Taormina}, M{\'o}nica and {Pietrzy{\'n}ski}, G. and {Kudritzki}, R.-P. and {Pilecki}, B. and {Thompson}, I.~B. and {Puls}, J. and {G{\'o}rski}, M. and {Zgirski}, B. and {Graczyk}, D. and {Urbaneja}, M.~A. and {Gieren}, W. and {Narloch}, W. and {Hajdu}, G.},
        title = "{Toward Early-type Eclipsing Binaries as Extragalactic Milestones. IV. Physical Properties of Three Detached O/B-type Systems in the LMC}",
      journal = {\apj},
         year = 2025,
        month = feb,
       volume = {980},
       number = {2},
          eid = {178},
        pages = {178},
          doi = {10.3847/1538-4357/ad9dda},
archivePrefix = {arXiv},
       eprint = {2501.04080},
 primaryClass = {astro-ph.SR},
       adsurl = {https://ui.adsabs.harvard.edu/abs/2025ApJ...980..178T}
}

@ARTICLE{Thurman23,
       author = {{Thurman}, Glenn M.},
        title = "{A photoelectric study of the eclipsing binary EG Serpentis}",
      journal = {\mnras},
         year = 2023,
        month = jun,
       volume = {522},
       number = {1},
        pages = {1310-1322},
          doi = {10.1093/mnras/stad787},
       adsurl = {https://ui.adsabs.harvard.edu/abs/2023MNRAS.522.1310T}
}

@ARTICLE{Torres10,
       author = {{Torres}, G. and {Andersen}, J. and {Gim{\'e}nez}, A.},
        title = "{Accurate masses and radii of normal stars: modern results and applications}",
      journal = {\aapr},
         year = 2010,
        month = feb,
       volume = {18},
       number = {1-2},
        pages = {67-126},
          doi = {10.1007/s00159-009-0025-1},
archivePrefix = {arXiv},
       eprint = {0908.2624},
 primaryClass = {astro-ph.SR},
       adsurl = {https://ui.adsabs.harvard.edu/abs/2010A&ARv..18...67T}
}

@ARTICLE{Torres12,
       author = {{Torres}, G. and {Clausen}, J.~V. and {Bruntt}, H. and {Claret}, A. and {Andersen}, J. and {Nordstr{\"o}m}, B. and {Stefanik}, R.~P. and {Latham}, D.~W.},
        title = "{Absolute dimensions of eclipsing binaries. XXIX. The Am-type systems SW Canis Majoris and HW Canis Majoris}",
      journal = {\aap},
         year = 2012,
        month = jan,
       volume = {537},
          eid = {A117},
        pages = {A117},
          doi = {10.1051/0004-6361/201117795},
archivePrefix = {arXiv},
       eprint = {1112.3974},
 primaryClass = {astro-ph.SR},
       adsurl = {https://ui.adsabs.harvard.edu/abs/2012A&A...537A.117T}
}

@ARTICLE{Torres17,
       author = {{Torres}, Guillermo and {McGruder}, Chima D. and {Siverd}, Robert J. and {Rodriguez}, Joseph E. and {Pepper}, Joshua and {Stevens}, Daniel J. and {Stassun}, Keivan G. and {Lund}, Michael B. and {James}, David},
        title = "{Absolute Dimensions of the Eccentric Eclipsing Binary V541 Cygni}",
      journal = {\apj},
         year = 2017,
        month = feb,
       volume = {836},
       number = {2},
          eid = {177},
        pages = {177},
          doi = {10.3847/1538-4357/836/2/177},
archivePrefix = {arXiv},
       eprint = {1612.02441},
 primaryClass = {astro-ph.SR},
       adsurl = {https://ui.adsabs.harvard.edu/abs/2017ApJ...836..177T}
}

@ARTICLE{Woody26,
       author = {{Woody}, Rebecca and {Conroy}, Charlie and {Cargile}, Phillip A. and {Dotter}, Aaron},
        title = "{Classical Cases of Non-solar Mixing Length in Low Mass Stars}",
      journal = {arXiv e-prints},
         year = 2026,
        month = sep,
          eid = {arXiv:2609.13384},
        pages = {arXiv:2609.13384},
          doi = {10.48550/arXiv.2609.13384},
archivePrefix = {arXiv},
       eprint = {2609.13384},
 primaryClass = {astro-ph.SR},
       adsurl = {https://ui.adsabs.harvard.edu/abs/2026arXiv260913384W}
}

@ARTICLE{Yakut25,
       author = {{Yakut}, Kadri and {Kalomeni}, Belinda and {Rappaport}, Saul},
        title = "{A novel approach to determining the age of stellar clusters: application to NGC 188}",
      journal = {\mnras},
         year = 2025,
        month = sep,
       volume = {542},
       number = {3},
        pages = {1817-1826},
          doi = {10.1093/mnras/staf1332},
archivePrefix = {arXiv},
       eprint = {2508.09255},
 primaryClass = {astro-ph.SR},
       adsurl = {https://ui.adsabs.harvard.edu/abs/2025MNRAS.542.1817Y}
}

@INCOLLECTION{Winn2010,
       author = {{Winn}, J.~N.},
        title = "{Exoplanet Transits and Occultations}",
    booktitle = {Exoplanets},
         year = 2010,
       editor = {{Seager}, S.},
        pages = {55-77},
          doi = {10.48550/arXiv.1001.2010},
       adsurl = {https://ui.adsabs.harvard.edu/abs/2010exop.book...55W}
}
\bibliographystyle{aasjournalv7}

%% This command is needed to show the entire author+affiliation list when
%% the collaboration and author truncation commands are used.  It has to
%% go at the end of the manuscript.
%\allauthors

%% Include this line if you are using the \added, \replaced, \deleted
%% commands to see a summary list of all changes at the end of the article.
%\listofchanges

\end{document}